\documentclass[a4paper,fleqn,usenatbib]{mnras}

\usepackage[T1]{fontenc}
\usepackage{ae,aecompl}
\usepackage[thinc]{esdiff}
\usepackage{chemformula}
\usepackage[utf8]{inputenc}

\usepackage{multirow}
\usepackage{hyperref}

\usepackage[normalem]{ulem}

\usepackage{graphicx}	
\usepackage{amsmath}	
\usepackage{amssymb}	

\usepackage{isotope}
\usepackage{chemformula}
\usepackage{siunitx}
\usepackage{wrapfig,framed}
\usepackage{outlines}

\usepackage{natbib}
\defcitealias{Booth_disint22}{Booth22}

\title[ ]{The evolution of catastrophically evaporating rocky planets}

\author[]{Alfred Curry$^{1}$, Richard Booth$^{2,1}$, James E. Owen$^{1}$, Subhanjoy Mohanty$^{1}$
\\
$^{1}$Astrophysics Group, Department of Physics, Imperial College London, Prince Consort Rd, London SW7 2AZ, UK\\
$^{2}${School of Physics and Astronomy, University of Leeds, Leeds, LS2 9JT, UK}}

\date{Accepted XXX. Received YYY; in original form ZZZ}

\pubyear{2022}

\begin{document}
\label{firstpage}
\pagerange{\pageref{firstpage}--\pageref{lastpage}}
\maketitle

\begin{abstract}
In this work, we develop a rocky planet interior model and use it to investigate the evolution of catastrophically evaporating rocky exoplanets. These planets, detected through the dust tails produced by evaporative outflows from their molten surfaces, can be entirely destroyed in a fraction of their host star's lifetime. To allow for the major decrease in mass, our interior model can simultaneously calculate the evolution of the pressure and density structure of a planet alongside its thermal evolution, which includes the effects of conduction, convection and partial melting. We first use this model to show that the underlying planets are likely to be almost entirely solid. This means that the dusty tails are made up of material sampled only from a thin dayside lava pool. If one wishes to infer the bulk compositions of rocky exoplanets from their dust tails, it is important to take the localised origin of this material into account. Secondly, by considering how frequently one should be able to detect mass loss from these systems, we investigate the occurrence of sub-Earth mass exoplanets, which is difficult with conventional planet detection surveys. We predict that, depending on model assumptions, the number of progenitors of the catastrophically evaporating planets is either in line with, or higher than, the observed population of close-in (substellar temperatures around 2200~K) terrestrial exoplanets.

\end{abstract}

\begin{keywords}
exoplanets - planets and satellites: interiors - planets and satellites: physical evolution - planets and satellites: composition
\end{keywords}
\section{Introduction}

Understanding rocky planet interiors is an essential part of exoplanet science. Firstly, their compositions lend insight into planet formation. Secondly, the interior chemistry determines the atmospheric chemistry through outgassing for planets below a few Earth masses that do not possess a \ch{H2}/He envelope. Atmospheric chemistry is influenced both through exchange with the early molten surface \citep[e.g.,][]{Abe1986,GAIL2014,Licht21verticalmagma}, and across the lifetime of the planet through volcanism \citep[e.g.,][]{Kite09,Noack2014,Tosi2017}. Furthermore, the thermal evolution of the interior affects the efficiency of outgassing processes and what species get outgassed, as well as providing thermal input into the base of the atmosphere. Thus, the interior is important to explain the range of possible exoplanetary atmospheres, including habitability. 

Chemical compositions of exoplanet interiors are most directly observed in the atmospheres of polluted white dwarfs \citep{Zuckerman2010}, which show metal lines due to planetary bodies that have recently been accreted and have been shown to have compositions that are consistent with Solar System rock \citep[e.g.,][]{DZDwarfs1,Harrison2021}. However, as these systems can only track the bulk composition of the pollutant, little information about the compositional structure or influence on the planet's atmosphere can be inferred. Another avenue of interest is the study of planets' atmospheres, particularly those of hot planets where the atmospheric composition is likely dominated by interaction with the surface and interior \citep[e.g.,][]{Zilinkas22}. However, the atmospheric interactions are still not fully understood and are complicated by atmospheric processes such as loss and photochemistry, even if such tenuous atmospheres become observable. The catastrophically evaporating planets, however, are less dependent on such atmospheric processes because they are observed directly through solid material.

Three catastrophically evaporating planets -- Kepler 1520b, \citep[formerly KIC1255b,][]{KIC1255-discov}, KOI-2700b \citep{KOI2700b-discov} and K2-22b \citep{K2-22b-discov} -- were discovered by the {\it Kepler}/{\it K2} missions \citep{KEPLER,K2-mission}. The distinguishing features of the lightcurves of these systems are (i) all three systems have highly variable transit depths indicating that the orbiting body is not a single solid object. (ii) there is an increase in flux directly before transit, which can be explained by the forward scattering of starlight by dust (iii) Kepler 1520b and KOI-2700b show a highly asymmetric averaged transit due to an extended tail \citep[see e.g.,][Figure 1]{vanLieshout16}. These features are well explained by the transit curves being due to tails of dusty material \citep{disint18}. 

The short periods of the dust tails imply high surface temperatures for any planet at that orbital distance (see \autoref{tab: systems}), sufficient to melt rocky materials. Therefore the origin of the dusty tails is thought to be the evaporation of rocky material from the molten surface of an underlying planet, which then expands and condenses as dust further from the surface. This idea was originally proposed by \citet{KIC1255-discov}, with further physical modelling by \citet{Perez-Becker13}. More recent models of the outflows have focused on photochemistry \citep{Ito-Ikoma21}, day to nightside flows (\citealt{Kang2021}, see also \citealt{castan-menou11}) and variability \citep{Bromely-Chiang23}. \citet{Booth_disint22} were notably able to show that dust can form self-consistently during this atmospheric escape under the right conditions. 

Since the material in the dusty tails comes directly from the solid portion of the planet, these systems are particularly interesting in the wider context of exoplanets because they may allow the study of interior composition. Thus far, attempts to constrain the composition of tail material have depended on modelling the light curves. For instance, \citet{vanLieshout14,vanLieshout16} found that the lightcurve of Kepler-1520b is consistent with corundum (\ch{Al3O2}) and KOI-2700b with corundum or fayalite (\ch{Fe2SiO4}), but both are inconsistent with pure iron or carbon, based on the species' sublimation rates. More recent modelling efforts \citep{CamposEstrada23} suggest that the tails are best explained by iron-rich silicate dust.
In addition, the size of the forward scattering peak can be used to infer grain sizes, which are required to break degeneracies with composition in modelling, and are constrained to a range of 0.1 - \SI{1}{\micro\metre} \citep{Budaj13}.

Further, multi-wavelength observations will allow inferences to become more detailed. {\it JWST} infra-red spectra may allow silicate minerals to be identified through their resonant features around 10$\mu$m \citep{Bodman18-JWST}. It may also be possible to detect atomic lines from the gas in the tails \citep{disint18}, although attempts to find atomic species in the tail of K2-22b have so far been inconclusive \citep{Ridden-Harper2019}.

As has been noted in previous works \citep[e.g.,][]{Perez-Becker13}, and as we will also show, the high mass loss rates inferred for these planets means that they can be entirely destroyed within several Gyrs. Models of the mass loss process agree that the mass loss rates should generally increase with decreasing mass, leading to an accelerating mass loss before total destruction. A consequence of this is that the observed systems must be within a region, of temperature, mass and time, where their mass loss rates are high enough to be observed, but the planet has not yet been destroyed. Therefore, the fact that any are observed at all may tell us about the occurrence rate of these planets' progenitors. This is of particular significance because,as we will show \citep[see also][]{Perez-Becker13}, these planets must have low current and initial masses ($\lesssim 0.3 M_\oplus$). Therefore, they are part of a distribution not observed by conventional detection techniques \citep[e.g.,][]{Christiansen2014}.

In order to fully understand the catastrophically evaporating planets, both in their own right and to make full use of their potential as probes of interior physics, it is necessary to have a model of their interiors. The evolution of the underlying rocky planet may affect the thermal state at the bottom of the mass outflow. Additionally, and more significantly, the interior evolution sets the surface composition, which can be probed observationally through the dusty tails. Different compositions, for instance, through different depths that melting reaches, can vastly affect the chemicals that would be present in the outflows \citep[e.g.,][]{schaefer_fegley2009}.

In this work, we present our evolutionary model for the interior of catastrophically evaporating planets. In particular, we investigate the evolution of melt, since the molten regions convect over much shorter timescales than solid regions, and so have an important influence on composition by circulating chemicals to the surface. We then use this model to investigate the population of the progenitors of the catastrophically evaporating planets. 

In \S\ref{sec: Method}, we describe our 1D numerical model for rocky interiors and its specific application to these systems. In \S\ref{sec: results}, we present the results of our thermal evolution calculations, with a discussion of their consequences and limitations in \S\ref{sec: discuss}. In \S\ref{sec: occurrence}, we investigate the implications for the the occurrence rate of low-mass planets. We conclude with a summary in \S\ref{sec: conclusion}.

\section{Interior Model} \label{sec: Method}
We have developed a 1D code for modelling the evolution of a rocky planet with a time-varying mass, which we shall apply to the evolution of catastrophically evaporating planets. The numerical scheme is based on stellar structure codes, as described in \citet{Boden} and \citet{Kippen}. We solve for the structure of the planet's rocky mantle and iron core. Here we summarise the physics included and its basic operation.

\subsection{Basic Equations} \label{sec: struct_eqns}
The essential equations for the internal structure of a spherical body, written with mass, $m$, as the independent variable, are:
\begin{enumerate}
    \item Hydrostatic Equilibrium
        \begin{equation}
           \diff{P}{m} = - \frac{Gm}{4 \pi r^4} \label{eq: hydrostat} 
        \end{equation}
   where $P$ is the pressure, $r$ the radius at the mass point $m$ and $G$ the gravitational constant.
    \item Mass conservation
        \begin{equation}
            \diff{r}{m} = \frac{1}{4 \pi r^2 \rho} \label{eq: masscon}
        \end{equation}
        where $\rho$ is the density.
    \item Energy conservation
        \begin{equation}
            \diff{L}{m} = H -T\diff{S}{t} = H -C_p \diff{T}{t} +\frac{\delta}{\rho} \diff{P}{t} \label{eq: lumin}
    \end{equation}
    where$L$ is the luminosity, and $H$ is the heat generation rate per unit mass (through radioactive decay in the case of our planets). $t$ is time, $T$ is the temperature and $S$ is the specific entropy, meaning $T\mathrm{d}S/\mathrm{d}t$ is the rate of heat exchange from a unit mass. The second equality follows from thermodynamic relations. $C_P$ is the specific heat capacity at constant pressure, and 
\begin{equation}
    \delta \equiv \left. -\diffp{{\ln{\rho}}}{{\ln{T}}}\right|_{P}
\end{equation} is a measure of thermal expansivity. 
\end{enumerate}
One also needs an equation for the temperature gradient, which will be related to heat transport. Following the conventions in stellar interiors, we define the temperature gradient
\begin{equation}
    \nabla \equiv \diff{\ln{T}}{\ln{P}} \label{eq: nabladef}
\end{equation}
and use hydrostatic equilibrium (\autoref{eq: hydrostat}) to write:
\begin{equation}
    \diff{T}{m} = - \frac{GmT}{4 \pi r^4 P} \nabla \label{eq: Tmnabla}\; .
\end{equation}
We will show in \S\ref{sec: heat_flow} how we determine  $\nabla$ from consideration of the heat transport.

Equations \ref{eq: hydrostat}--\ref{eq: lumin} and \autoref{eq: Tmnabla} form a closed system of differential equations when combined with an equation of state and other material properties, which may be functions of temperature and pressure (\S\ref{sec: physical properties}). The only other information required to solve for the four dependent variables ($P$, $r$, $L$ and $T$) as functions of $m$ is four boundary conditions. The two inner boundary conditions are simply $R=0$ and $L=0$ at $m=0$. At the outer boundary, we use a fixed $P=P_0$ and a relation between the outer temperature $T_0$ and the other outer properties. We explain these outer boundary conditions for the cases we investigate here in \S\ref{sec: BCs}.

In order to solve this system of equations, we follow a Henyey scheme, the details of which can be found in \citet{Boden}, chapter 5, for example. In summary, we set up a mass grid such that the differential equations become a series of simultaneous equations. We solve these through Newton's method, up to a given tolerance threshold. The tolerance criterion we use is that the difference in a dependent variable across a mass grid cell must have a fractional accuracy of $10^{-6}$ or better. The time dependence in the energy equation is solved implicitly.

\subsection{Melting}\label{sec: melting}
Rocks are chemical mixtures and thus can be partially molten even in thermodynamic equilibrium at fixed temperature and pressure. This process is complicated because the melt composition will generally differ from that of the solid rock, and will also depend strongly on both the local conditions and the overall composition of the rock. To simplify the full problem, we introduce a parameter for the mass fraction of melt, $\phi$, which is simply a function of $P$ and $T$. 

Following other works \citep[e.g.,][]{abe1993thermal}, we use the simple linear function
\begin{equation}
    \phi = \frac{T - T_\text{sol}(P)}{T_\text{liq}(P) - T_\text{sol}(P)} \label{eq: melt_frac}
\end{equation}
where $T_\text{sol}$ is the solidus, the temperature below which the material is completely solid (i.e., $\phi$=0 for $T$$<$$T_{sol}$), and $T_\text{liq}$ is the liquidus, the temperature above which the material is completely liquid (i.e., $\phi$=1 for $T>T_{liq}$); both $T_{sol}$ and $T_{liq}$ are functions of $P$.Following e.g., \citet{abe1993thermal,Bower17}, we consider these fixed functions, but again, the reality is far more complex and highly composition-dependent.

For the mantle solidus and liquidus profiles, we use the fit to the Simon and Glatzel equation \citep{Simon-Glatzel} from \citet{Andrault11} for high pressures, and the curves in \citet{LITASOV2002} for low pressures. 
For use in our code, we tabulate these functions and access values using cubic Hermite interpolation.

\subsection{Heat flow in the mantle}\label{sec: heat_flow}
We include heat flow in our model by finding how the temperature gradient $\nabla$ depends on the energy flux. In practice, we find how the energy flux, $F = L/(4\pi r^2)$, depends on the temperature gradient and invert this function numerically.\footnote{Specifically, we use the \texttt{TOMS 748} algorithm \citep{toms748}.}

\subsubsection{Conduction and Convection}
Heat transport occurs in rocky interiors through conduction and convection. We model conduction using Fick's law, so conductive heat flux is given by:
\begin{equation}
    F_\text{cond} = - k\diff{T}{r} \label{eq: conduction}
\end{equation}
assuming a constant conductivity, $k$.

Convection is extremely important for the evolution of rocky planets, both in the liquid and solid phases. To model convection we use mixing length theory \citep[see e.g.][]{abe1995basic,Kippen} and use the equation:
\begin{equation}
    F_\text{conv} = - \rho l u C_P \left( \nabla - \nabla_\text{Ad} \right) \frac{T}{P} \diff{P}{r} \label{eq: Fconv}
\end{equation}
where $l$ is the mixing length, $u$ is the speed of convection and $\nabla_\text{Ad}$ is the adiabatic, logarithmic temperature-pressure gradient (see \autoref{eq: nabladef}). 

The total flux is given by
\begin{equation}
    F = F_\text{conv} + F_\text{cond} \label{eq: totalF} \; .
\end{equation}

The mixing length prescription requires an estimate of the convective velocity $u$. To find the velocity, we consider the forces $ \mathcal{F}_i$ acting on a parcel of the material moving due to convection. If the parcel is moving at a constant speed, the buoyancy force must be balanced by any drag forces. The drag forces are ram pressure, which is most important in the low viscosity limit, and viscous drag, which is more important in the high viscosity limit. These three forces may be given by the following formulae:
\begin{subequations}
  \begin{alignat}{5}
  &\mathcal{F}_\text{buoy} &=& \; V \frac{-\delta g \rho l}{P}\left( \nabla - \nabla_\text{Ad} \right) \diff{P}{r} \\
     &\mathcal{F}_\text{RAM} \; &=& \;  \rho u^2 A \\
     &\mathcal{F}_\text{visc} &=& \; 6\pi \nu\rho R u
    \end{alignat}
\end{subequations}
where $g$ is the gravitational acceleration, $\nu \equiv \eta/\rho$ is the  kinematic viscosity, and $\eta$ is the dynamic viscosity.  Here $R$, $A$ and $V$ are the fluid parcel's radius, cross-sectional area and volume, respectively. 

Combining these results in a quadratic equation for $u$, the (positive) solution to which is
\begin{equation}
	u = \left( -1 + \sqrt{1- \frac{A V \delta g l \left( \nabla - \nabla_\text{Ad} \right)  }{\left(3\pi \nu R\right)^2 P}\diff{P}{r}}\right)\frac{3\pi\nu R}{A} \; . \label{eq: quad_form}
\end{equation}
This has the limits
\begin{subequations}
    \begin{alignat}{4}
        u_\textit{visc} &=& -\frac{V \delta g l \left( \nabla - \nabla_\text{Ad} \right) }{6\pi \nu R P} \,\diff{P}{r} \\
        u_\textit{invisc} &=& \; \sqrt{-\frac{V \delta g l \left( \nabla - \nabla_\text{Ad} \right) }{PA} \, \diff{P}{r}}
    \end{alignat} 
\label{eq: our visc lims}
\end{subequations}
for high and low viscosity respectively.

We then choose geometric factors relating $R$, $A$ and $V$ to $l$, $l^2$ and $l^3$ in order to produce the velocity limits used by \citet{abe1995basic} for high and low viscosity. These are
\begin{subequations}
    \begin{alignat}{4}
        u_\textit{visc}  &=& \frac{-\delta g l^3}{18 \nu P}\left( \nabla - \nabla_\text{Ad} \right) \diff{P}{r} \label{eq: highviscv}\\
	    u_\textit{invisc} &=& \sqrt{\frac{-\delta g l^2}{16 P}\left( \nabla - \nabla_\text{Ad} \right) \diff{P}{r}} \label{eq: lowviscv}
    \end{alignat}
\end{subequations}
which it can be seen are equivalent to our \autoref{eq: our visc lims} up to numerical factors.

Our formulation using \autoref{eq: quad_form} makes the transition smooth, which aids numerical convergence, as opposed to a switch at a critical value used in other works \citep[e.g.,][]{abe1995basic,Bower17}. 

\subsubsection{Application in partial melt}\label{sec: partial_conv}
A general formula for the adiabatic gradient in a medium is
\begin{equation}
    \nabla_\text{Ad} = \frac{P}{T} \frac{\delta}{C_P \rho} \; .\label{eq: gen_adiabat}
\end{equation}
Under the assumption that the melt fraction $\phi$ is always in equilibrium, the adiabatic gradient is given by the ``wet adiabat'', where the latent heat alters the values of $\delta/\rho$ and $C_P$, and thus changes the value of $\nabla_\text{Ad}$ (\autoref{eq: gen_adiabat}). These properties are also adjusted throughout the other equations. 

The adjustments to the density $\rho$, thermal expansivity $\delta$ and heat capacity $C_P$ may be easily derived by assuming additive volumes and entropies of the melt and solid phases, i.e., $V = V_l\phi + V_s(1-\phi)$ and $S = S_l\phi + S_s(1-\phi)$, where $V$ and $S$ are specific volume and entropy and subscripts $l$ and $s$ denote liquid and solid properties, respectively. Consequently, the density is given by
\begin{equation}
    \rho = \frac{1}{\phi /\rho_l + (1-\phi) / \rho_s} \; .
\end{equation}

To find $\delta$ and $C_P$ under partial melting one only requires the definitions
\begin{equation*}
    \delta = \frac{1}{V} \left.\diffp{V}{T}\right|_P  = -\frac{1}{\rho} \left.\diffp{\rho}{T}\right|_P \;\text{and  }\; C_p = T\left.\diffp{S}{T}\right|_P \; .
\end{equation*}
The results are
\begin{eqnarray}
    \frac{\delta}{\rho} &=& \phi \frac{\delta_l}{\rho_l} + (1-\phi)\frac{\delta_s}{\rho_s} + T\Delta V \left. \diffp{\phi}{T} \right|_P \label{eq: delv_prime} \\
    C_p &=& \phi C_{p,l} + (1-\phi)C_{p,s} + T\Delta S  \left. \diffp{\phi}{T} \right|_P \label{eq: Cp_prime}
\end{eqnarray}
where $\Delta V \equiv \left(\frac{1}{\rho_l} - \frac{1}{\rho_s}\right)$ is the specific volume change of melting, and $\Delta S$ is the specific entropy change of melting. 

Consequently, \autoref{eq: gen_adiabat} becomes
\begin{equation}
    \nabla_\text{Ad} = \frac{P}{T} \left( \frac{ \phi \delta_l / \rho_l + (1-\phi)\delta_s / \rho_s + T\Delta V \left. \diffp{\phi}{T} \right|_P}{\phi C_{p,l} + (1-\phi)C_{p,s} + T\Delta S  \left. \diffp{\phi}{T} \right|_P } \right) \; . \label{eq: expanded ad_grad}
\end{equation}

For this work, we assume that the timescales of melting are shorter than those of mixing, so melt equilibrium is maintained, and our mixing length formulation fully captures the heat flow in the system. 

\subsection{Physical properties of the mantle}\label{sec: physical properties}
In order to solve the equations in the previous three sections we need to know the values of the relevant physical properties, density, heat capacity etc., for solid and molten rocky materials. In this subsection, we describe how we calculate them.

\subsubsection{Equations of state}
$\rho, C_P$ and $\delta$ are all obtained as functions of $P$ and $T$. For the solid phase, we use the equation of state for enstatite, the Earth's main mantle component, from \citet{Stixtrude2}. For the melt, we use the \ch{Mg Si O3} equation of state in \citet{RTpress}. 

We precalculate these properties on a linear grid of temperature and pressure and retrieve values using bicubic Hermite interpolation. Under partial melting, density, thermal expansivity and heat capacity are given as a combination of the melt and solid properties, as described above in \S\ref{sec: partial_conv}.

\subsubsection{Volume and entropy change for melting}
In order to describe melting, it is necessary to know the change in specific volume, $\Delta V$, and entropy, $\Delta S$, between the solid and melt (see \autoref{eq: delv_prime} - \ref{eq: expanded ad_grad}). $\Delta V$ can be calculated from the densities given by the two equations of state. However, calculating $\Delta S$ from an equation of state requires an absolute entropy scale, which does not emerge naturally from our prescriptions. Instead, we calculate the entropy change through
\begin{equation}
    \Delta S = \Delta V \left.\diffp{P}{T}\right|_\phi \label{eq: DeltaS}
\end{equation}
which is derived in Appendix \ref{app: CC}. We use the estimate:
\begin{equation}
    \left.\diffp{P}{T}\right|_\phi = \phi \,/\, \diff{T_\text{liq}}{P} + (1-\phi) \,/\, \diff{T_\text{sol}}{P}
\end{equation}
which simply interpolates between the $T-P$ gradient of the $\phi=0$ line, the solidus, and the $\phi=1$ line, the liquidus.

\subsubsection{Viscosity}\label{sec: viscosity}
The effective viscosity varies by many orders of magnitude as the melt fraction changes. A steep change from more liquid to solid-like behaviour is often taken to occur at a critical melt fraction, $\phi_c$ \citep[e.g.][]{SOLOMATOV-chapter}. This melt fraction is essentially the fraction of melt at the point when closely packed spheres of the typical size of crystals can no longer move past each other. We use an experimentally measured value of $\phi_c=0.4$ from \citet{Lejeune1995}.

Additionally, the viscosity of the solid component is temperature and pressure-dependent, and we model it using an Arrhenius law for diffusion creep \citep{Tackley13} 
\begin{equation}
    \eta_s(P,T) = \eta_0 \exp\left( \frac{E_0 + PV_0}{R_\text{gas}T} - \frac{E_0}{R_\text{gas}T_0}\right) \label{eq: arrhenius visc}
\end{equation}
where the $s$ subscript once again denotes solid, and the values of parameters are shown in \autoref{tab: viscosity params}.

For low melt fractions ($ 0 < \phi < \phi_c$), we use an exponential parameterisation for the dynamic viscosity \citep{kelemen1997}
\begin{equation} 
\eta = \eta_s(P,T_\text{sol}(P)) \exp(-\alpha_\eta\phi) \; .
\end{equation} 
with $\eta_s(P,T)$ as given in \autoref{eq: arrhenius visc} and evaluated at the solidus, i.e., just before partial melting occurs. We assume that diffusion creep is the dominant mechanism of solid deformation \citep[e.g.,][]{Tackley13}, and so use $\alpha_\eta = 26$, as found in \citet{Mei2002}. 

For high melt fractions ($\phi > \phi_c $), we use the formula from \citet{Roscoe52}
\begin{equation}
    \eta = \eta_l\left(\frac{1 - \phi_c }{\phi - \phi_c}\right)^{2.5} \label{eq: Roscoe}
\end{equation}
where the $l$ subscript denotes liquid. We take the liquid viscosity $\eta_l$ to be a constant, as its temperature dependence is small relative to the solid phase and to changes in melt fraction \citep{SOLOMATOV-chapter}. 

We combine these formulae into a smooth function, with no singularity at $\phi=\phi_c$, given by
\begin{equation}
    \eta = 
    \begin{cases}
        \eta_s(P,T) \;&, \; \phi = 0 \\
        \eta_s(P,T_\text{sol}) \exp(-\alpha_\eta\phi) \;&, \; 0 < \phi \leq \phi_c \\
        \frac{1}{\eta_s(P,T_\text{sol})^{-1}\exp(\alpha_\eta \phi) \, + \, \eta_l^{-1}\left(\frac{\phi - \phi_c}{1 - \phi_c }\right)^{2.5}} \;&, \; \phi_c < \phi < 1 \\
        \eta_l \;&, \; \phi = 1 
    \end{cases} \label{eq: full_viscosity}
\end{equation}
Since the solid viscosity is much greater than the liquid (see \autoref{tab: viscosity params}) the $\phi > \phi_c$ formula is essentially the same as \autoref{eq: Roscoe}, unless the melt fraction is very close to $\phi_c$.
\begin{table}
\resizebox{\columnwidth}{!}{%
\begin{tabular}{|p{0.35\columnwidth}|c|c|p{0.37\columnwidth}|}
\hline
{\bf Parameter} & {\bf Symbol [unit]} &  {\bf Value} & {\bf Reference }\\ \hline
Solid reference viscosity & $\eta_0$ [Pa s] & $\SI{1e21}{}$ & \citet{Tackley13}  \\ 
Viscosity activation energy & $E_0$ [kJ mol$^{-1}$] & $\SI{300}{}$ & \citet{Tackley13} \\
Activation temperature & $T_0$ [K]& $\SI{1600}{}$ & \citet{Tackley13} \\ 
Activation volume & $V_0$ [cm$^3$ mol$^{-1}$] & $\SI{5}{}$ & \citet{Tackley13} \\
Liquid viscosity & $\eta_l$ [Pa s]& $\SI{0.1}{}$ & \citet{SOLOMATOV-chapter}\\
Diffusion creep parameter& $\alpha_\eta$ & $26$ & \citet{Mei2002} \\
Critical melt fraction& $\phi_c$ & 0.4 & \citet{Lejeune1995} \\
Gas constant& $R_\text{gas}$ [JK$^{-1}$mol$^{-1}$]& 8.3145 &  \\\hline

\end{tabular}%
}
\caption{Values in our viscosity model (\S\ref{sec: viscosity}).}\label{tab: viscosity params}
\end{table}

\subsection{The iron core}
Planets above a few thousand km in radii are likely to have formed an iron core through gravitational settling of the denser iron from the mantle \citep[e.g.,][]{Elkins-Tanton12-review}. We concentrate on planets with a core mass fraction of 0.3, similar to the Earth. Were planets to have significantly different initial silicate-to-iron ratios to the Earth, or undergo a major collision that might strip the mantle, then it is possible that the core mass fraction would be different for exoplanets. However, we find that differences in evolution are generally not significant enough to alter the conclusions we draw.

We assume the core is pure iron and solve for its structure using our Henyey scheme, like the mantle. For iron, we use physical properties for the $\gamma$ phase of iron from \citet{Dorogokupets2017}, as it is the appropriate phase at the pressures in the cores of the low mass planets we will consider \citep[e.g.,][]{TSUJINO2013}.

For these calculations, we neglect the fact that at early times the mantle will be molten and set the temperature structure in the core to be adiabatic. The reason for these simplifications is that we are mainly interested in the amount of energy provided to the mantle, which is likely not affected by these assumptions, as will be discussed in \S\ref{sec: discuss-other}. Modelling the core in full would add extra complexity, especially as its evolution is not well constrained \citep[see e.g.,][]{Zhang22}.

\subsection{Radioactive heating}\label{sec: radioactivity}
For Gyr old planets, the energy generated by the decay of long-lived radioactive isotopes is an important contributor to the total luminosity. The significant elements in the Earth are \isotope[232]{Th}, \isotope[238]{U}, \isotope[40]{K} and \isotope[235]{U} \citep{turcotte2002geodynamics}. These elements are lithophiles, which means they preferentially dwell in the mantle. Over shorter period(less than a few Myrs), short-lived radioisotopes are more important; in the Solar System, the most significant of these were \isotope[26]{Al} and \isotope[60]{Fe}. Our models use the same concentrations of long-lived radioisotopes as found in the Earth's mantle, including their decay over time \citep[data from][]{turcotte2002geodynamics}. We also include \isotope[60]{Fe} in the core and \isotope[26]{Al} in the mantle at the concentration estimated for the Solar System in \citet{lugaro18-review}. 

\subsection{Outer boundary conditions}\label{sec: BCs}
\begin{figure}
    \centering
    \includegraphics[width=\linewidth]{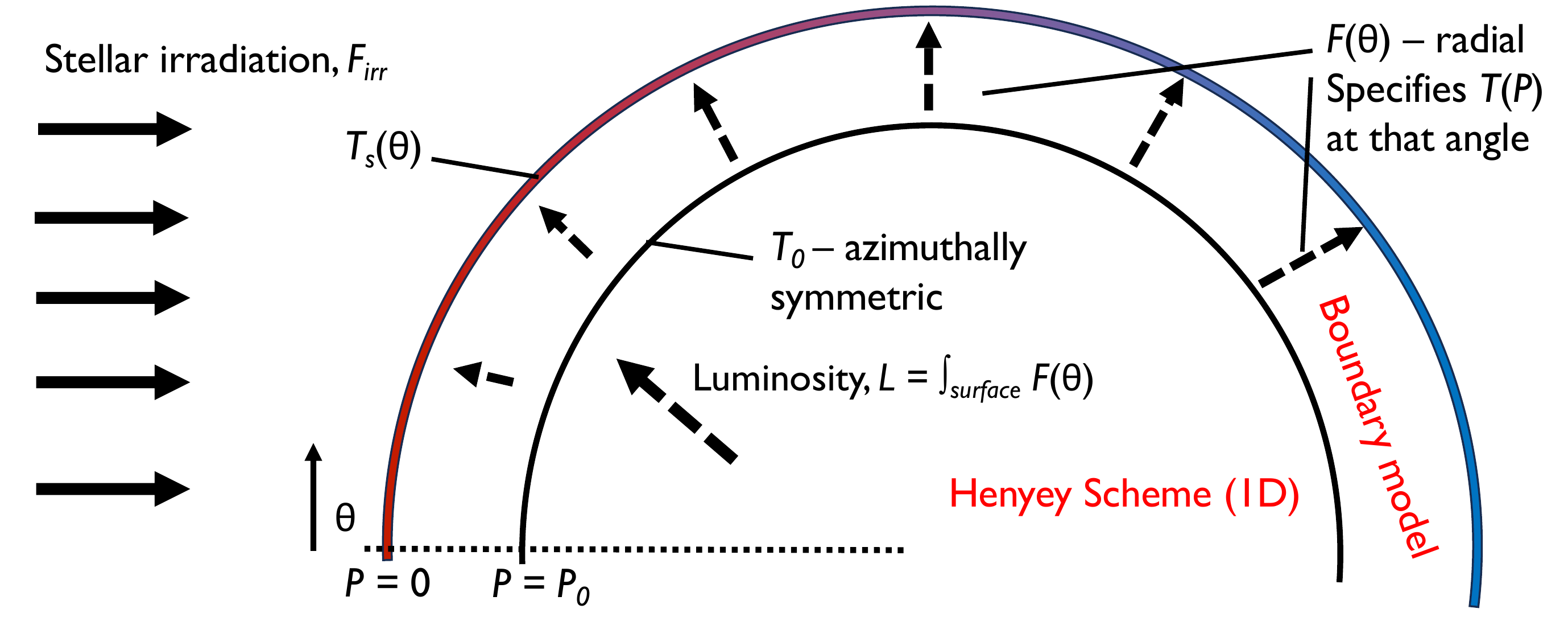}
    \caption{Schematic of our boundary model (\S\ref{sec: BCs}). The inner part of the planet is considered spherically symmetric, and its structure is solved for by our Henyey scheme. Its edge is at a pressure $P_0$, which has a spherically symmetric temperature, $T_0$. The planet's surface has a day-to-nightside temperature gradient, $T_s(\theta)$, due to the stellar irradiation determined by Equations  \ref{eq: F_surface BB} and \ref{eq: F*}. For the outer layer ($P<P_0$), we consider radial fluxes $F(\theta)$ such that the flux integrated over the surface is equal to the luminosity for the interior, $L$.}
    \label{fig: BC_schematic}
\end{figure}

\begin{figure}
    \centering
    \includegraphics[width=\linewidth]{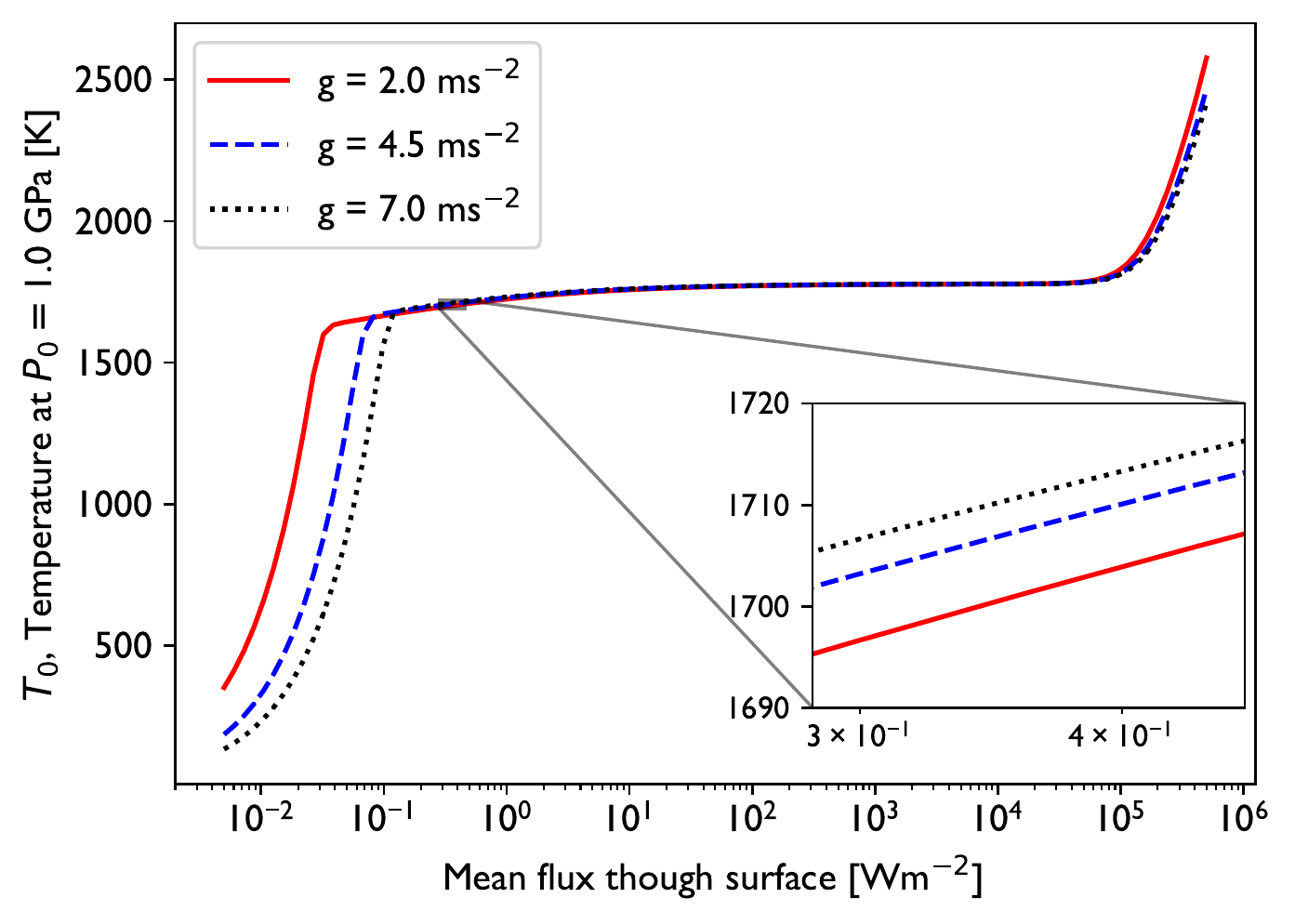}
    \caption{Temperature $T_0$ at a pressure $P_0 = 1$ GPa, for boundary conditions with no redistribution and $T_{ss} = 2320$K, as a function of the mean flux from the planet (luminosity divided by surface area), shown for three different surface gravities $g$. See \S\ref{sec: BCs}. In this figure, we display the smooth fitted function described in Appendix \ref{sec: fit BC}.}
    \label{fig: T0_func}
\end{figure}

The catastrophically evaporating systems we are investigating are very close to their host stars and are likely tidally locked. Therefore, they have permanent daysides that are highly irradiated. We take this lack of symmetry into account in our evolutionary models by adapting the outer boundary conditions. The full problem of heat transport in the planet is complex due to the mixture of radial and angular heat transport and the effect of heat transport in any magma pool; we, therefore, make a series of assumptions as discussed below. We explain the principles of the approach here, with some additional details supplied in Appendix \ref{app: BC}.

The first and most important assumption is that there is some depth within the planet below which the planet can be assumed to be spherically symmetric, and thus the 1D structure model described in previous sections can be applied unchanged (see schematic in \autoref{fig: BC_schematic}). We designate the pressure this occurs at as $P_0$. For regions at lower pressures, i.e., above this boundary, we allow the symmetry to be broken, and quantities then depend upon the angle $\theta$ from the substellar point (the point on the planet's surface closest to the star). This assumption is justified if the inward energy flux, due to external heating from the star, that penetrates deeper than $P_0$ is much less than the outward flux from the planet cooling. We will discuss this further in \S\ref{sec: BC caveats}, with reference to our results, but the essential reason this is true in most cases is that heat can only be transported inwards by conduction, not convection, and thus the efficiency of inward heat transport is low.


What we require is a relation between the temperature at the boundary of the spherically symmetric interior, $T_0$, and the interior luminosity $L$ (see \autoref{fig: BC_schematic}). For this, we must consider the region of the planet close to the surface ($P < P_0$) where there is an angular dependence. In order to model this region, several assumptions must be made.

We begin by considering the surface ($P=0$). We assume that there is no angular redistribution of the star's energy here. The first reason for this is that since the planets are tidally locked, there can be no redistribution due to rotation. Secondly, any atmosphere generated through outgassing volatiles will likely be thin (for instance, the atmospheric escape models of \citealt{Booth_disint22} generate atmospheres with maximum pressures of $\sim 10^{-5}$ bar) and so we expect them to be unable to transport heat efficiently. Thirdly, \citet{Kite16} showed that a surface magma ocean cannot transport enough heat laterally to decrease the temperature gradient imposed by irradiation. Consequently, we treat the surface as a local black body; thus, its temperature is given by
\begin{equation}
    T_s(\theta) = \left(\frac{F(\theta,T_0) + F_{irr}(\theta)}{\sigma}\right)^\frac{1}{4} \label{eq: F_surface BB}
\end{equation}
where $F_{irr}(\theta)$ is the irradiation from the star, and $F(\theta,T_0)$ is the angle-dependent heat flux from the interior, which depends on $T_0$.) {\it A priori}, the value of both $F(\theta,T_0)$ and $T_s(\theta)$ are unknown, and the aim is to find both together.

Assuming the planet is far enough from the star, the stellar irradiation is plane-parallel and given by
\begin{equation}
    F_{irr}(\theta) = \begin{cases}
        \sigma T_*^4\frac{R_*^2}{a^2} \cos{\theta} \; &, \; 0 \leq \theta < \pi / 2 \,\,\,\,\,\text{(dayside)}\\
        0 \; &, \; \pi / 2 \leq \theta \leq \pi \,\,\,\,\,\text{(nightside)}
    \end{cases} \label{eq: F*}
\end{equation}
where $R_*$, $T_*$ and $a$ are the stellar radius, stellar effective temperature and planet's semi-major axis, respectively. For much of the planet's lifetime $F < F_{irr}$ and so the surface temperature is proportional to $F_{irr}(\theta)^\frac{1}{4}$, yielding a steep temperature gradient from day to nightside.

Considering now the sub-surface part of the $P<P_0$ region, we assume that the angular flux is much smaller than the radial flux in this region as well. This will be justified in \S\ref{sec: BC caveats}. This means that rather than a full 2D problem, the heat transport within the $P<P_0$ layer is simply a series of 1D heat transport equations at any angle $\theta$ (see \autoref{fig: BC_schematic}). In fact, at any given $\theta$ the conduction/convection equations (Equations \ref{eq: conduction}-\ref{eq: totalF}) become just one ordinary differential equation for $T(P)$ -- \autoref{eq: dTdP const g}/\ref{eq: dTdP const g inviscid} -- if gravity $g$ and heat flux $F$ are assumed to be vertically constant (see Appendix \ref{sec: cond_conv for BC}.) This assumption is reasonable if the $P<P_0$ region is thin. Thus one only needs $F(\theta,T_0)$ to specify the temperature--pressure structure at any angle $\theta$.

The magnitude of any assumed inward heat transport by conduction would depend on the specific choice of $P_0$. We believe that any inward heat flux will be small due to the inefficiency of conduction. So, to avoid the dependence on the arbitrary choice of $P_0$, we set the heat flux at any angle where the surface temperature is higher than $T_0$ to 0. Doing so should not affect the overall evolution (see \S\ref{sec: BC caveats}.)

The desired boundary condition for the Henyey scheme is the temperature at $P_0$ as a function of the luminosity from the interior, i.e., $T_0(L)$. One must find the function $F(\theta,T_0)$ that produces a given spherically symmetric $T_0$. $F(\theta,T_0)$ is linked to the luminosity, $L$, by integrating over the surface:
\begin{equation}
    L = 2\pi R^2 \int^\pi_0 F(\theta,T_0) \sin{\theta} \, \mathrm{d}\theta \label{eq: luminosity integral}
\end{equation}
where $R$ is the total radius of the planet. 

We now have all the information required to find $F(\theta,T_0)$, under the previous assumptions. If one guesses the interior heat flux at a given angle, $F(\theta)$, then \autoref{eq: F_surface BB} gives the surface temperature. One can then use this as a boundary condition for integrating $\mathrm{d}T/\mathrm{d}P$ (\autoref{eq: dTdP const g}/\ref{eq: dTdP const g inviscid}) to find $T_0$, the temperature at $P_0$. This gives a mapping between $T_0$ and the flux, and so for a specified $T_0$ the corresponding angular function $F(\theta,T_0)$ can be found. \autoref{eq: luminosity integral} allows us to find the function $L(T_0)$, which is the inverse of the relation we seek.

The function $F(\theta,T_0)$ has to be computed numerically, and we do so by solving \autoref{eq: dTdP const g}/\ref{eq: dTdP const g inviscid} to find $T_0$ for a grid of $F$ and $\theta$. We use 5th-order Runge-Kutta integration, with an adaptive step size and a relative tolerance of $10^{-6}$. This grid is then fit to a spline and $F(\theta,T_0)$ is found using Brent's method \citep{Brent1974AlgorithmsFM}.

We compute the boundary conditions for various gravitational accelerations, $g$, and substellar temperatures, $T_{ss}$. The substellar temperature is defined by:
\begin{equation}
    T_{ss}^4 \equiv \frac{F_{irr}(0)}{\sigma} = T_*^4 \frac{R_*^2}{a^2}\label{eq: Tss}
\end{equation}
and so is a measure of the stellar flux a planet receives. It is also the temperature of the point on the planet closest to the star given no redistribution of energy or heat flow from the interior. As the assumption of no redistribution is likely well justified for the short-period planets we consider, it is approximately the maximum temperature of the planet at late times when the internal heat flux is small.

The resulting relation between luminosity and temperature at the edge of the 1D domain is shown in \autoref{fig: T0_func} for a particular $T_{ss}$.\footnote{The curve shown in \autoref{fig: T0_func} is actually a smooth function we fit to our calculated model to aid the convergence of the Henyey scheme}, as detailed in Appendix \ref{sec: fit BC}. The basic features of the plot are that at high and low fluxes, the flux is a strong function of temperature, essentially due to black body cooling. Meanwhile, for a large range of intermediate fluxes, the temperature changes little. This is because, when $T_0$ is close to the critical melt fraction, the viscosity at that point, and thus the amount of energy that can be transported, changes by orders of magnitude for very small temperature changes (\S\ref{sec: viscosity}).

\subsection{Grid and timestepping}\label{sec: grid/time}
We solve the equations in \S\ref{sec: struct_eqns} on a mass grid where the amount of mass enclosed below cell $j$ is $m_j \propto j^\alpha$. $\alpha$ is chosen to give a compromise between cells getting smaller in radius towards the edge, where high resolution is required to deal with the final crystallisation of the mantle, and maintaining a reasonable resolution near the centre of the planet. As density is close to constant, the first condition requires $\alpha \lesssim 3$ ($\alpha = 3$ would correspond to cells of constant radius if the density is constant). For the second condition, it helps to have $\alpha > 1$. We find $\alpha = 1.5$ works well.

The timescales over which processes occur in planetary evolution differ vastly. We must therefore adapt the timestep in order to maintain accuracy and also allow the models to run in a reasonable amount of time. We employ a simple algorithm to do this. When undergoing the convergence steps of the Henyey scheme the timestep is fixed and is chosen at the start. As a first suggestion for the $n^\text{th}$ timestep, we use the formula:
\begin{equation}
    \Delta \tilde{t}_n  = \Delta t_{n-1} \frac{L_{n-1}}{L_{n-2}-L_{n-1}}  f_L
\end{equation}
where $L$ is again the luminosity and subscripts denote the timestep number, so $\Delta t_{n-1}$ is the previous timestep. This extends or shrinks the timestep such that the fractional change in the luminosity gets closer to $f_L$. We use $f_L = 30\%$.

We then also predict the luminosity for this step using linear interpolation
\begin{equation}
    L_n = L_{n-1} + \frac{\Delta\tilde{t}_n}{\Delta t_{n-1}} \left(L_{n-1}-L_{n-2}\right)
\end{equation}
and use this to predict $T_0$ for the current step, which is a function of $L$ (\S\ref{sec: BCs}). We then impose that $T_0$ does not change by more than a fraction $f_T$, and the new timestep is given by
\begin{equation}
    \Delta t_n = \Delta\tilde{t}_n  \min\left(1, \frac{T_{0,n-1}}{T_{0,n-1}-T_{0,n}(L_n)} f_T \right) \; .
\end{equation}
We use $f_T = 2\%$. The timestep may be shrunk further to limit the amount of mass lost in this step, as described in the next section.  Finally, for the first timestep, since there is no previous timestep to compare to, we use 0.1\% of the Kelvin-Helmholz timescale.

\subsection{Mass Loss}\label{sec: massloss}
\begin{figure}
    \centering
    \includegraphics[width=\linewidth]{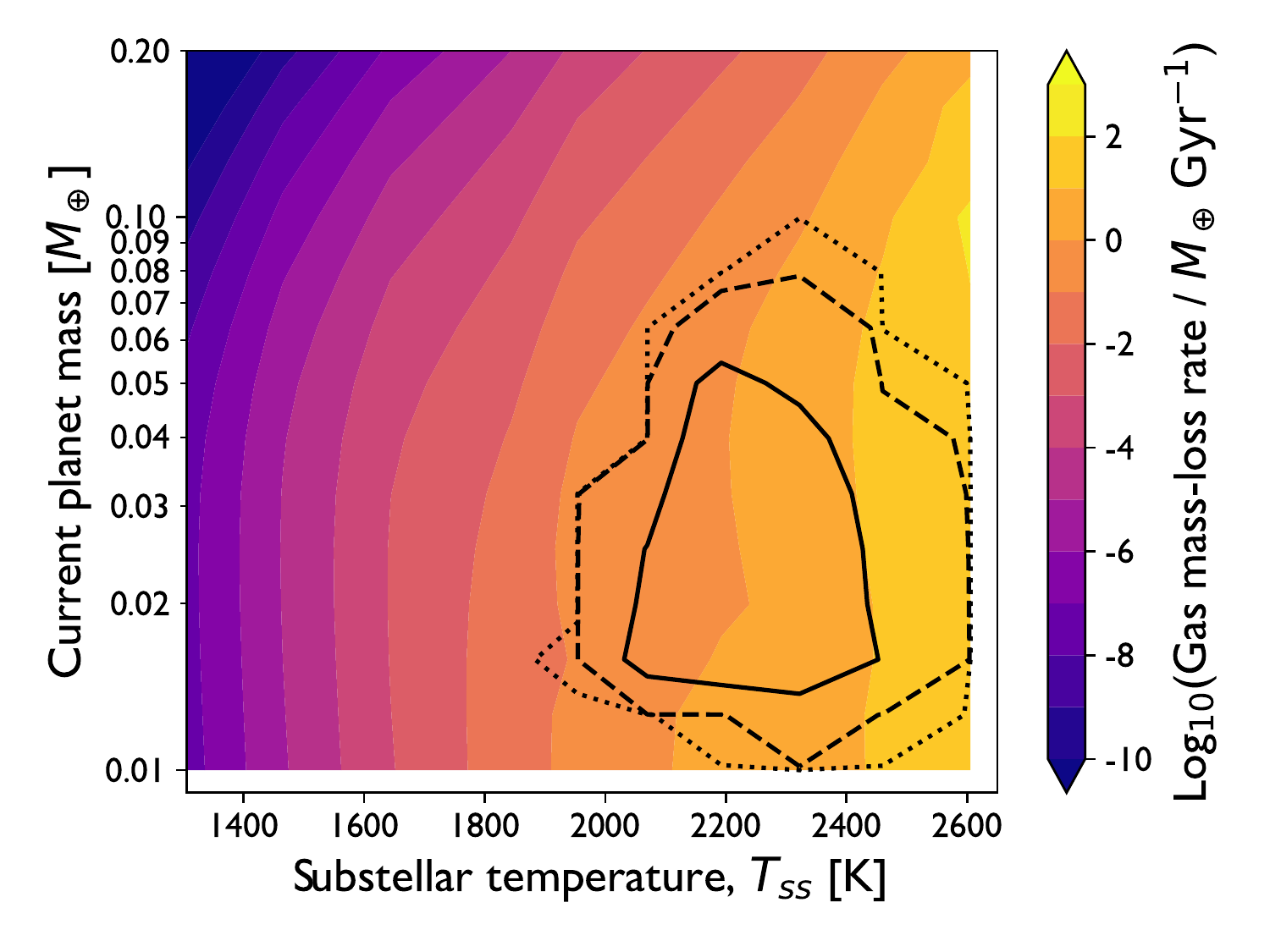}
    \caption{Contours of gas mass loss rate for planets of different mass and substellar temperatures and fixed bulk density of $0.67\rho_\oplus$, similar to that of Mars,} as calculated by \citet{Booth_disint22}. It differs slightly from their Figure 5, which is calculated for a fixed mass-radius relation. Dust mass loss rates of $10^{-6}$, $10^{-4}$ and $10^{-2}$ $M_\oplus \text{Gyr}^{-1}$ are shown with the dotted, dashed and solid black contours. The irregular shape is due to the finite number of data points.
    \label{fig: mdot_grid}
\end{figure}
\begin{figure*}
    \centering
    \includegraphics[width=\linewidth]{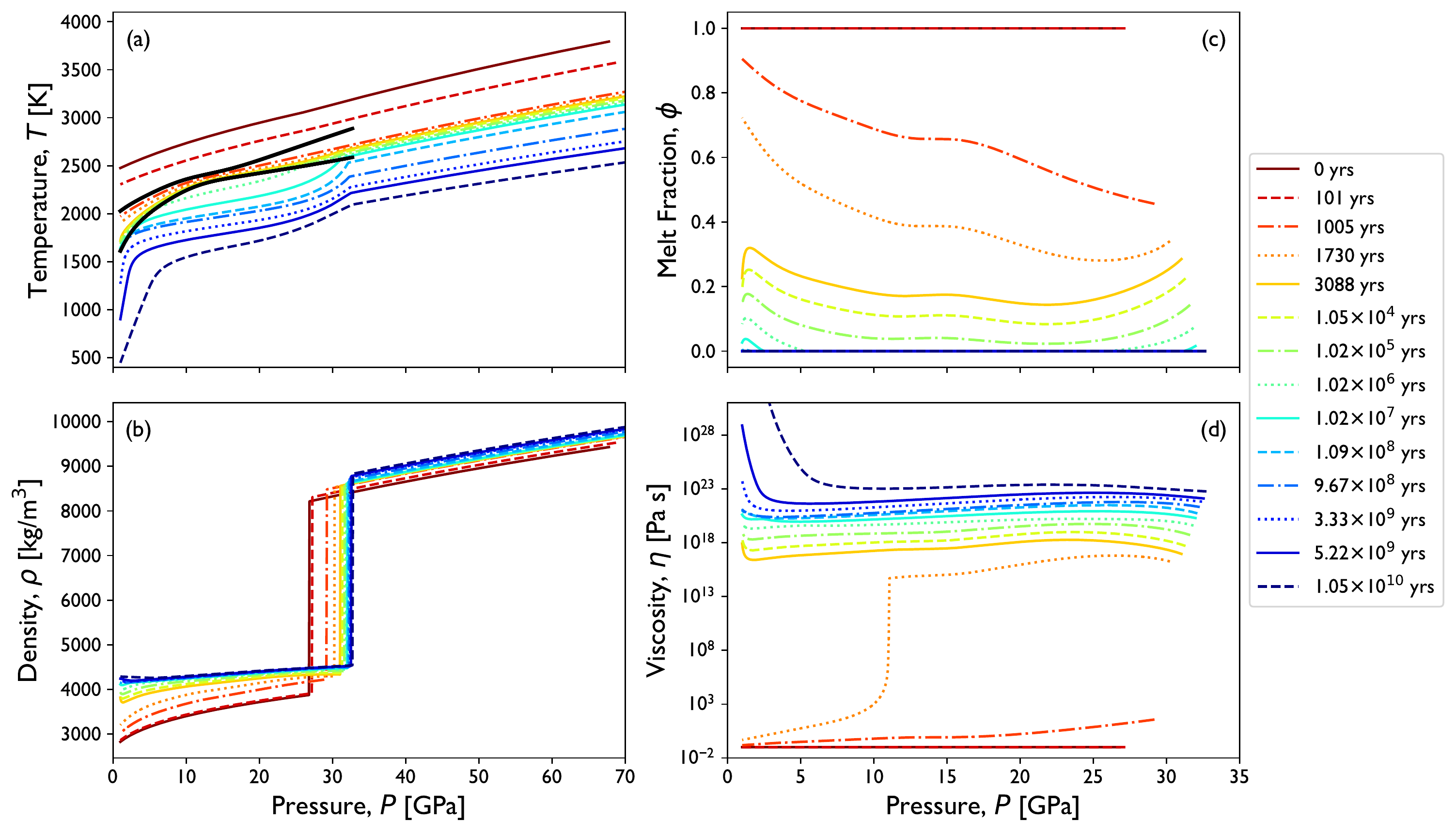}
    \caption{Evolution of the internal structure of a 0.15 $M_\oplus$ planet with core mass fraction 0.3, and a substellar temperature of 2320K, but no mass loss. Note that the right-hand panels show only the mantle, so the pressure scale is different. The black solid lines on the temperature plot mark the liquidus (upper line) and solidus (lower line), which only apply to the mantle. Time snapshots are roughly spaced logarithmically in time but with extra snapshots between $10^3 - 10^4$ yrs to demonstrate the crystallisation of the magma ocean and between $10^9 - 10^{10}$~yrs to show the late time evolution.}
    \label{fig: internal no mdot}
\end{figure*}
\begin{figure}
    \centering
    \includegraphics[width=\linewidth]{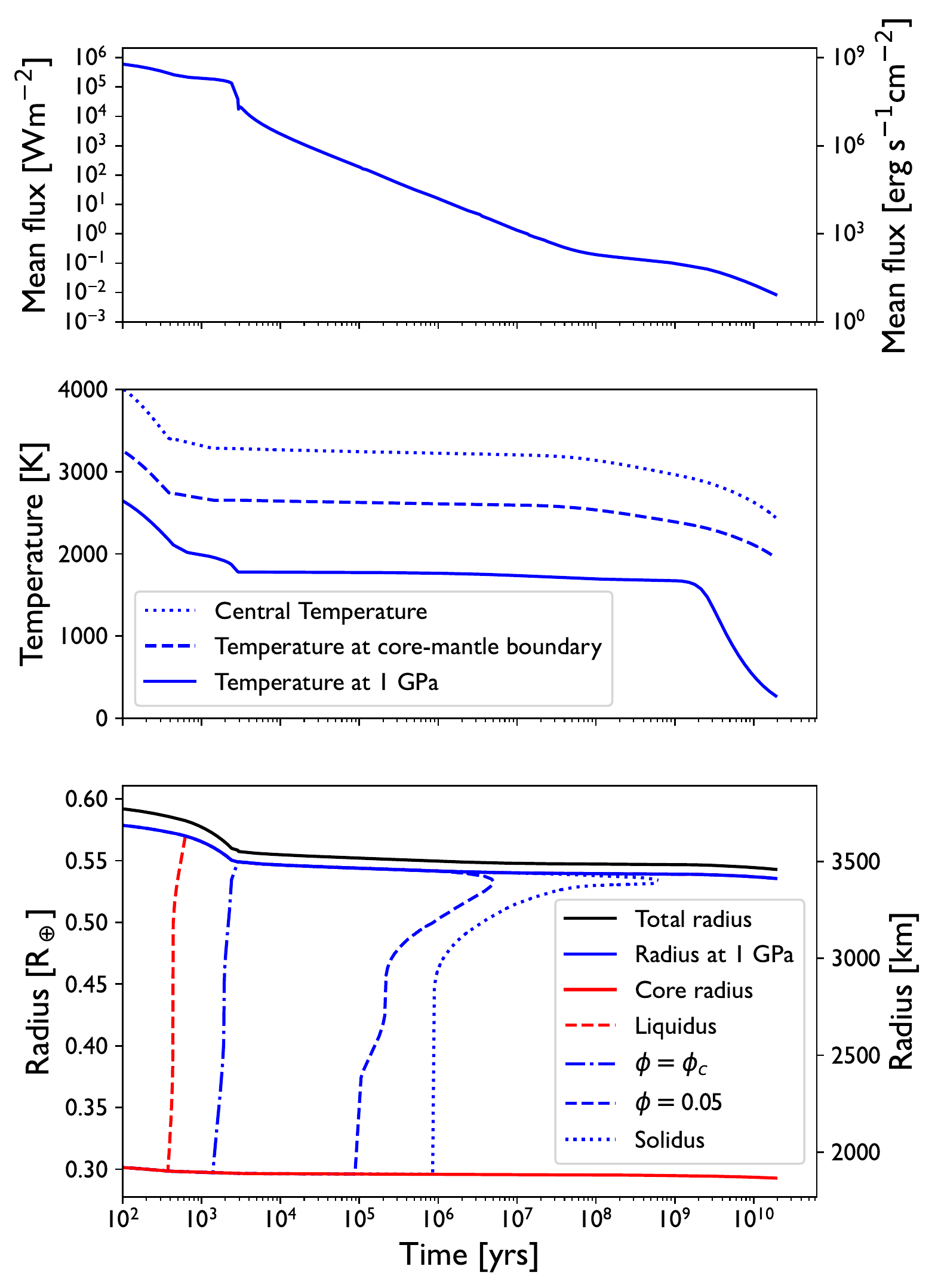}
    \caption{Evolution of the mean surface flux ({\it top panel}), temperature ({\it middle panel}) and the radius and molten state ({\it bottom panel}) of a 0.15 $M_\oplus$ planet with a core mass fraction of 0.3 and a substellar temperature of 2320K, but no mass loss. In the bottom panel, regions to the left of the liquidus are entirely liquid, to the left of the $\phi = \phi_c$ line are partially molten, to the left of the solidus behave like a solid but have some partial melting and to the right of the solidus are fully solid. The region above the 1~GPa line is unresolved by the 1D model and will have melting on the dayside surface but be cold and solid on the night side. We also do not plot any partial melting at the core-mantle boundary, which does occur in our models, at certain times, due to the thermal boundary layer there (see \autoref{fig: internal no mdot}) because we are more interested in the state towards the surface.} 
    \label{fig: bulk no mdot}
\end{figure}
Since our aim is to model the evolution of rocky planets undergoing evaporative mass loss, we incorporate mass loss as follows. 

At each timestep, we ascribe a mass loss rate determined by the current mass, radius and surface temperature. We use mass loss rates computed using the method described in \citet{Booth_disint22}, which assumes that the dusty outflows form $\SI{1}{\micro m}$ grains with properties similar to forsterite.Example mass loss rates are shown in \autoref{fig: mdot_grid}. We pre-tabulate mass loss models for a given substellar temperature (\autoref{eq: Tss}) for a grid of planet masses and densities. 

When the mass loss rate has been determined, we then calculate the mass change to the next timestep according to our suggested timestep (see \S\ref{sec: grid/time}). In order to make the numerical problem easier to solve, rather than removing the mass from only the outermost cell, as would be the closest to the physical reality, we instead shrink a finite number of the outermost cells and keep the mass contained in cells interior to this the same. This prevents a drastic change in the size of the outermost cell, which might be numerically unstable, allowing more mass to be removed per timestep. Before doing so, we check that the mass loss results in the mass contained in these cells being reduced by no more than a certain fraction, and if the mass step is too great, we reduce the timestep so that this condition is satisfied. Our default setup is that the outermost $5\%$ of total cells are shrunk and by no more than $1\%$. This procedure requires us to merge cells so that we can continue to remove mass after the cells shrink and to split cells, to ensure that individual cells do not become too large a fraction of the total planet's mass. We make these grid changes at the beginning of each timestep, when appropriate thresholds are reached and use piecewise cubic Hermite interpolation \citep[PCHIP,][]{PCHIP}, where interpolation of values is required.

During mass loss, the gravitational potential of the planet is altered. Since mass loss occurs at the beginning of the timestep, this energy change is not included in \autoref{eq: lumin}. We, therefore, introduce an extra source term to that equation, using the prescription used in {\it MESA} \citep[see][\S3.3]{MESA19}. This approach considers the mass moving through and out of cells and the energy deposited by it, and we also assume that the mass loss timescale is longer than the thermal timescale. We generally find this energy contribution is small.

\section{Results} \label{sec: results}
\subsection{No mass loss}

To demonstrate our model, we first show a fiducial case without mass loss, as the features of the thermal evolution are clearer in this case. We choose a substellar temperature of 2320 K as it is the fiducial value used in \citet{Booth_disint22}\footnote{\citet{Booth_disint22} choose this value as it is the substellar temperature of Kepler 1520b, for the stellar parameters they use. In \autoref{tab: systems}, we show a temperature with more up-to-date parameters.}, which is the source of our mass loss model (\S\ref{sec: massloss}).

Snapshots of the internal structure are shown in \autoref{fig: internal no mdot} and the evolutions of some overall properties are shown in \autoref{fig: bulk no mdot}. As the planet's temperature decreases (\autoref{fig: internal no mdot}, {\it Panel a}), it crystallises from the inside out ({\it Panel c}), which is essentially due to the shape of the solidus/liquidus relative to the temperature structure, which is close to adiabatic.\footnote{This is a well-known feature, e.g., \citealt{walker1975differentiation}, although, depending on the unconstrained shape of the liquidus, `middle-out' crystallisation has also been proposed, e.g., \citealt{Middle-out-Stix2009}.} As the melt fraction passes through the critical point of $\phi = \phi_c = 0.4$, a many order of magnitude change in viscosity is observed ({\it Panel d} and see \S\ref{sec: viscosity}). The increase in melt fraction close to the core-mantle boundary and decrease towards the surface seen at later times is because the mixing length is equal to the distance to the nearest boundary, generating thermal boundary layers, which can also be seen in {\it Panel a}. In addition, the solid viscosity is highly temperature dependent (\autoref{eq: arrhenius visc}). This results in a positive feedback: the viscosity increases towards the surface, due to lower temperatures there. This makes convection less efficient and the thermal boundary layer stronger, so the temperature at the surface falls, causing the viscosity to further increase. The effect is most noticeable at late times in {\it Panels a} and {\it d}. 


The first two panels of \autoref{fig: bulk no mdot} show the evolution of flux and temperature, which are best understood with reference to \autoref{fig: T0_func}. There is an extended period of time where the temperature at $P_0$ does not change, but the luminosity continues to drop due to the large changes in viscosity around $\phi_c$. Note also that the flux levels off slightly at $\sim 10^8$~yrs before decaying over the following $10^{10}$yrs, which is due to the radioactive decay of long-lived radioisotopes becoming the dominant heat source over primordial energy.

The final panel shows the planet shrinking, which is due to the density increase over time (see {\it Panel b} of \autoref{fig: internal no mdot}). Both the core and mantle shrink over the whole evolution, but the largest effect on the radius is the change from liquid to solid in the mantle, causing a marked decrease in total radius in the first few thousand years. As can be seen from the horizontal extent of the lines in {\it Panel b} of \autoref{fig: internal no mdot}, the density increase also results in a pressure increase at the core--mantle boundary and centre of the planet.

As well as the planet shrinking, the final panel of \autoref{fig: bulk no mdot} also shows the molten state of the planet. One sees that, within a few hundred years, the planet is no longer fully molten (dashed red line) and, within a few thousand, the whole mantle is below the critical melt fraction (dot-dashed line) meaning that its behaviour is close to that of a solid. Partial melting does persist for Gyrs, but, beyond a few Myrs, it is very low (see blue dashed line that marks $\phi = 5\%$).

In short, despite the high substellar temperature, the planet is almost entirely solid for the majority of its life, because it can cool from the night side. What is not included in this one dimensional plot is that a magma pool will persist on the dayside, since the temperature there, which is approximately the substellar temperature (see \S\ref{sec: BCs}), is sufficient to fully melt the rock (i.e., it is above the liquidus.) It is this magma pool that evaporation occurs from, leading to the observed mass loss.

\subsection{Mass loss}
\begin{figure}
    \centering
    \includegraphics[width=\linewidth]{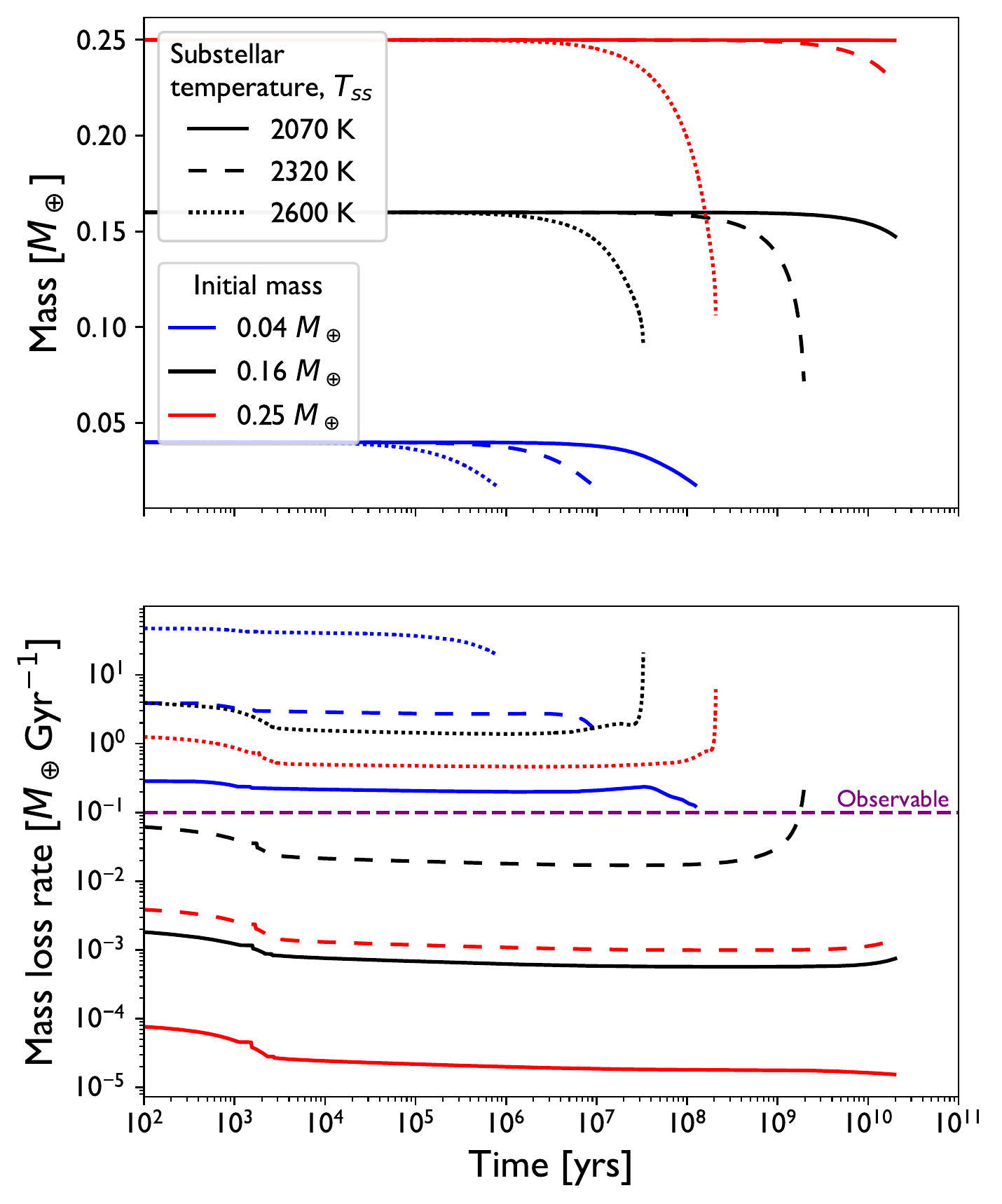} 
    \caption{Evolution of planets' mass and mass loss rates with different initial masses and substellar temperatures. Mass loss rates are calculated using \citet{Booth_disint22}. All planets start with a core mass fraction of 0.3.}
    \label{fig: mmdot}
\end{figure}

\begin{figure*}
    \centering
    \includegraphics[width=\linewidth]{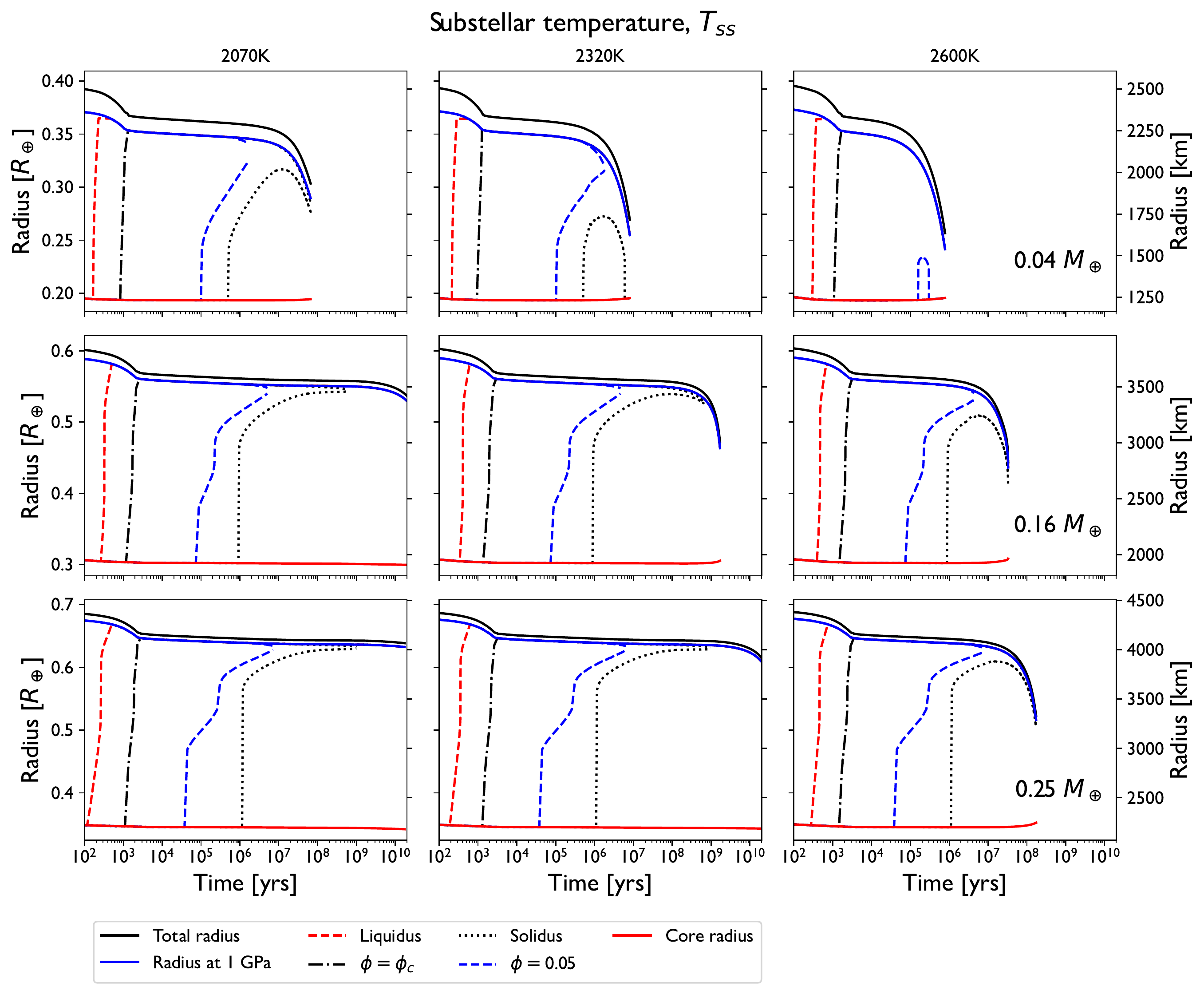}
    \caption{Evolution of the molten state of planets with different initial masses and substellar temperatures. Mass loss rates are calculated using \citet{Booth_disint22}.
    All planets start with a core mass fraction of 0.3.}
    \label{fig: crystallisation mdot}
\end{figure*}

\begin{figure}
    \centering
    \includegraphics[width=\linewidth]{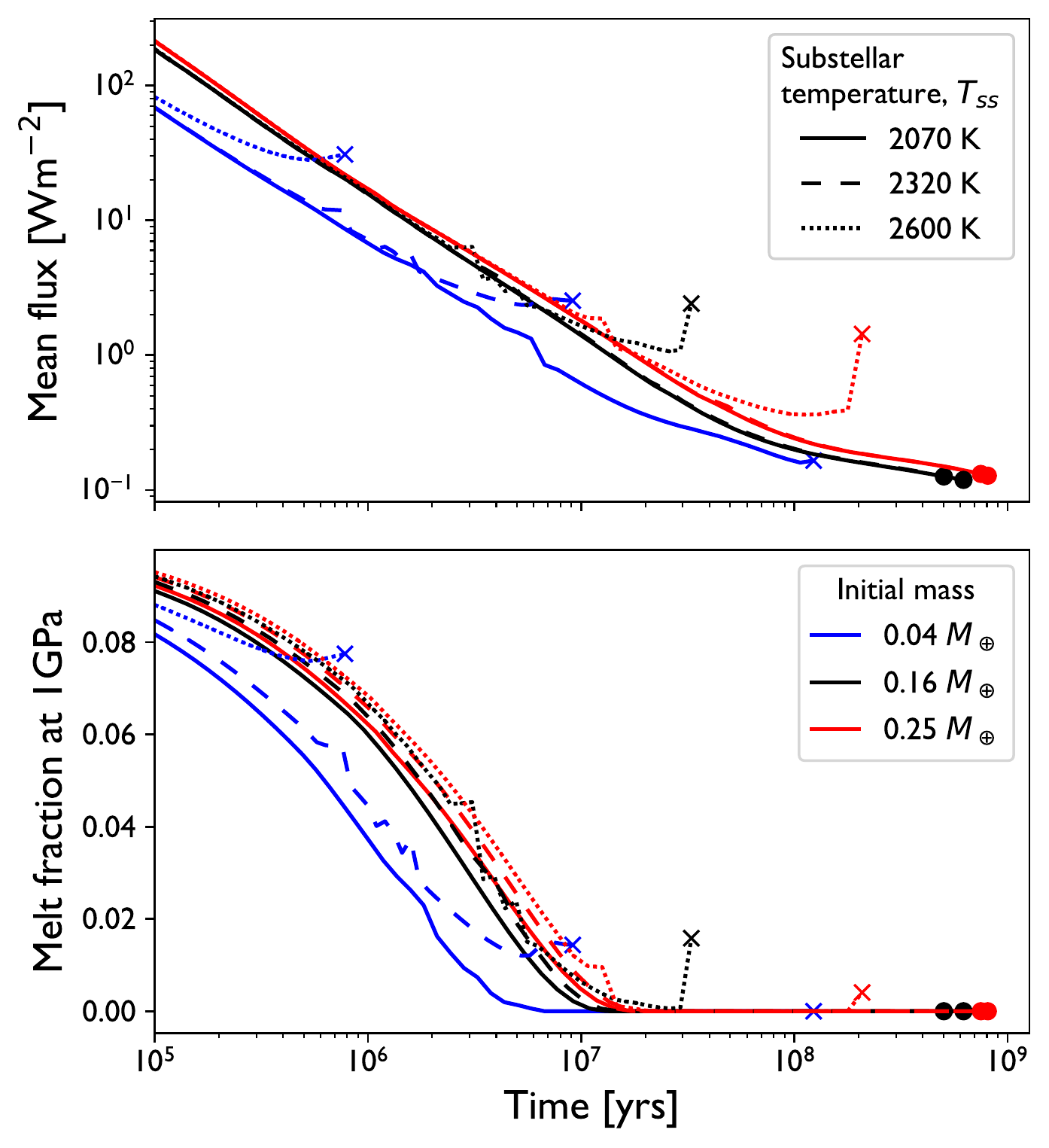}
    \caption{Mean (outward) energy flux and melt fraction at 1~GPa as a function of time for various initial planet masses and substellar temperatures for planets undergoing mass loss according to the models of \citet{Booth_disint22}. All masses have an initial core mass fraction of 0.3. The increase in melt fraction at late times for some simulations is due to the increase in flux, as discussed in the text. Simulations that end with a dot had full crystallisation up to 1GPa. Those that end with a cross reached the maximum bulk planet density that mass loss rates were computed to. }
    \label{fig: depth}
\end{figure}
\autoref{fig: mmdot} shows the evolution of the total mass, radius and mass loss rate for a number of models with different initial planet masses and substellar temperatures. Mass loss rates increase with increasing substellar temperature and decreasing initial planet mass (as seen in \autoref{fig: mdot_grid}), which is seen in the faster mass and radius decrease for such planets. We mark on the mass loss rate panel an `observable' cut-off mass loss rate of $10^{-1} M_\oplus$~Gyr$^{-1}$, motivated by calculations by \citet{Perez-Becker13}. We return to considering when the catastrophically evaporating planets are observable in more detail in \S\ref{sec: evap_times}. However, even with this simple condition, some interesting features can be noted about observed planets. Some planets are observable throughout their lifetimes, but their lifetimes are short ($\lesssim 10^8$yrs). These are planets with sufficiently high substellar temperatures (dotted lines) or low initial masses (blue lines). At the other extreme, much cooler planets (solid lines) or massive planets (red curves) are never observable within the age of the universe. There is also an intermediate regime of planets that become observable late in their lifetimes once they have lost sufficient mass. This was also noted by \citet{Perez-Becker13} from their earlier mass loss models. We discuss the consequence of this on the number of systems we might observe in \S\ref{sec: occurrence}.

In \autoref{fig: crystallisation mdot}, we show the evolution of the molten state of the mantle. In all the models, the mantle still entirely crystallises to a melt fraction below $\phi_c$ within $\sim 10^4$yrs, meaning the planet's interiors should be essentially solid by the time they are observed around ~Gyrs old main sequence stars. Furthermore, in many models, crystallisation occurs fully by around 1 Gyr. The exceptions are the highest temperature and lowest mass cases, where the planet is evaporated before the melt fraction is zero. 

However, even if the melt fraction is not zero, it is very small in all cases, and the planet still essentially behaves like a solid, as shown in the lower panel of \autoref{fig: depth}. At the same time, note that the melt fraction increases slightly at late times in some cases. The reason for this is subtle. The amount the planet's centre can cool is limited by the amount of energy that convection can carry away. The decrease in mass means there is less material above the lower mantle and core, allowing it to cool more easily. Thus the temperature of the core and lower mantle can decrease, causing an increase in luminosity and thus mean flux at the surface, as shown in the top panel of \autoref{fig: depth}. This increased flux induces more melting at 1GPa. This increased partial melting may have some effect on any volcanism on the planet and may alter the detailed composition through solid-melt partitioning, but since the amount of melting is small, it does not affect the overall conclusion that the catastrophically evaporating planets are essentially solid.


To summarise, these highly irradiated planets solidify significantly within a few thousand years and are completely solid within a few Gyrs. This is because they cool easily from their nightsides. Consequently, major melting occurs solely on the dayside, as proposed by e.g., \citet{Kite16}. Mass loss does not have a large effect on this result.

\section{Discussion} \label{sec: discuss}

\subsection{Depth of the magma pool and implications for chemistry}\label{sec: moltenmass}
\begin{figure}
    \centering
    \includegraphics[width=\linewidth]{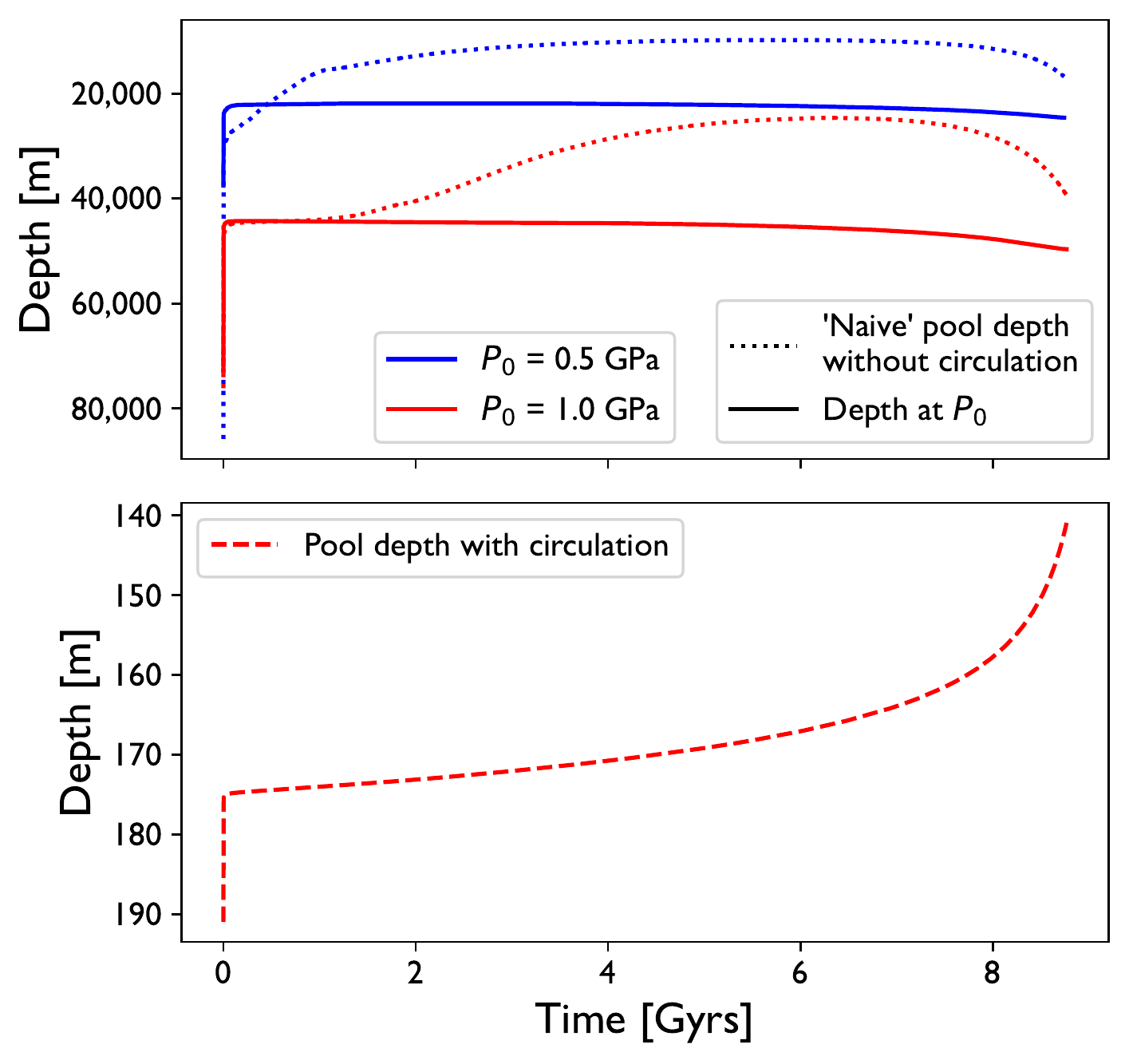}
    \caption{The depth of a dayside magma pool under different assumptions, and defining the boundary of the 1D portion of the code at different pressures, $P_0$. For both cases, the planet has a mass of 0.2 $M_\oplus$, a substellar temperature of $T_{ss} = 2320$K and undergoes mass loss according to \citet{Booth_disint22}. Note the scale change in the $y$-axis. In the first case, without circulation (upper panel), a maximum depth is found by calculating the position of the critical melt fraction assuming conduction to the edge of the 1D portion of the code. This naturally depends strongly on the value of $P_0$. We also show the depth of $P_0$. In the second case (bottom panel), magma pool circulation is included, which produces a much shallower pool. See \S\ref{sec: moltenmass} for further details of the calculations.}
    \label{fig: pooldepth}
\end{figure}
\begin{figure}
    \centering
    \includegraphics[width=\linewidth]{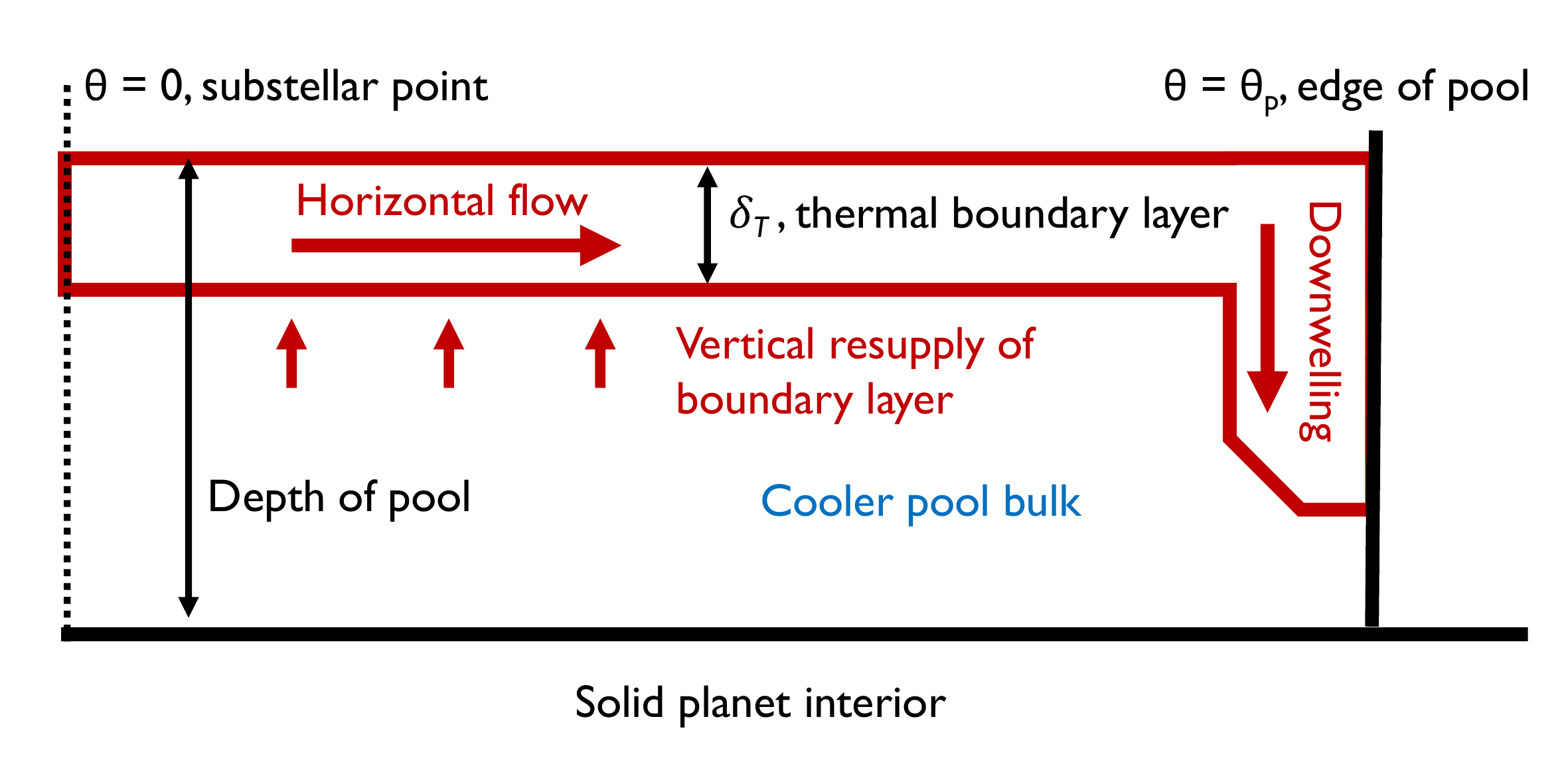}
    \caption{Schematic of convection in a surface lava pool generated by stellar irradiation. The strong angular temperature gradient at the surface drives a horizontal flow in a thin boundary layer that flows into the cooler bulk of the pool at its edge.}
    \label{fig: pool_schematic}
\end{figure}

We found in \S\ref{sec: results} that catastrophically evaporating planets are likely to have almost entirely crystallised by the time they are observed. However, the peak surface temperature is sufficient to melt rock (higher than the liquidus), so there must be a molten region on the dayside. 

The most naive way to calculate the depth of this pool is to assume a constant temperature gradient between the surface and a point in the interior where the temperature is known, i.e., conduction into the planet with constant conductivity. This neglects any convective heat transport in the magma pool (which we address below.) The depth of the pool in this case is the point along this temperature gradient where the material becomes solid, best defined as when the melt fraction reaches the critical value ($\phi = \phi_c$). We use $T_0$, the temperature at a pressure $P_0$, the edge of our 1D model (see \S\ref{sec: BCs}) as the interior temperature, and thus the depth is dependent on what we choose for $P_0$, as shown in \autoref{fig: pooldepth} (top panel). Even this naive calculation shows that the pool is shallow relative to the planet's size ($\gtrsim 1000$ km), but, as we shall describe below, including circulation makes the pool even shallower.

\citet{Kite16} argue that lateral circulation in the lava pool greatly reduces the depth of the pool, compared to the above estimate. The reason for this is the thermal structure created by horizontally driven convection. Convection in the pool is driven by the day-to-nightside temperature gradient at the surface.\footnote{As was noted in \S\ref{sec: BCs}, however, this convection cannot smooth the day-to-nightside temperature gradient.} This configuration is known through experiments and theory to create a thin thermal boundary layer at the surface, as shown in \autoref{fig: pool_schematic} \citep[e.g.,][]{Vallis2006,Hughes+Griffiths2008}. Using scaling laws from \citet{Vallis2006}, the depth of this boundary layer is
\begin{equation}
    \delta_T = \left(\frac{k}{\rho C_P} \frac{4 \Omega \sin(\theta_p/2)\tan(\theta_p/2) R^2\theta_p}{(\Delta\rho/\rho) g} \right)^\frac{1}{3} \label{eq: pool BL}
\end{equation}
\citep[adapted from][Eq. 8]{Kite16}. Here $R$ is the planetary radius, $\Omega$ its orbital frequency, $\theta_p$ the angular size of the pool centred on the substellar point, and $\Delta\rho$ the density change across the boundary layer. Since there is a large temperature drop across the thermal boundary layer at the surface, the point where $\phi = \phi_c$ will be shallower than in the naive case. We adopt the assumption in \citet{Kite16} that the total depth of the pool should not be more than $\sim 10$ times this boundary layer.

$\theta_p$ may be found, for a given substellar temperature, $T_{ss}$, by using the fact that the surface temperature varies approximately as $\cos^{1/4}\theta$ (see \S\ref{sec: BCs}), and finding the point that the critical melt fraction is reached on the surface. $T_{ss}\sim 2100$ K, representative of the known systems KOI 2700b and K2-22b (\autoref{tab: systems}), gives, using our melting curves (\S\ref{sec: melting}), a very wide pool with $\theta_p \approx 0.4\pi$. If one takes surface values of the physical quantities, assuming $\Delta\rho = 10\%$, a typical value for melt-to-solid density change, at a one-day orbit, then one finds the scaling
\begin{equation}
     \delta_T \approx 18\text{m} \left(\frac{R}{R_\oplus}\right)^\frac{2}{3}\left(\frac{g}{g_\oplus}\right)^{-\frac{1}{3}} \; . \label{eq: pool scaling}
\end{equation}
Since for an individual planet, the other terms in \autoref{eq: pool BL} do not change, \autoref{eq: pool scaling} may be used to track the evolution of the pool's depth for that planet.

We plot the evolution of the pool depth, assuming it is equal to 10 $\delta_T$, for one case in \autoref{fig: pooldepth} (bottom panel). This demonstrates the scale difference between this case, including circulation, and the simple argument without, as noted by \citet{Kite16}.

It is interesting to note the different evolutions for the two cases. At early times the depth of the pool with both estimates decreases due to cooling and contraction of the planet. The contraction causes the gravity to increase; thus, the pressure at which the surface lava becomes solid is at a shallower depth. At later stages, the naive estimate increases, whereas the estimate with circulation decreases. This is because, at this point, mass loss is the most important factor. For the naive estimate, the corresponding decrease in gravity means that the depth to a certain pressure increases, meaning the pool depth increases. On the other hand, for the estimate with circulation, the pool's depth decreases as the planet gets smaller and denser. This is because the estimate is coupled to the size of the pool's thermal boundary layer which scales with mass and radius in a different way (\autoref{eq: pool scaling}). However, it should be noted that the fractional changes for both cases are relatively small, so the true time evolution may well be altered by effects we have not captured, for instance, compositional changes in the pool.

The important point, however, is that, in either case, it is likely that the planets have only a shallow molten region on their dayside. This has consequences for the composition of the dusty tails, which are the observational signatures of catastrophically evaporating planets. The evaporation that generates the tails must come from the molten region of the planet. Our results show that this is only a shallow region at the surface at any time, which gradually moves into the planet. Were the planet fully molten, relatively volatile elements such as \isotope[]{Na} would be lost. However, with a shallow pool, it is possible that they are locked in the solid portion of the planet, which is gradually melted away, and so are still present in the winds. A more detailed consideration of the chemistry is required to determine the compositional evolution, which will come in a future study.

\subsection{Assumptions in the boundary conditions} \label{sec: BC caveats}
As discussed in \S\ref{sec: BCs}, due to the complexity of the radial and angular heat distribution of the star's energy, we made some assumptions about the outer layers of the planets. We argued that the inward heat fluxes should be small compared to the outward ones, which justified having the inner regions of the planet evolve as a spherically symmetric structure, unaffected by the star. At very late times, when all internal energy has dissipated, this assumption is clearly untrue. Therefore, in the following, we shall check whether including any inward or angular fluxes would affect our conclusions.

At very late times, for tidally locked planets, the temperature should decrease from the dayside to the nightside, and energy should flow through the planet. Heat flux, in this case, should be of the order
\begin{equation}
    F = k \frac{T_{ss}-T_n}{2R}
\end{equation}
where $T_n$ is the nightside temperature, and $k$ is again the conductivity. For a 1000 km planet with $T_{ss} = 2320$ K, and a negligible nightside temperature, this gives a value of $\sim \SI{5e-3}{Wm^{-2}}$. The blackbody temperature required to emit this on the nightside is 17 K, justifying the assumption of the nightside temperature being negligible. One can see in \autoref{fig: bulk no mdot} that even after 10 Gyrs, the outward fluxes do not get this low, meaning the assumption of no inward flux is justifiable up to this point. Furthermore, the planet crystallised significantly before this point when the outward flux was at least 20 times higher. Therefore, the additional flux contribution would not be enough to melt the planet.

We also assumed that angular fluxes were small, which allowed us to estimate the flux from the nightside simply as that coming from the interior, not the dayside. Angular fluxes in the solid planet should also be of a similar order to the late-time day-to-nightside flux, estimated above; thus those fluxes become important at a similar time. Consequently, any adjustment to the nightside temperature would only become significant once crystallisation has occurred, meaning angular fluxes also cannot affect our conclusion about the planet having solidified.

We have so far only shown examples where the pressure at the edge of the 1D interior model is fixed at $P_0 = 1$ GPa. We chose this pressure as regions deeper than it are unlikely to be affected by stellar heating. A demonstration of this is that assuming conduction inwards from the surface, the flux into $P_0$ will be approximately:
\begin{equation}
    F_{in} \sim \frac{k(T_{ss}-T_0)}{P_0} \rho g
\end{equation}
with quantities as defined in \S\ref{sec: Method}, and evaluated at the surface. Taking a typical value for our planets of $g=\SI{4.5}{\m\s^{-2}}$ and $T_{ss}=\SI{2300}{K}$ and assuming $T_0=\SI{1500}{K}$, which is below the solidus, gives $\sim \SI{0.05}{W\m^{-2}}$. This is less than the amount of outward flux at the point when the planet has fully crystallised, meaning at least up to that point $P_0 = 1$ GPa is deep enough to assume inward heat flux to deeper in the planet is small. Furthermore, there is heat redistribution in the pool, meaning this is likely an overestimate of any inward flux, as addressed in the previous section.

We have justified that this value of $P_0$ is valid in terms of having a low enough inward heat flux through it. However, the exact choice is still arbitrary, so it would be useful to test the effect that choosing a different valid value would have. If the value was much deeper, the assumptions of constant gravity and flux in the outer region become less accurate since more than 10\% of the total mass of smaller planets would be included in the boundary region.\footnote{For instance, for a $0.04M_\oplus$ planet, 8\% of the mass is already at pressures below $P_0 = 1$~GPa.} Therefore, to investigate the effect of choosing this arbitrary pressure, we instead compared to a lower pressure case of $P_0 = 0.5$ GPa. We found that, in this case, deep mantle temperatures (close to the core-mantle boundary) change by less than 0.5\%, and upper mantle temperatures (around 1GPa) change by less than 5\%. However, because the temperature of the upper mantle does not change much between 10,000 years and several Gyrs (\autoref{fig: bulk no mdot}, middle panel), and the viscosity is strongly dependent on temperature, the time at which crystallisation occurs is nevertheless sensitive to the small change. Crystallisation of the mantle below $\phi_c$ takes a few hundred years longer, and full crystallisation (the whole mantle having $\phi=0$) takes a few hundred Myrs longer (differences of $\sim10\%$.) This effect is a consequence of some simplifications made in the outer layer with pressure below $P_0$. The outer layer does not include any energy production due to cooling or radioactive decay. Therefore, in the 0.5 GPa case, the planet has more energy available due to the outer layer being a smaller proportion of the planet and so cools slightly slower. 

It should be noted, however, that beyond 10 Myrs, while the melt fraction in the mantle is non-zero, it is very low (\autoref{fig: bulk no mdot}, bottom panel and \autoref{fig: depth}), so the planets are essentially solid. In Figures \ref{fig: bulk no mdot} and \ref{fig: crystallisation mdot}, we plotted $\phi = 5\%$, to represent a low melt fraction that could probably be considered solid. This would have to be delayed to 1~Gyr to alter the conclusion that planets are solid when they are observed evaporating around main sequence stars. This point is delayed by a few 10s of Myrs by halving $P_0$, so we deem it unlikely that our model differs sufficiently from the physical reality to change this overall conclusion.

\subsection{Further caveats}\label{sec: discuss-other}
We have used a relatively simple model to investigate the evolution of the catastrophically evaporating planets. Thus there is necessarily some physics missing. For instance, the composition is fixed and uniform in the mantle. In reality, there may be planet-to-planet variation and compositional evolution within the planet. However, these are unlikely to change the density and heat capacity properties enough to alter our conclusions significantly. The composition would also affect the solidus and liquidus, which may make a more significant difference due to the difference between melt and solid having such a large effect on viscosity (see \S\ref{sec: viscosity}). A full investigation of this is beyond the scope of this work, but given the whole mantle passes below the critical melt fraction fairly early, such differences are likely unimportant over the planets' full lifetimes. 

We also assumed that the melting is in equilibrium, meaning temperature and pressure directly determine the melt fraction. Non-equilibrium effects are potentially important for compositional evolution during the early magma ocean phase \citep[e.g.,][]{SOLOMATOV-chapter}, but once again, since it passes through the critical melt fraction point early on, these effects are unlikely to be important for long term evolution.

Another simplification we made is in the treatment of the iron core. Since we are most interested in the mantle's evolution, the core's precise state is not important, only the energy it provides to the mantle. In reality, the core will start molten and solidify over time, which generates extra energy through the latent heat of fusion and through gravitational potential energy as the iron density increases through the phase transition, resulting in the contraction of the core. However, we do not include such a phase transition.

For our example in \autoref{fig: internal no mdot}, the melting point of iron at the core-mantle boundary is around 2450K \citep{SLUITER2012}. This means that the planet must be between $10^8 - 10^9$ years old, and the mantle is already essentially crystallised by the time any of these extra heat sources start to come into effect.

The total energy released as latent heat may be estimated as $E_L \sim M_c \Delta L_{Fe} $, where $\Delta L_{Fe}$ is the latent heat of fusion of iron, and $M_c$ the mass of the core. The energy from gravitational contraction can be estimated with $E_g \sim GM_c^2/R_c(\frac{1}{3}\Delta \rho_c/\rho_c)$, with $R_c$ the core radius, $\rho_c$ the density of iron and $\Delta \rho_c$ the solid melt density contrast. Typical values for, for instance, a $0.15 M_\oplus$ planet with a core mass fraction of 0.3 give total values of several $10^{28}$ J for these two energy sources. This is of order the amount of energy lost by the core by cooling, as calculated by our models. Thus these additional energy sources may have some effect on the mantle at late times. However, it seems unlikely that the luminosity of the core should increase at this point, as the energy would instead re-melt the core. Furthermore, the mantle itself typically has energies an order of magnitude above that of the core, both in thermal energy and radioactive elements. Therefore, it seems unlikely that including these core freezing effects would affect the fact that the mantle has crystallised first. 

Another energy source we have neglected is tides. The short orbits of these planets mean that tidal forces are potentially strong, although heating may not be long-lived because circularisation timescales are consequently short. \citet{Jackson2008}, suggest that heating rates can be high even for small but non-zero eccentricities, perhaps even high enough to keep the mantle molten. Considering these effects would require consideration of the orbital evolution, which is beyond the scope of this work.

Orbital evolution could also occur due to angular momentum exchange between the outflow and the planet, thus changing the planet's irradiation level. This might result in the planet spiralling outwards or inwards, cutting off mass loss or causing a runaway. Neglecting this effect is justified if the material from the planet is expelled from the system by the star, retaining its own angular momentum, meaning no angular momentum is exchanged with the planet. This fits with the current understanding of the dusty tails \citep[e.g.,][]{disint18}. Furthermore, \citet{Perez-Becker13} showed that there is little orbital evolution even when trying to maximise the amount of angular momentum exchanged.

\section{Occurrence rate of low mass planets}\label{sec: occurrence}

Our evolutionary models demonstrate that there is only a small period of time, relative to the main sequence lifetime of stars, that catastrophically evaporating planets have detectable mass loss, but have not yet completely evaporated, as also noted by \citet{Perez-Becker13}. The three systems that were found in the Kepler/K2 data must be in this phase. In \S\ref{sec: occ_will_evap} we estimate the occurrence rate of planets that evaporate at detectable rates, at some point in their lifetimes, needed to explain the number of detections.

{The planets with detectable mass loss must be part of a wider distribution of low-mass planets, which is unlikely to be peaked at the temperature and mass that gives detectable mass loss. In \S\ref{sec: occ_general} we assume this distribution is a power law in order to extend our occurrence rate to a wider range of low-mass planets. This allows comparison of the sub-Earth mass planet population, which we can probe but is generally undetectable by conventional exoplanet surveys \citep[e.g.][]{Christiansen2014,Fulton17valley}, with observed occurrence rates for higher mass planets.

The inferred occurrence rates will depend on the mass loss model used since it changes the duration that planets are observed and the threshold of detectability. We present calculations with the same mass loss model we have used throughout \citep{Booth_disint22}.

\subsection{Occurrence rate of planets with detectable mass loss}\label{sec: occ_will_evap}
We can crudely estimate the number of planets that must be detectable at some point in their lifetimes, in order to produce the number of observed evaporating systems, using the equation 
\begin{equation}
    N_\text{detect} \sim n_p N_* f_\text{evap} P_\text{trans} \label{eq: rough_detect}
\end{equation}
\citep[see][]{Perez-Becker13}. Here $N_\text{detect}$ is the number of planets detected, $n_p$ is the unknown number of planets per star, $N_*$ is the number of stars surveyed and $f_\text{evap}$ is the fraction of the host star's lifetime that the planet is observable. $P_\text{trans}$ is the probability that the system is transiting in the right orientation to be observed from the Earth and is given by \citep{Borucki1984}
\begin{equation}
    P_\text{trans}(a,R_*) = \frac{R_*}{a} \label{eq: trans_prob}
\end{equation}
where $R_*$ is the stellar radius and $a$ the orbital separation.

To find $f_\text{evap}$ one must identify the phase in a planet's lifetime for which it is detectable. We address this over the following two subsections and then use it to estimate the occurrence rate from \autoref{eq: rough_detect}.

\subsubsection{Fitted function to the mass loss rate}\label{sec: fitted_mdot}

Whether a planet is observable is dependent on its mass loss rate, and so to find $f_\text{evap}$ we must determine the mass loss rates of planets uniformly over temperature and mass. Our mass loss models were only run for a finite number of temperatures and initial masses, so we require a method of interpolating between them.  We choose to fit the evolution of our models in mass- mass loss rate space ($M-\dot{M}$) using the function
\begin{equation}
    \dot{M}(t) = \left[ \left\{A M(t)^{a} \right\}^{-\gamma} + \left\{ B M_\textit{init}^{b} \exp( - \lambda M(t)^c) \right\}^{-\gamma} \right]^{-\frac{1}{\gamma}} \label{eq: fit_mass_loss}
\end{equation}
where $M_\textit{init}$ is the initial mass of the planet and all other parameters are fit for and depend on substellar temperature. The physical reasoning behind this form is as follows. The mass loss rate at any time, for a given substellar temperature, depends on the instantaneous mass and core mass fraction. In turn, the core mass fraction depends only on the current mass and initial mass, if the initial core mass fraction is assumed to be the same for all planets. Thus the mass loss rate is just a function of $M$ and $M_\textit{init}$. At low masses, the mass loss rate increases with mass, which we assume to act as a power law (first term in curly brackets, with $a > 0$.) At high masses, the mass loss decreases with increasing mass, approximately exponentially, hence the second term in curly brackets. These low and high mass behaviours are then combined using the smoothing parameter $\gamma$. In principle, more dependence on $M$ or $M_\textit{init}$ could be introduced, for instance, the first term could include an $M_\textit{init}$ dependence, but we found that the above form could capture the behaviour without introducing extra degeneracy.

The advantage of using this physically motivated form is that it not only allows interpolation but also extrapolation to higher core mass fractions with more confidence than an interpolation method such as a spline. The mass loss models become more numerically difficult to run at higher core mass fraction, and therefore we did not run them to a core mass fraction of 1 when the entire mantle has evaporated.

For each substellar temperature, we fit \autoref{eq: fit_mass_loss} to a selection of our numerical outputs with different initial masses and at different evolutionary stages using \texttt{scipy curve\_fit}. These are shown in \autoref{tab: mdot_fit_params}. To find the mass and mass loss rate evolution for a given initial mass and substellar temperature, we integrate this function numerically. We find that this can reproduce our full calculations to better than 8\%, even for initial masses that the function was not fitted with. For the purposes of our occurrence rate calculation this is sufficient given the other uncertainties and assumptions. To interpolate between substellar temperatures we use spline interpolation of the parameters in \autoref{eq: fit_mass_loss}/\autoref{tab: mdot_fit_params}.

\begin{table}\centering
\resizebox{!}{1.5cm}{%
\begin{tabular}{|l|c|c|c|c|}\hline
\multirow{2}{*}{\textbf{Parameter}} & \multicolumn{4}{c|}{\textbf{Substellar tempertature [K]}}                  \\ \cline{2-5}
                                    & \textbf{2070} & \textbf{2190} & \textbf{2320} & \textbf{2460} \\ \hline
$A$       &  4.99 &   4.1 &   100 &  78.9 \\\hline
$\alpha$   & 0.911 &  0.59 & 0.959 & 0.569 \\\hline
$B$       & 0.046 & 0.102 & 0.327 & 0.706 \\\hline
$a$       & -1.15 & -1.47 & -1.67 & -2.16 \\\hline
$b$       &  0.84 & 0.778 & 0.688 &   0.6 \\\hline
$\lambda$ &  30.8 &  26.5 &  21.3 &  16.3 \\\hline
$\gamma$  &  7.32 &  1.43 &  1.01 &  1.35 \\\hline
\end{tabular}
}
\caption{Best fit parameters for our boundary condition function \autoref{eq: fit_mass_loss}}\label{tab: mdot_fit_params}
\end{table}
\subsubsection{Length of time that systems are observable} \label{sec: evap_times}
\begin{figure}
    \centering
    \includegraphics[width=\linewidth]{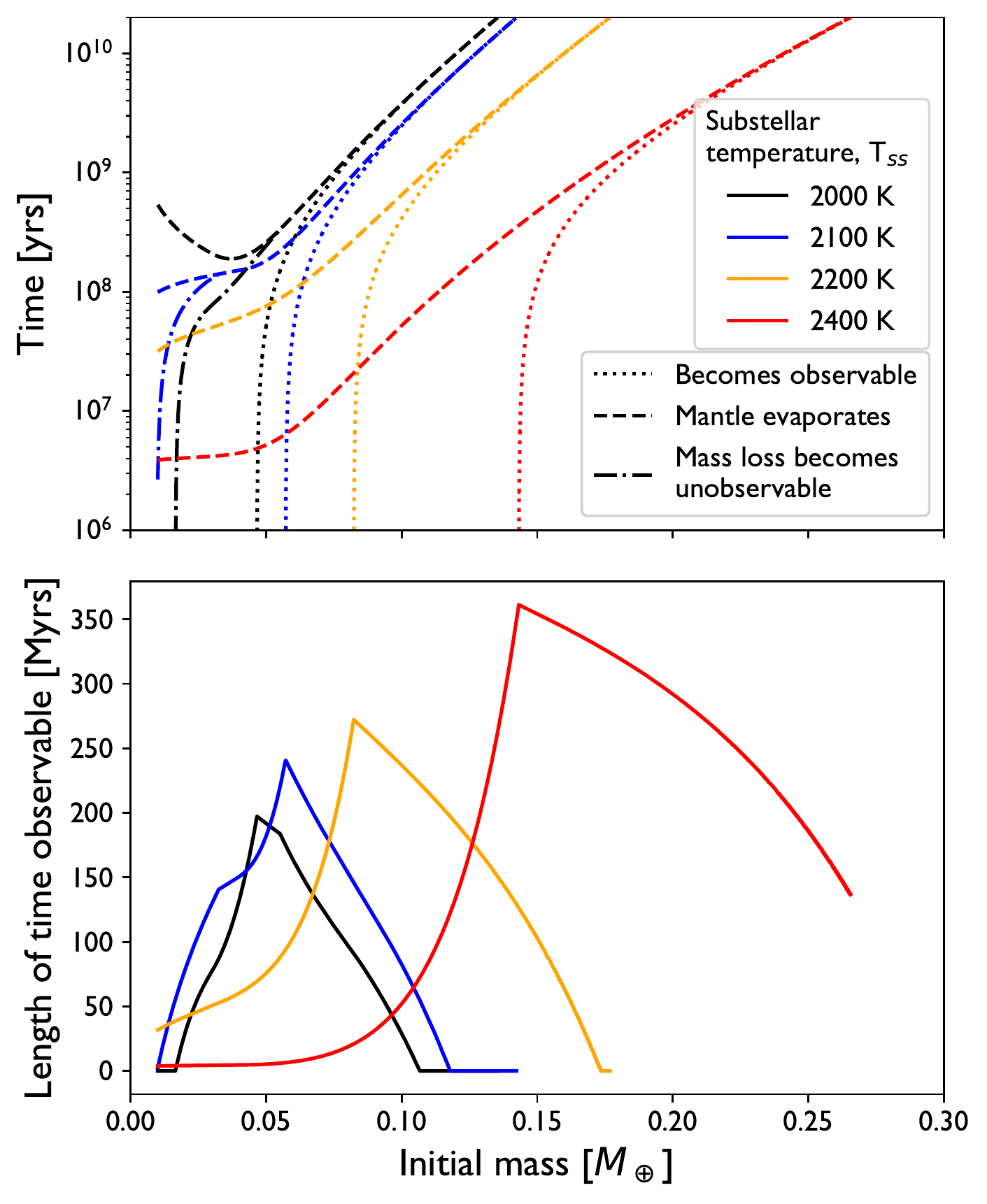}
    \caption{The time period of observable mass loss for planets with core mass fractions of 0.3, mass loss rates calculated using the method of \citet{Booth_disint22} and assuming that a planet becomes observable if its mass loss rate exceeds 0.1~$M_\oplus \text{Gyr}^{-1}$ \citep{Perez-Becker13}.  {\it Top panel:} The dotted line shows the age at which a planet of a given substellar temperature gains a mass loss rate high enough to be observable. The dashed line shows the time when the whole of the planet's mantle evaporates, while the dash-dot shows when the mass loss rate becomes unobservable, which mostly coincides with the mantle evaporation line. {\it Bottom panel:} The length of time that a planet has an observable mass loss rate as a function of initial mass and substellar temperature.}
    \label{fig: optimisticTime_min_max}
\end{figure}
\begin{figure}
    \centering
    \includegraphics[width=\linewidth]{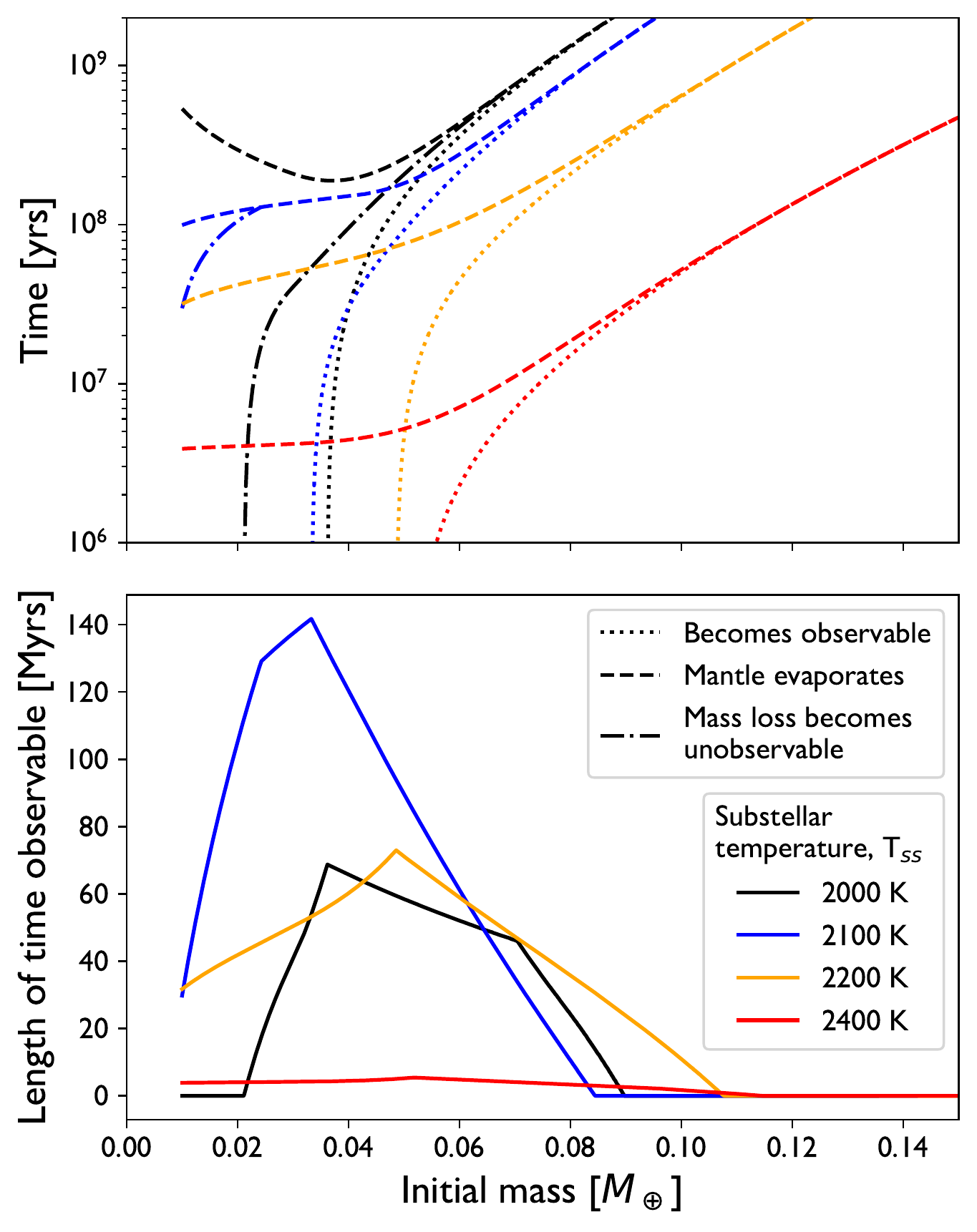}
    \caption{The same as \autoref{fig: optimisticTime_min_max}, but instead planets are considered observable if the models of \citet{Booth_disint22} predict dust production. Note the difference in the range of initial masses compared to \autoref{fig: optimisticTime_min_max}.}
    \label{fig: BoothTime_min_max}
\end{figure}

To work out the times when the planets are observable, we solve the ODE in \autoref{eq: fit_mass_loss} over a range of substellar temperatures and initial masses to find the mass loss rate evolution using two definitions of the detectable mass loss rate. Firstly, an optimistic case (henceforth {\it Case 1}) where we consider planets detectable if their mass loss rate exceeds 0.1$M_\oplus \text{Gyr}^{-1}$, which is considered a lower limit on detectability by \citet{Perez-Becker13}. In contrast, we also consider the threshold to be the region of temperature and current mass where dust is produced in the models of \citet{Booth_disint22} (henceforth {\it Case 2}), shown in \autoref{fig: mdot_grid}. This should, in principle, be the more physical case.

We do not consider the evaporation of the core after the mantle has evaporated because the composition of the dust tails is thought to be inconsistent with iron \citep{vanLieshout14,vanLieshout16}. Therefore, we are technically only considering the progenitors of planets with observed evaporating mantles. That being said, iron likely has a much higher mass loss rate \citep{Perez-Becker13} due to its higher vapour pressure, and so a planet's evaporation should accelerate once the iron core is exposed. Therefore the lifetimes we calculate are likely reflective of planets having both their mantles and cores evaporated. 

The times at which systems start and cease to be observable as well as the length of time that they are observable are shown in Figures \ref{fig: optimisticTime_min_max} and \ref{fig: BoothTime_min_max} for {\it Case 1} and {\it 2} respectively. In both cases for lower-mass planets, the mass loss is observable from birth (see top panels), and so the length of time a planet is observable is limited by the time it takes either to evaporate or for the mass loss rate to decrease below detectability. As a result, the length of time that the planet is detectable increases with initial mass (lower panels).

Once planets are massive enough, they take time to become observable, and the detectability is now limited by the length of time it takes to become observable, which increases with initial mass. In both {\it Case 1} and {\it 2} the time it takes to become observable and the time it takes to lose the entire mantle seem to converge (top panels), which is conceptually because at high masses it is possible to lose the entire mantle without ever reaching a high enough mass loss rate to be observed. The two lengths of time are only actually equal before the age of the universe for the lowest temperatures. The peak length of time occurs when a planet starts small enough that it has a detectable mass loss from birth but it is large enough that it evaporates for a long time. Planets that start with higher masses spend less time observable because at the point when they reach this mass, the mass of the mantle is lower since all planets start with the same core mass fraction. For {\it Case 1} the peak length of time moves to higher masses as the temperature increases. This is because the higher temperatures allow larger-mass planets to have observable mass loss rates. This effect is drowned out by other temperature-dependent effects for {\it Case 2}, which we will discuss below.

A major difference between {\it Case 1} and {\it 2} is that for {\it Case~1} the timescales that systems are observable for increase with increasing temperature (\autoref{fig: optimisticTime_min_max}), whereas for {\it Case 2} the timescales decrease for high temperatures (\autoref{fig: BoothTime_min_max}). This is due to the different assumptions behind the two cases. Firstly, in {\it Case 2} dust is no longer produced at higher temperatures. The physical reason for this is that the dust condensation rate is too low compared to the flow rate. Hence dust simply doesn't have time to form before the gas leaves the planet, as discussed further in \citet{Booth_disint22}. Furthermore, for higher temperatures, mass loss rates increase, meaning planets spend less time in the observable region (see \autoref{fig: mdot_grid}). These two effects result in the timescale decrease with temperature for {\it Case 2}, seen in \autoref{fig: BoothTime_min_max}. In contrast in {\it Case 1} the observable timescale increases with temperature. This is because higher temperatures allow more massive planets to have observable mass loss. Larger mass planets take longer to fully evaporate, so the overall effect is that the timescales increase. This also means that the range of initial masses that can be detectable is much higher for {\it Case 1}, where there is no high-temperature cut-off.

A final difference for {\it Case 2} is that the observable timescales are overall shorter since the mass loss rate required to be detectable is, at its lowest, similar to but generally higher than the fixed value for {\it Case 1}.

For the calculations we show, we assume that the progenitor planets all have core mass fractions of 0.3. If there were a range of core masses the amount of time the mantle is evaporating would change. It is not immediately clear whether it would increase or decrease because a larger mantle has more material to evaporate, but evaporates quickly as it has a lower gravity for the same mass. We briefly investigated this by using models with a very low core mass fraction of 0.1. We found that these two effects almost cancel out and that the timescales of observable evaporation are very similar, although the lower core mass fraction examples become observable earlier, which would slightly decrease their observability around old stars. Consequently, we do not believe this will greatly affect our analysis in the following sections.

\subsubsection{Basic estimate of the number of planets that will have detectable mass loss} \label{sec: rough estimate}

$f_\text{evap}$ in \autoref{eq: rough_detect} may be estimated as the ratio of the observable period to a typical stellar age. Taking a typical age to be $\sim 5$~Gyrs, and a typical observable time (Figures \ref{fig: optimisticTime_min_max} and \ref{fig: BoothTime_min_max}) to be around 200~(80)~Mrys for {\it Case 1} ({\it 2}) gives $f_\text{evap} \sim 0.04$ (0.016).

The primary Kepler survey observed $\sim 180,000$ stars. Of these $\sim 145,000$ are main sequence stars, among which two catastrophically evaporating systems have been identified (Kepler 1520b and KOI-2700b -- we disregard K2-22b as it was part of the separate K2 survey). Therefore, we take $N_\text{detect} = 2$ and $N_* = 145,000$. We further use a value of $P_\text{trans} = 0.25$, for the archetypal system Kepler 1520b (properties can be found in \autoref{tab: systems}). \autoref{eq: rough_detect} then implies that the required number of planets per star that undergo observable evaporation at some point in their lifetimes, in order to reproduce the detected number of these planets, is $n_p =$ 0.14\% (0.3\%) for {\it Case 1 }({\it 2}). This is essentially the same as the estimate of 0.1--1\% in \citet{Perez-Becker13} since the detectable timescales are very similar.

\subsubsection{Probability of detection across the stellar population}\label{sec: prob across pop}

\begin{figure}
    \centering
    \includegraphics[width=\linewidth]{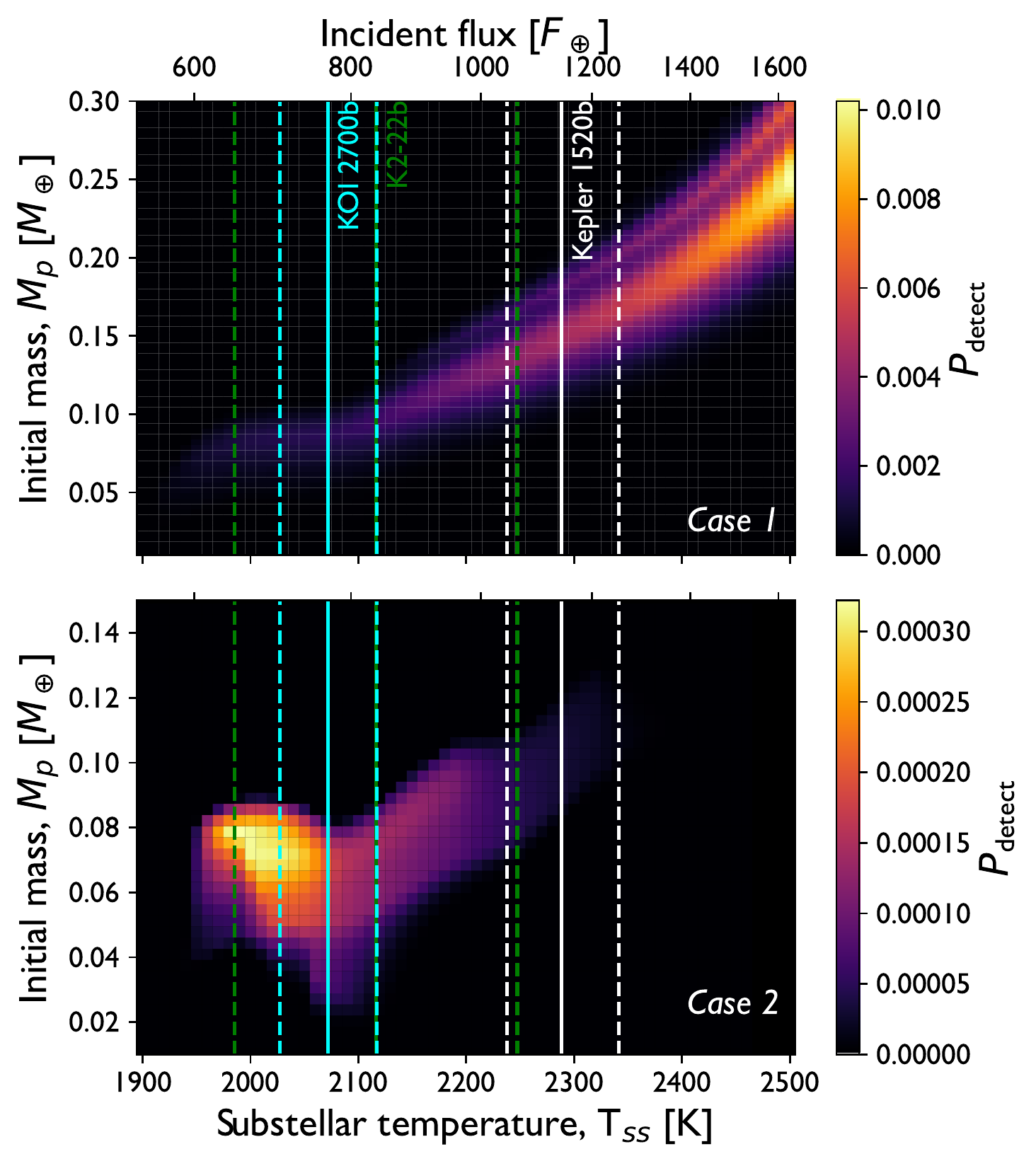}
    \caption{The probability that a planet of a given initial mass would be observable around a star, marginalised over the stellar mass and age distribution of stars observed in the Kepler primary mission, $P_\text{detect}(M_p,T_{ss})$ in \autoref{eq: P_detect_a_T}. In the top panel, we consider a planet observable if its mass loss rate exceeds 0.1~$M_\oplus \text{Gyr}^{-1}$ \citep{Perez-Becker13}, {\it Case 1} in the text. In the bottom panel, planets are considered detectable if the models of \citet{Booth_disint22} predict dust production (see \autoref{fig: mdot_grid}, {\it Case 2}). Note the different y-axis scales and range of $P_\text{detect}$ values. Solid lines show the substellar temperatures of the known systems, with associated error bars shown with dashed lines. Note that the best fit substellar temperature of K2-22b coincides with the upper error bar of KOI 2700b and that K2-22b has significantly higher errors, as the stellar parameters are less well constrained.}
    \label{fig: bothProbTss_Mp}
\end{figure}

The calculation above uses just a single value for $P_\text{trans}$, which is a function of stellar properties. Here we improve our estimate by taking into account the full variation of stars that were observed. 

Kepler's primary mission observed stars with a distribution of masses ($M_*$) and ages ($t$) which we will write as $n_*(M_*,t)$, defined as the number of stars in the range $[M_*,M_*+\mathrm{d}M_*]$ and $[t,t+\mathrm{d}t]$.

One can work out the probability of detecting a planet catastrophically evaporating around the average star, as a function of the planet's initial mass and substellar temperature (\autoref{eq: Tss}), by integrating over the stellar population:
\begin{equation} 
    \begin{split}
    P_\text{detect}(M_p,T_{ss}) =  \\ \int_{M_{*,\text{min}}}^{M_{*,\text{max}}}\int_{t_{\text{obs},\text{min}}(T_{ss},M_p)}^{t_{\text{obs},\text{max}}(T_{ss},M_p)} \hat{n}_*(M_*,t) P_\text{trans}(a,R_*) \; \mathrm{d}t \mathrm{d}M_* \; .
    \end{split} \label{eq: P_detect_a_T}
\end{equation}
Here $\hat{n}_*(M_*,t)$ is the normalised distribution of stellar ages and masses in the sample, i.e., $n_*(M_*,t) / N_*$, where $N_*$ is again the total number of stars. $P_\text{trans}(a,R_*)$ is as defined in \autoref{eq: trans_prob}. $t_{\text{obs},\text{min/max}}(T_{ss},M_p)$ denote the earliest and latest times that the evaporation of a planet is observable, which are naturally functions of the planet's properties and were calculated in \S\ref{sec: evap_times} (Figures \ref{fig: optimisticTime_min_max} and \ref{fig: BoothTime_min_max}). This integral is a generalisation of the product $f_\text{evap} P_\text{trans}$ in \autoref{eq: rough_detect}.

To find the distribution of stellar masses and ages in the Kepler sample ($N_*(M_*,t)$) we use the stellar properties derived from {\it GAIA} in the GKS sample \citep{Berger-GKS20} and apply the colour cut in \citet{Fulton17valley} to remove giant stars. To calculate $R_*$ and $L_*$ we use the {\it MIST} stellar isochrones \citep{MIST-Dotter16,MIST-Choi16,MESA1,MESA13,MESA15} with solar metallicity. While this neglects metallicity variation, our simple model does not warrant this complication. In order to evaluate \autoref{eq: P_detect_a_T} we binned the stars in the GKS sample according to stellar age and first integrated over the distribution of stellar masses for each age bin before integrating over the age distribution. Each distribution was normalised, thus giving the probability per star.

$P_\text{detect}(M_p,T_{ss})$ is plotted in \autoref{fig: bothProbTss_Mp}, for both the cases we consider. In the top panel, we see the result of the trends identified for {\it Case 1} in \S\ref{sec: evap_times}. Higher-mass planets are observable at higher temperatures and it takes longer for the higher-mass planets to evaporate, thus they are more likely to be observed. However, the contours show two ridges for a given temperature. The ridge at lower initial mass is due to the peaks in the length of time observed, as seen in \autoref{fig: optimisticTime_min_max}. However, stars that are several Gyrs old are more common, which means at a population level it is more likely to observe a larger initial mass planet because they are more long-lived, even if the time they are observable might be shorter. This creates a second ridge at higher initial mass. The upper value of $P_\text{detect} \sim 0.01$ also corresponds to the rough estimate in \S\ref{sec: rough estimate} since $P_\text{detect} \sim f_\text{evap} P_\text{trans} \sim 0.04 \times 0.25$.

The bottom panel of \autoref{fig: bothProbTss_Mp}, meanwhile, reflects the trends for {\it Case 2} (\S\ref{sec: evap_times}, \autoref{fig: BoothTime_min_max}), with detectability decreasing for higher temperatures. The consequence of this is that there are cuts of observability at both low and high temperatures. In this case, even the peak probability of detection, $\sim \SI{3e-4}{}$, is lower than that roughly estimated in \S\ref{sec: rough estimate} of $P_\text{detect} \sim f_\text{evap} P_\text{trans} \sim \SI{4e-3}{}$. This is because the observable planets are limited to cooler temperatures, and are thus further away and so less likely to be transiting (\autoref{eq: trans_prob}).

One can use these refined probabilities to redo the calculation in \S\ref{sec: rough estimate}. For {\it Case 1} the value of $n_p$ is unchanged at 0.14\%; for {\it Case 2}, taking a typical value of $P_\text{detect} \sim \SI{2e-4}{}$, gives a requirement of $\sim 6.8\%$ of stars having a planet which can have observable catastrophic evaporation in its lifetime. These values of $n_p$, by definition, refer just to planets that have observable mass loss in their lifetimes, i.e., just the non-zero probability regions in \autoref{fig: bothProbTss_Mp}. Thus for {\it Case 1} the value refers approximately to planets of $0.1< M_p/M_\oplus < 0.3$ and $\SI{2200}{K} < T_{ss} <\SI{2600}{K}$ and for {\it Case 2} it refers to $0.04< M_p/M_\oplus < 0.08$ and $\SI{1950}{K} < T_{ss} <\SI{2200}{K}$. This will be an important consideration in the next section.

\subsection{General occurrence rate of low mass, short-period planets}\label{sec: occ_general}
Currently, our calculated occurrence rates only encompass planets that will drive observable mass loss. However, it is unlikely that low-mass planets only exist at the temperature and sizes that produce observable mass loss. Thus, the observed catastrophically evaporating planets are likely to be part of a wider population of low-mass planets. Therefore, we can use our models of when planets do and do not produce observable mass loss to estimate the general occurrence rate of low mass, short-period planets\footnote{As discussed in Section~\ref{sec:other_occurence}, some planets with very low masses are evaporated away before they can be observed. Thus, our general occurrence rate can be thought of as the occurrence rate of the \it initial population} of low-mass, short-period planets, assuming their orbits remain unchanged.. Since not all low-mass, short-period planets will give rise to detectable mass loss, this {\it general} occurrence rate is necessarily larger than our previously calculated values.  At the most basic level, this increase in planetary occurrence rate from just the observable population to the general population will scale with the ratio of parameter space volumes occupied by the general population compared to the observable population. Since the parameter space volume occupied by the observable planets is small (i.e. we require a narrow range of planet masses and substellar temperatures), the occurrence of the general population will be significantly larger than just the observable catastrophically evaporating population. 

\subsubsection{Calculation}
In order to deduce the general occurrence rate of low-mass planets from the catastrophically evaporating planets, one needs to know the region of parameter space that gives rise to detectable mass loss and how this ``observable'' region compares to the underlying population. We have already worked out the shape of the observable region for two different cases in \S\ref{sec: prob across pop} (\autoref{fig: bothProbTss_Mp}). However, the distribution of the underlying population is unknown, but it is probably fair to assume it is reasonably smooth and not peaked exactly at the region where mass loss is observable.

We define the number distribution of the general low mass, short-period planet population, $n_p(M_p,T_{ss})$, as the number of planets per star with masses and substellar temperatures in the ranges $[M_p,M_p+\mathrm{d}M_p]$  and $[T_{ss},T_{ss}+\mathrm{d}T_{ss}]$. Following both observed populations \citep[e.g.,][]{Petigura2022} and population models \citep[e.g.,][]{Makino98-runaway}, we choose to model it as a power law:
\begin{equation}
    n_p(M_p,T_{ss}) \equiv \diffp{N_p}{{T_{ss}}{M_p}} =  C M_p^\alpha T_{ss}^\beta \label{eq: n_p dist}
\end{equation}
where $C$ is a normalisation constant, and $\alpha$ and $\beta$ are indices that we will investigate different values of. This is implicitly assumed to be universal over stellar type and age. We choose this parameterisation because $M_p$ and  $T_{ss}$ are independent variables in our model.

In a generalisation of \autoref{eq: rough_detect}, the number of detections can be written as
\begin{equation} 
    \begin{split}
    N_\text{detect} = N_* \int\int n_p(M_p,T_{ss}) P_\text{detect}(M_p,T_{ss}) \; \mathrm{d}M_p\mathrm{d}T_{ss} \; .
    \end{split}\label{eq: n_detect}
\end{equation}

where we highlight now that $P_{\rm detect}$ can be zero when a planet with a given mass and surface temperature never produces observable mass loss. Thus, if we assume some $\alpha$ and $\beta$, we can evaluate this integral and infer the normalisation factor, $C$, that is needed to match the number of detections, $N_\text{detect}$, allowing us to infer the number of underlying planets as a function of the assumed shape of the distribution.

As explained at the start of this section, the idea is to extrapolate from planets in the region of parameter space with observable mass loss to the general population, so the detectable mass loss region must be well defined. This is not the case for {\it Case 1}, where the detectable region continues to extend to higher substellar temperatures (\autoref{fig: bothProbTss_Mp}, upper panel). However, if the stellar irradiation is too high, dust should never be able to condense. Therefore, to perform the calculation for {\it Case 1}, we impose a maximum sub-stellar temperature, above which it is unlikely for dust particles to exist. In reality, this cut-off should be compositionally dependent, but we simply set it to be 2500 K. It is important to note that our occurrence rate calculation for {\it Case 1} explicitly depends on this choice.

It is instructive to convert our inferred number distribution into a measure that can be compared with other occurrence rate calculations. We choose to compare our results to calculations in \citet{Petigura2022}, which is an up-to-date analysis of the {\it Kepler} data, using {\it GAIA} derived stellar properties. They present occurrence rates as a function of planetary irradiation, $F_{irr}$, which is equivalent to substellar temperature through \autoref{eq: Tss} ($F_{irr} \equiv F_{irr}(0)$), and also give an analytic fit to their occurrence rate. Their occurrence rate is defined as 0.5~$\partial N_p / \partial \log{F_\text{irr}}$ for planets between 1--1.7 $R_\oplus$, which is roughly a 0.23 dex bin in radius.  Thus, we choose to integrate over the same logarithmic range in radius, but centred on $0.2 M_\oplus$ ($\sim 0.6 R_\oplus$) for {\it Case 1} and $0.06 M_\oplus$ ($\sim 0.4 R_\oplus$) for {\it Case 2}. These values are chosen because they are close to the peaks of $P_\text{detect}$ for each case (\autoref{fig: bothProbTss_Mp}). To convert from a radius range to a mass range, we assume a simple mass-radius relation of $M \propto R^4$ \citep[e.g.,][]{Valencia_Sasselov2007}, which means that the \citet{Petigura2022} range of 0.23 dex in planet radius corresponds to 0.92 dex in mass \footnote{We only use this mass-radius relation for the conversion of the radius range to an equivalent mass range. This is significantly easier with a simple relation such as this. The specific values of radius that we quote for a given mass are calculated using our interior model for a planet with a core mass fraction of 0.3.}.

The resultant values for the occurrence rate are shown as a function of the power law indices in \autoref{fig: both_alpha_beta}. The axes of \autoref{fig: both_alpha_beta} attempt to cover a range that should encompass the physical reality. The $x$-axis, $\alpha$, varies the dependence on initial planetary mass, and we ensure that we encompass values of $\sim -8/3$ suggested for runaway growth through planetesimal accretion \citep[e.g.,][]{Makino98-runaway}.
The $y$-axis shows the dependence on substellar temperature, $T_{ss}$. \citet{Petigura2022} measure the dependence on incident flux to be $\sim F_\text{irr}^{-2.6}$~to~$F_\text{irr}^{-3.1}$ corresponding to $\beta \sim -7.4$~to~$-9$, so we extend to values beyond this.

A very notable feature of \autoref{fig: both_alpha_beta} is the strong variation in the occurrence rates with $\alpha$ and $\beta$ for {\it Case 1}, but not for {\it Case 2}. The reason for this is the area of $M_p$--$T_{ss}$ parameter space that the two cases cover. The occurrence rate, 0.5~$\partial N_p / \partial \log{F_\text{irr}}$, is essentially just a scaling of $n_p$ evaluated at the peak of $P_\text{detect}$. Since the non-zero region of $P_\text{detect}$ is so small for {\it Case 2} (bottom panel of \autoref{fig: bothProbTss_Mp}), $P_\text{detect}$ effectively acts like a delta-function. This means that the integral \autoref{eq: n_detect} reduces to approximately $N_\text{detect} \propto n_p$, with $n_p$ evaluated at the peak of $P_\text{detect}$. Thus, the occurrence rate has a weak $\alpha$ and $\beta$ dependence. In contrast, for {\it Case 1}, the non-zero area of $P_\text{detect}$ is much larger, meaning the variation of $n_p$ over that area is much more important, so the occurrence rate has a much stronger dependence on $\alpha$ and $\beta$. 

As expected, the general occurrence rates ({\it Case 1}: 0.5--20\%, {\it 2}: 60--120\%) are much higher than those calculated in \S\ref{sec: prob across pop} ({\it Case 1}: 0.14\%, {\it 2}: 6.8\%), since we are now considering the number of planets per star in a larger region of parameter space. The values in \S\ref{sec: prob across pop}  are only true for the non-zero $P_\text{detect}$ region (\autoref{fig: bothProbTss_Mp}), whereas our more general measure, 0.23 dex in radius and 0.5 dex in $F_\text{irr}$ corresponds to 0.07--0.58$M_\oplus$ and 2115--2693K for {\it Case 1} and 0.02--0.17$M_\oplus$ and 1775--2367K for {\it Case 2}, which is a region at least an order of magnitude larger than the region populated by observable catastrophically evaporating planets. 

\subsubsection{Comparison to other occurrence rate values}
\label{sec:other_occurence}
In \autoref{fig: compare_occ} we compare our occurrence rates with that found by \citet{Perez-Becker13} for the catastrophically evaporating planets; with super-Earth (1--1.7 $R_\oplus$, $\sim$ 1--8$M_\oplus$) observations from \citet{Petigura2022}; and with sub-Earth mass planets of radii 0.5--0.75 $R_\oplus$ (0.6--0.3 $M_\oplus$) from \citet{Hsu2019}.  We convert the values from \citet{Hsu2019} (their Table 2) to the same measure as \citet{Petigura2022}, assuming a $1M_\odot$ star (the median stellar mass for the GKS sample), with solar metallicity, as they only present results as a function of orbital period. The value of 0.1--1\% derived by \citet{Perez-Becker13} is only true for planets that will have detectable mass loss at some point. Therefore, for a valid comparison, we need to define what the detectable region of $T_{ss}$--$M_p$ space is, just as we have done for our own values above. We scale the \citet{Perez-Becker13} values assuming the {\it Case 1} observable region. Specifically, we scale them according to the ratio between the parameter space areas used by \citet{Petigura2022} (0.23 dex in radius, 0.5 dex in $F_\text{irr}$) and that where $P_\text{detect} > 10^{-4}$ (\autoref{fig: bothProbTss_Mp}, upper panel, limited again to $T_{ss} < 2500$ K.)

Comparing firstly our general occurrence rates to that found by \citet{Perez-Becker13}, we see in \autoref{fig: compare_occ} that for {\it Case 1}, the estimate is very similar. This is because we have simply scaled the values for planets that will have observed mass loss found in \S\ref{sec: prob across pop}, which were very similar, to a larger area of parameter space. The only difference is the large variation caused by the uncertainty in the power law distribution, which we do not take into account for the \citet{Perez-Becker13} values. {\it Case 2}, on the other hand, is around an order of magnitude higher. This was addressed in \S\ref{sec: prob across pop}, where we also found the number of planets per star that will undergo catastrophic evaporation to be an order of magnitude higher (7\% compared to 0.14\%). This was mainly due to the fact that the dust model we use suggests detectability only at lower temperatures, which have lower transit probability, which means the occurrence rate must be higher to produce the same number of detections.

Compared to the observationally derived occurrence rates for both super-Earths and sub-Earths, the rates derived by both \citet{Perez-Becker13} and our {\it Case 1} are a little higher. However, the difference is insignificant and would suggest that the catastrophically evaporating planets are just the low-mass analogues of the same population. For {\it Case 2}, however, our inferred occurrence rate is significantly higher than that observed.

This very high occurrence rate prediction prompts the question of whether, if it is true, this population of low-mass planets should already have been detected by transits. It is difficult to detect sub-Earth-sized planets, but not impossible. Using the methodology of \citet{Fulton17valley}, which considers the signal to noise over the {\it Kepler} sample and includes the geometric probability of a transit (\autoref{eq: trans_prob}), we estimate the chances of finding a planet of $0.4 R_\oplus$ to be $\sim \num{3e-3}$ at our typical substellar temperatures of $\sim 2000$~K. According to our models, many of these planets should have been destroyed within 0.1 Gyrs, with the exact fraction depending on the shape of the number distribution. However, even taking the most extreme values of $\alpha$ and $\beta$, the number of planets that should be detectable in transit is only reduced by $\sim$70\%. If our {\it Case 2} occurrence rate is true, this should result in at least 120 of the $\sim140,000$ main sequence stars in the Kepler sample having detectable Mercury-sized planets at around 1 day (where the catastrophically evaporating planets lie). However, there are no confirmed planets in the Exoplanet Archive \citep{ExoArchive}, excluding the catastrophically evaporating planets themselves, within the temperature and radius range. Since the {\it Case 1} and \citet{Perez-Becker13} estimates are both around 10--100 times lower, they do not imply such a strong inconsistency. That being said, the {\it Case 1} model is clearly not perfect as it predicts many more planets at higher temperatures than lower, which is contrary to what is observed, which we return to in \S\ref{sec: systems in context}.

There are two main ways of reconciling the high value implied by {\it Case 2}. Firstly, it could be that the limited range of temperatures and masses that dust is produced at in the models is too small. While the models of \citet{Booth_disint22} are an advance on previous models of dust production in the systems' outflows, there remain various simplifications, such as a singular dust grain size.

On the other hand, if we take the models as true, the high occurrence rate can be explained by the observed evaporating planets forming further out, then being scattered into their observed orbits or reaching their current orbital period via tidal decay. Under our assumptions above, the observed catastrophically evaporating planets form at their current locations and at the same time as their host stars. Our occurrence rate refers to the progenitor population, and so if the orbits have changed, the planets can, firstly, have originated at a larger range of temperatures. Furthermore, they can also have started with lower initial masses since, under the current assumptions, low-mass planets are destroyed before their mass loss can be observed. This means that the overall parameter space that the progenitor population exists over would be increased, meaning the inferred occurrence rate would not be so high.  We leave the investigation of scattering and tidal inspiral to future works. 

\begin{figure}
    \centering
    \includegraphics[width=\linewidth]{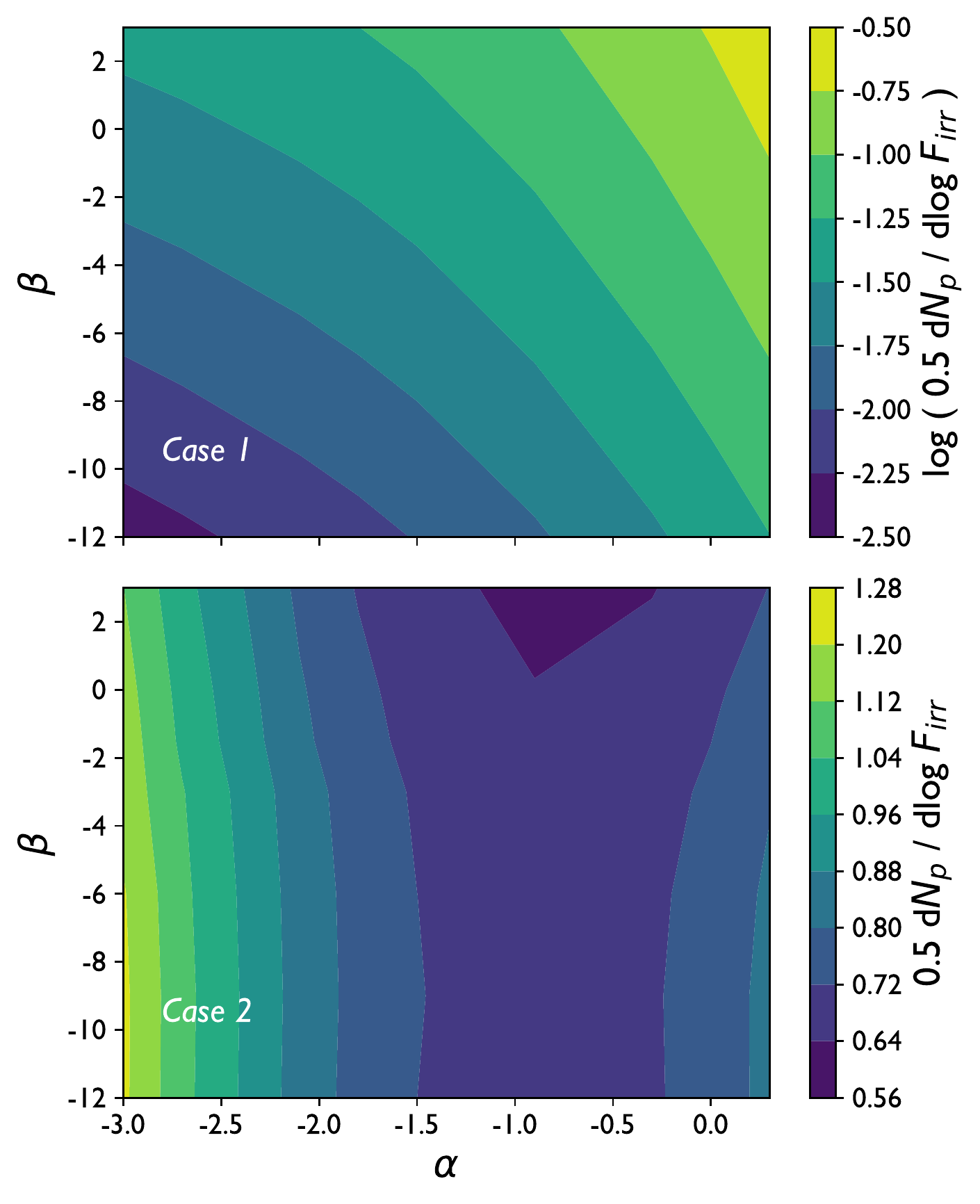}
    \caption{Planetary occurrence rate as a function of $\alpha$ and $\beta$ (the power law indices of $M_p$ and $T_{ss}$ in \autoref{eq: n_p dist}) for {\it Case 1} and {\it 2}. We define our occurrence rate in the same way as \citet{Petigura2022} as 0.5~$\partial N_p / \partial \log{F_\text{irr}}$ for a 0.23 dex bin in radius (see text). $N_p$ is the number of planets per star and $F_\text{irr}$ the stellar irradiation at the planet's orbital radius. Note that for {\it Case 1} the scale is logarithmic, while for {\it Case 2} it is not. Our measure is evaluated  at $T_{ss} = 2400$~K, $M_p = 0.2 M_\oplus$ for {\it Case 1} ($\sim 0.7 R_\oplus$, $F_\text{irr} \sim 1400 F_\oplus$) and $T_{ss} = 2050$~K, $M_p = 0.06 M_\oplus$ ($\sim 0.5 R_\oplus$, $F_\text{irr} \sim 740 F_\oplus$) for {\it Case 2}, which are near the peaks of the detection probability distributions (\autoref{fig: bothProbTss_Mp}). }
    \label{fig: both_alpha_beta}
\end{figure}
\begin{figure}
    \centering
    \includegraphics[width=\linewidth]{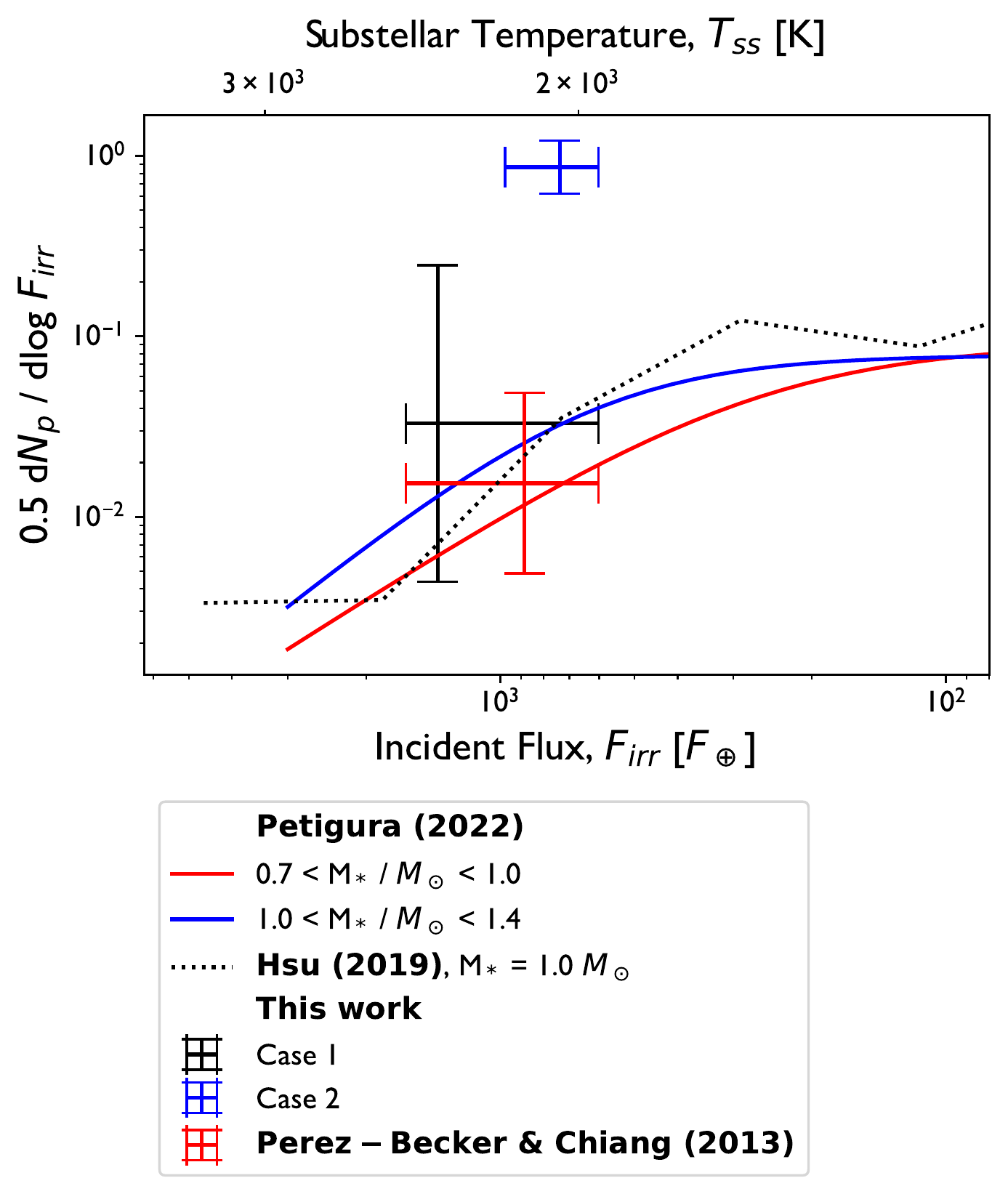}
   \caption{Comparison of our occurrence rate with those measured for super-Earths by \citet{Petigura2022} and sub-Earths by \citet{Hsu2019}, and that inferred for the catastrophically evaporating planets by \citet{Perez-Becker13}. The $y$-axis is the definition of occurrence rate used by \citet{Petigura2022}, integrated over 0.23 dex in planet radius. We scale the occurrence rate found by \citet{Perez-Becker13} according to the area of substellar temperature--initial mass parameter space where mass loss should be observable ($P_\text{detect} > 10^{-4}$) for our {\it Case 1} (see \autoref{fig: bothProbTss_Mp}). Horizontal error bars approximate this flux range and are centred on the temperature that \citet{Perez-Becker13} consider (2145 K), whereas for our values we use temperatures close to the peak of $P_\text{detect}$. Note that while the logarithmic range of radius/mass is the same for all measurements, they are centred at different values. The mass range for \citet{Petigura2022} is $\sim 1 M_\oplus$--$8 M_\oplus$, for \citet{Hsu2019} it is $\sim 0.6 M_\oplus$--$0.3 M_\oplus$ and for ours it is $\sim$ 0.07–-0.58 $M_\oplus$ for {\it Case 1} and $\sim 0.02$--$0.17 M_\oplus$ for {\it Case 2}. The vertical error bars associated with our measurement are due to the range in considered power-law indices of the number distribution, as seen in \autoref{fig: both_alpha_beta}. The vertical error bars on the measurement from \citet{Perez-Becker13} are due to the uncertainty they incorporate.}
    \label{fig: compare_occ}
\end{figure}

\subsection{The known systems in the context of our calculations} \label{sec: systems in context}
In \autoref{tab: systems}, we show the properties of the three known catastrophically evaporating planets and their host stars. We also plot their substellar temperatures in \autoref{fig: bothProbTss_Mp}. As ought to be the case, our model suggests they all have detectable mass loss ($P_\text{detect}>0$). However, for {\it Case 1}, the probability of systems being detectable is strongly peaked towards high temperatures, whereas no such effect can be seen in the observed systems. This may, therefore, act as support towards the dust production models of \citet{Booth_disint22} being closer to the physical picture since such models predict a higher chance of detecting lower temperature systems, so they more closely resemble the data. The driving effect is the cut-off of dust production at high temperatures, which is probably independent of the details of the calculations.

This explanation for the temperatures of the systems is, of course, not unique. At least one alternative explanation could be that planets at very high substellar temperatures are intrinsically rare. This could be because hotter stars host fewer rocky planets, but the period distribution is the same. Alternatively, it could be because the occurrence rate decreases at lower orbital periods, thus higher substellar temperatures. This fits with the observed period distribution of Super-Earths \citep{Sanchis-Ojeda_2014,Petigura2022}, which has a drop off at lower orbital periods, and makes sense theoretically as short period planets can inspiral onto their stars due to tidal interactions \citep[e.g.][]{lee&chiang2017}.

Another potentially interesting feature of the observed systems is that they are all K or M-dwarfs. The chances of picking two stars with $M_* < 0.8 M_\odot$ at random out of the initial {\it Kepler} sample (Kepler 1520b and KOI-2700b) is 2.4\%, based on the mass distribution. This is a low probability, but not so low as to justifiably rule out it being a coincidence. If there is a bias towards smaller stars, it could be because of the fact that, at the same planet temperature, a planet around a smaller star is closer to the star and thus more likely to be transiting (\autoref{eq: trans_prob}). Furthermore, the transit depth will be larger due to the lower stellar radius, and the period will be shorter, which may make it easier to identify the systems. These are both effects we did not take into account. Thus, the fact that the systems are around low-mass stars might not be surprising.

\begin{table}
\resizebox{\columnwidth}{!}{%
\begin{tabular}{|l|c|c|c|}
\hline
{\bf Property} & {\bf Kepler 1520b} &  {\bf KOI-2700b} & {\bf K2-22b}\\ \hline
Stellar mass [$M_\odot$] & $0.712\pm0.031$ & $0.728 \pm^{0.033}_{0.03}$ & $0.6\pm 0.07$ \\\hline 
Stellar effective & \multirow{2}{*}{$4622.2\pm^{85.2}_{78.1}$} & \multirow{2}{*}{$4617.6\pm^{83.1}_{78.5}$} & \multirow{2}{*}{$3830\pm 100$} \\
 temperature [K] & & & \\\hline
Stellar radius [$R_\odot$] & $0.694\pm 0.017$ & $0.716\pm 0.015$ & $0.57 \pm 0.06$ \\ \hline
Orbital period [hrs] & 15.68 & 21.84 & 9.146 \\ \hline
Semi-major axis [a.u.] & $0.0132\pm 0.0002$ & $0.0165\pm 0.0002$ & $0.0087\pm0.0003$\\\hline
Planet substellar & \multirow{2}{*}{$2289\pm ^{53}_{50}$} & \multirow{2}{*}{$2072 \pm ^{45}_{44}$} & \multirow{2}{*}{$2116\pm ^{130}_{130}$} \\ 
temperature, $T_{ss}$ [K] & & & \\\hline

\end{tabular}%
}
\caption{Properties of the observed systems. For Kepler 1520b and KOI-2700b, we take stellar properties from the GKS catalogue \citep{Berger-GKS20} and orbital periods from \citet{KIC1255-discov} and \citet{KOI2700b-discov} respectively. Properties for K2-22b are taken from \citet{K2-22b-discov}. Errors on the orbital period are small ($\sim1$s).}\label{tab: systems}
\end{table}

\section{Summary} \label{sec: conclusion}
In this work, we have presented a model for the thermal evolution of catastrophically evaporating planets. The model computes heat transport through conduction and convection, including the effects of melting, alongside hydrostatic equilibrium, allowing the mass of the model to be evolved self consistently. We have used this model to show that the catastrophically evaporating planets are likely almost entirely crystallised, other than a shallow molten region on the dayside of the planet. This result is robust to the details of the redistribution of heat from the star, which is challenging to model in full. Therefore, we suggest that the composition of the observed dusty tails samples only the particular region of the mantle that the planet has evaporated to. Future work should include modelling of the chemical evolution of the mantle in order to find the composition of the outer layer at a particular point in time. Additionally, further modelling of the chemical evolution of the lava pool, as well as the outflow itself, may be required to make a full link between the mantle composition and that of the dust. 

We also use our model to investigate the occurrence rate of the progenitors of the catastrophically evaporating planets and low-mass planets in general, assuming they have remained at their current locations since birth. When using a simple cut-off for the mass loss rate required to be observable, we obtain an estimate of 0.14\% of stars having a planet that will have observable evaporation, in line with the estimates of \citet{Perez-Becker13}. However, if we instead consider the theoretical predictions of dust production from \citet{Booth_disint22}, the value is an order of magnitude higher at $\sim 7\%$. Extending this to general planets in the temperature and mass region close to those which can have detectable mass loss rates, by assuming a power law distribution, the simple cut-off model implies that 0.5\%--20\% have a planet in the range 0.07--0.58$M_\oplus$ and 2115--2693~K. This is consistent with observed planet demographics. However, for the dust production model, we find around 1 planet per star in the range 0.02--0.17$M_\oplus$ and 1775--2367~K. This is up to 100 times more common than Super-Earths and sub-Earths in the same substellar temperature range, derived by {\it Kepler}, and is likely inconsistent with the lack of Mercury radius planets observed. This may be evidence that the catastrophically evaporating planets were moved onto their current orbits at later times or imply that the narrow range of temperatures and planet masses that the \citet{Booth_disint22} models predict dust is produced at is too small.

However, when considering the detectability of systems in general, we found that the substellar temperatures of the detected systems are better explained by a dust production model, such as that of \citet{Booth_disint22}, than a simple model where planets become observable above a certain mass loss rate. In a simpler model, the hottest planets are most observable as they have the highest mass loss rates. However, in the model of \citet{Booth_disint22}, dust production is reduced at high temperatures. Therefore, our findings motivate further theoretical work to better determine what range of planetary parameters can produce observable dusty outflows. This, along with an improved understanding of any orbital evolution of these systems, will allow us to approach the true occurrence of these systems.

\section{Data availability}
The code and simulated data underlying this article will be shared on reasonable request to the corresponding author. Any observational or experimental data used are available at the references in the main text.

\section{Acknowledgements}
 A.C. acknowledges the support of an STFC PhD studentship. R.B. and J.E.O. are supported by the Royal Society through University Research Fellowships. J.E.O. has also received funding from the European Research Council (ERC) under the European Union’s Horizon 2020 research and innovation programme (Grant agreement No. 853022, PEVAP) and a 2020 Royal Society Enhancement Award. For the purpose of open access, the authors have applied a Creative Commons Attribution (CC-BY) licence to any Author Accepted Manuscript version arising. This work makes use of the \texttt{Boost} \citep{boost2014} and \texttt{Eigen} \citep{eigen} \texttt{C++} libraries for the interior model. It also makes use of the Python packages \texttt{Scipy} \citep{scipy2020}, \texttt{Numpy} \citep{numpy} and \texttt{Matplotlib} \citep{matplotlib}. We thank the reviewer, Eugene Chiang, for their helpful comments throughout the review process, which greatly improved the final paper.



\bibliographystyle{mn2e}
\bibliography{refs}

\begin{thebibliography}{}

\bibitem[\protect\citeauthoryear{Abe}{Abe}{1993}]{abe1993thermal}
Abe Y.,  1993, Evolution of the Earth and planets(A 95-18465 03-91),
  Washington, DC/Brussels, Belgium, American Geophysical Union/International
  Union of Geodesy and Geophysics(Geophysical Monograph,, pp 41--54

\bibitem[\protect\citeauthoryear{Abe}{Abe}{1995}]{abe1995basic}
Abe Y.,  1995, The Earth's central part: its structure and dynamics, pp
  215--235

\bibitem[\protect\citeauthoryear{{Abe} \& {Matsui}}{{Abe} \&
  {Matsui}}{1986}]{Abe1986}
{Abe} Y.,  {Matsui} T.,  1986, \jgr, 91, E291

\bibitem[\protect\citeauthoryear{{Akeson}, {Chen}, {Ciardi}, {Crane}, {Good},
  {Harbut}, {Jackson}, {Kane}, {Laity}, {Leifer}, {Lynn}, {McElroy}, {Papin},
  {Plavchan}, {Ram{\'\i}rez}, {Rey}, {von Braun}, {Wittman}, {Abajian}, {Ali},
  {Beichman}, {Beekley}, {Berriman}, {Berukoff}, {Bryden}, {Chan}, {Groom},
  {Lau}, {Payne}, {Regelson}, {Saucedo}, {Schmitz}, {Stauffer}, {Wyatt} \&
  {Zhang}}{{Akeson} et~al.}{2013}]{ExoArchive}
{Akeson} R.~L.,  {Chen} X.,  {Ciardi} D.,  {Crane} M.,  {Good} J.,  {Harbut}
  M.,  {Jackson} E.,  {Kane} S.~R.,  {Laity} A.~C.,  {Leifer} S.,  {Lynn} M.,
  {McElroy} D.~L.,  {Papin} M.,  {Plavchan} P.,  {Ram{\'\i}rez} S.~V.,  {Rey}
  R.,  {von Braun} K.,  {Wittman} M.,  {Abajian} M.,  {Ali} B.,  {Beichman} C.,
   {Beekley} A.,  {Berriman} G.~B.,  {Berukoff} S.,  {Bryden} G.,  {Chan} B.,
  {Groom} S.,  {Lau} C.,  {Payne} A.~N.,  {Regelson} M.,  {Saucedo} M.,
  {Schmitz} M.,  {Stauffer} J.,  {Wyatt} P.,    {Zhang} A.,  2013, \pasp, 125,
  989

\bibitem[\protect\citeauthoryear{Alefeld, Potra \& Shi}{Alefeld
  et~al.}{1995}]{toms748}
Alefeld G.~E.,  Potra F.~A.,    Shi Y.,  1995, ACM Trans. Math. Softw., 21,
  327–344

\bibitem[\protect\citeauthoryear{{Andrault}, {Bolfan-Casanova}, {Nigro},
  {Bouhifd}, {Garbarino} \& {Mezouar}}{{Andrault} et~al.}{2011}]{Andrault11}
{Andrault} D.,  {Bolfan-Casanova} N.,  {Nigro} G.~L.,  {Bouhifd} M.~A.,
  {Garbarino} G.,    {Mezouar} M.,  2011, Earth and Planetary Science Letters,
  304, 251

\bibitem[\protect\citeauthoryear{{Berger}, {Huber}, {van Saders}, {Gaidos},
  {Tayar} \& {Kraus}}{{Berger} et~al.}{2020}]{Berger-GKS20}
{Berger} T.~A.,  {Huber} D.,  {van Saders} J.~L.,  {Gaidos} E.,  {Tayar} J.,
  {Kraus} A.~L.,  2020, \aj, 159, 280

\bibitem[\protect\citeauthoryear{Bodenheimer, Laughlin, Rozyczka, Plewa, Yorke
  \& Yorke}{Bodenheimer et~al.}{2006}]{Boden}
Bodenheimer P.,  Laughlin G.,  Rozyczka M.,  Plewa T.,  Yorke H.,    Yorke H.,
  2006, Numerical Methods in Astrophysics: An Introduction.
Series in Astronomy and Astrophysics, CRC Press

\bibitem[\protect\citeauthoryear{{Bodman}, {Wright}, {Desch} \&
  {Lisse}}{{Bodman} et~al.}{2018}]{Bodman18-JWST}
{Bodman} E. H.~L.,  {Wright} J.~T.,  {Desch} S.~J.,    {Lisse} C.~M.,  2018,
  \aj, 156, 173

\bibitem[\protect\citeauthoryear{{Booth}, {Owen} \& {Schulik}}{{Booth}
  et~al.}{2023}]{Booth_disint22}
{Booth} R.~A.,  {Owen} J.~E.,    {Schulik} M.,  2023, \mnras, 518, 1761

\bibitem[\protect\citeauthoryear{Borucki}{Borucki}{2016}]{KEPLER}
Borucki W.~J.,  2016, Reports on progress in physics. Physical Society (Great
  Britain), 79, 036901

\bibitem[\protect\citeauthoryear{{Borucki} \& {Summers}}{{Borucki} \&
  {Summers}}{1984}]{Borucki1984}
{Borucki} W.~J.,  {Summers} A.~L.,  1984, \icarus, 58, 121

\bibitem[\protect\citeauthoryear{{Bower}, {Sanan} \& {Wolf}}{{Bower}
  et~al.}{2018}]{Bower17}
{Bower} D.~J.,  {Sanan} P.,    {Wolf} A.~S.,  2018, Physics of the Earth and
  Planetary Interiors, 274, 49

\bibitem[\protect\citeauthoryear{Brent}{Brent}{1974}]{Brent1974AlgorithmsFM}
Brent R.~P.,  1974, IEEE Transactions on Automatic Control, 19, 632

\bibitem[\protect\citeauthoryear{{Bromley} \& {Chiang}}{{Bromley} \&
  {Chiang}}{2023}]{Bromely-Chiang23}
{Bromley} J.,  {Chiang} E.,  2023, arXiv e-prints, p. arXiv:2302.04898

\bibitem[\protect\citeauthoryear{{Budaj}}{{Budaj}}{2013}]{Budaj13}
{Budaj} J.,  2013, \aap, 557, A72

\bibitem[\protect\citeauthoryear{{Campos Estrada}, {Owen}, {Jankovic}, {Wilson}
  \& {Helling}}{{Campos Estrada} et~al.}{2023}]{CamposEstrada23}
{Campos Estrada} B.,  {Owen} J.~E.,  {Jankovic} M.~R.,  {Wilson} A.,
  {Helling} C.,  2023, arXiv e-prints, p. arXiv:2311.02477

\bibitem[\protect\citeauthoryear{{Castan} \& {Menou}}{{Castan} \&
  {Menou}}{2011}]{castan-menou11}
{Castan} T.,  {Menou} K.,  2011, \apjl, 743, L36

\bibitem[\protect\citeauthoryear{{Choi}, {Dotter}, {Conroy}, {Cantiello},
  {Paxton} \& {Johnson}}{{Choi} et~al.}{2016}]{MIST-Choi16}
{Choi} J.,  {Dotter} A.,  {Conroy} C.,  {Cantiello} M.,  {Paxton} B.,
  {Johnson} B.~D.,  2016, \apj, 823, 102

\bibitem[\protect\citeauthoryear{{Christiansen}, {Clarke}, {Burke}, {Jenkins}
  \& {Kepler Completeness Working Group}}{{Christiansen}
  et~al.}{2014}]{Christiansen2014}
{Christiansen} J.~L.,  {Clarke} B.~D.,  {Burke} C.~J.,  {Jenkins} J.~M.,
  {Kepler Completeness Working Group} 2014, in {Haghighipour} N.,  ed.,
  Formation, Detection, and Characterization of Extrasolar Habitable Planets
  Vol.~293, {The Kepler Completeness Study: A Pipeline Throughput Experiment}.
pp 88--93

\bibitem[\protect\citeauthoryear{{Dorogokupets}, {Dymshits}, {Litasov} \&
  {Sokolova}}{{Dorogokupets} et~al.}{2017}]{Dorogokupets2017}
{Dorogokupets} P.~I.,  {Dymshits} A.~M.,  {Litasov} K.~D.,    {Sokolova} T.~S.,
   2017, Scientific Reports, 7, 41863

\bibitem[\protect\citeauthoryear{{Dotter}}{{Dotter}}{2016}]{MIST-Dotter16}
{Dotter} A.,  2016, \apjs, 222, 8

\bibitem[\protect\citeauthoryear{{Elkins-Tanton}}{{Elkins-Tanton}}{2012}]{Elkins-Tanton12-review}
{Elkins-Tanton} L.~T.,  2012, Annual Review of Earth and Planetary Sciences,
  40, 113

\bibitem[\protect\citeauthoryear{Fritsch \& Carlson}{Fritsch \&
  Carlson}{1980}]{PCHIP}
Fritsch F.~N.,  Carlson R.~E.,  1980, SIAM Journal on Numerical Analysis, 17,
  238

\bibitem[\protect\citeauthoryear{{Fulton}, {Petigura}, {Howard}, {Isaacson},
  {Marcy}, {Cargile}, {Hebb}, {Weiss}, {Johnson}, {Morton}, {Sinukoff},
  {Crossfield} \& {Hirsch}}{{Fulton} et~al.}{2017}]{Fulton17valley}
{Fulton} B.~J.,  {Petigura} E.~A.,  {Howard} A.~W.,  {Isaacson} H.,  {Marcy}
  G.~W.,  {Cargile} P.~A.,  {Hebb} L.,  {Weiss} L.~M.,  {Johnson} J.~A.,
  {Morton} T.~D.,  {Sinukoff} E.,  {Crossfield} I. J.~M.,    {Hirsch} L.~A.,
  2017, \aj, 154, 109

\bibitem[\protect\citeauthoryear{Gaillard \& Scaillet}{Gaillard \&
  Scaillet}{2014}]{GAIL2014}
Gaillard F.,  Scaillet B.,  2014, Earth and Planetary Science Letters, 403, 307

\bibitem[\protect\citeauthoryear{Guennebaud, Jacob et~al.,}{Guennebaud
  et~al.}{2010}]{eigen}
Guennebaud G.,  Jacob B.,    et~al.,, 2010, Eigen v3,
  http://eigen.tuxfamily.org

\bibitem[\protect\citeauthoryear{{Harris}, {Millman}, {van der Walt},
  {Gommers}, {Virtanen}, {Cournapeau}, {Wieser}, {Taylor}, {Berg}, {Smith},
  {Kern}, {Picus}, {Hoyer}, {van Kerkwijk}, {Brett}, {Haldane}, {del R{\'\i}o},
  {Wiebe}, {Peterson}, {G{\'e}rard-Marchant}, {Sheppard}, {Reddy}, {Weckesser},
  {Abbasi}, {Gohlke} \& {Oliphant}}{{Harris} et~al.}{2020}]{numpy}
{Harris} C.~R.,  {Millman} K.~J.,  {van der Walt} S.~J.,  {Gommers} R.,
  {Virtanen} P.,  {Cournapeau} D.,  {Wieser} E.,  {Taylor} J.,  {Berg} S.,
  {Smith} N.~J.,  {Kern} R.,  {Picus} M.,  {Hoyer} S.,  {van Kerkwijk} M.~H.,
  {Brett} M.,  {Haldane} A.,  {del R{\'\i}o} J.~F.,  {Wiebe} M.,  {Peterson}
  P.,  {G{\'e}rard-Marchant} P.,  {Sheppard} K.,  {Reddy} T.,  {Weckesser} W.,
  {Abbasi} H.,  {Gohlke} C.,    {Oliphant} T.~E.,  2020, \nat, 585, 357

\bibitem[\protect\citeauthoryear{Harrison, Bonsor, Kama, Buchan, Blouin \&
  Koester}{Harrison et~al.}{2021}]{Harrison2021}
Harrison J. H.~D.,  Bonsor A.,  Kama M.,  Buchan A.~M.,  Blouin S.,    Koester
  D.,  2021, Monthly Notices of the Royal Astronomical Society, 504, 2853

\bibitem[\protect\citeauthoryear{{Hollands}, {Koester}, {Alekseev}, {Herbert}
  \& {G{\"a}nsicke}}{{Hollands} et~al.}{2017}]{DZDwarfs1}
{Hollands} M.~A.,  {Koester} D.,  {Alekseev} V.,  {Herbert} E.~L.,
  {G{\"a}nsicke} B.~T.,  2017, \mnras, 467, 4970

\bibitem[\protect\citeauthoryear{{Howell}, {Sobeck}, {Haas}, {Still},
  {Barclay}, {Mullally}, {Troeltzsch}, {Aigrain}, {Bryson}, {Caldwell},
  {Chaplin}, {Cochran}, {Huber}, {Marcy}, {Miglio}, {Najita}, {Smith},
  {Twicken} \& {Fortney}}{{Howell} et~al.}{2014}]{K2-mission}
{Howell} S.~B.,  {Sobeck} C.,  {Haas} M.,  {Still} M.,  {Barclay} T.,
  {Mullally} F.,  {Troeltzsch} J.,  {Aigrain} S.,  {Bryson} S.~T.,  {Caldwell}
  D.,  {Chaplin} W.~J.,  {Cochran} W.~D.,  {Huber} D.,  {Marcy} G.~W.,
  {Miglio} A.,  {Najita} J.~R.,  {Smith} M.,  {Twicken} J.~D.,    {Fortney}
  J.~J.,  2014, \pasp, 126, 398

\bibitem[\protect\citeauthoryear{{Hsu}, {Ford}, {Ragozzine} \& {Ashby}}{{Hsu}
  et~al.}{2019}]{Hsu2019}
{Hsu} D.~C.,  {Ford} E.~B.,  {Ragozzine} D.,    {Ashby} K.,  2019, \aj, 158,
  109

\bibitem[\protect\citeauthoryear{{Hughes} \& {Griffiths}}{{Hughes} \&
  {Griffiths}}{2008}]{Hughes+Griffiths2008}
{Hughes} G.~O.,  {Griffiths} R.~W.,  2008, Annual Review of Fluid Mechanics,
  40, 185

\bibitem[\protect\citeauthoryear{Hunter}{Hunter}{2007}]{matplotlib}
Hunter J.~D.,  2007, Computing in Science \& Engineering, 9, 90

\bibitem[\protect\citeauthoryear{{Ito} \& {Ikoma}}{{Ito} \&
  {Ikoma}}{2021}]{Ito-Ikoma21}
{Ito} Y.,  {Ikoma} M.,  2021, \mnras, 502, 750

\bibitem[\protect\citeauthoryear{{Jackson}, {Barnes} \& {Greenberg}}{{Jackson}
  et~al.}{2008}]{Jackson2008}
{Jackson} B.,  {Barnes} R.,    {Greenberg} R.,  2008, \mnras, 391, 237

\bibitem[\protect\citeauthoryear{{Kang}, {Ding}, {Wordsworth} \&
  {Seager}}{{Kang} et~al.}{2021}]{Kang2021}
{Kang} W.,  {Ding} F.,  {Wordsworth} R.,    {Seager} S.,  2021, \apj, 906, 67

\bibitem[\protect\citeauthoryear{Kelemen, Hirth, Shimizu, Spiegelman \&
  Dick}{Kelemen et~al.}{1997}]{kelemen1997}
Kelemen P.~B.,  Hirth G.,  Shimizu N.,  Spiegelman M.,    Dick H.,  1997,
  Philosophical Transactions of the Royal Society of London. Series A:
  Mathematical, Physical and Engineering Sciences, 355, 283

\bibitem[\protect\citeauthoryear{{Kippenhahn}, {Weigert} \&
  {Weiss}}{{Kippenhahn} et~al.}{2012}]{Kippen}
{Kippenhahn} R.,  {Weigert} A.,    {Weiss} A.,  2012, {Stellar Structure and
  Evolution}

\bibitem[\protect\citeauthoryear{{Kite}, {Fegley} Bruce, {Schaefer} \&
  {Gaidos}}{{Kite} et~al.}{2016}]{Kite16}
{Kite} E.~S.,  {Fegley} Bruce J.,  {Schaefer} L.,    {Gaidos} E.,  2016, The
  Astrophysical Journal, 828, 80

\bibitem[\protect\citeauthoryear{{Kite}, {Manga} \& {Gaidos}}{{Kite}
  et~al.}{2009}]{Kite09}
{Kite} E.~S.,  {Manga} M.,    {Gaidos} E.,  2009, \apj, 700, 1732

\bibitem[\protect\citeauthoryear{{Lee} \& {Chiang}}{{Lee} \&
  {Chiang}}{2017}]{lee&chiang2017}
{Lee} E.~J.,  {Chiang} E.,  2017, \apj, 842, 40

\bibitem[\protect\citeauthoryear{{Lejeune} \& {Richet}}{{Lejeune} \&
  {Richet}}{1995}]{Lejeune1995}
{Lejeune} A.-M.,  {Richet} P.,  1995, \jgr, 100, 4215

\bibitem[\protect\citeauthoryear{{Lichtenberg}, {Bower}, {Hammond},
  {Boukrouche}, {Sanan}, {Tsai} \& {Pierrehumbert}}{{Lichtenberg}
  et~al.}{2021}]{Licht21verticalmagma}
{Lichtenberg} T.,  {Bower} D.~J.,  {Hammond} M.,  {Boukrouche} R.,  {Sanan} P.,
   {Tsai} S.-M.,    {Pierrehumbert} R.~T.,  2021, Journal of Geophysical
  Research (Planets), 126, e06711

\bibitem[\protect\citeauthoryear{Litasov \& Ohtani}{Litasov \&
  Ohtani}{2002}]{LITASOV2002}
Litasov K.,  Ohtani E.,  2002, Physics of the Earth and Planetary Interiors,
  134, 105

\bibitem[\protect\citeauthoryear{Lugaro, Ott \& Kereszturi}{Lugaro
  et~al.}{2018}]{lugaro18-review}
Lugaro M.,  Ott U.,    Kereszturi {\'A}.,  2018, Progress in Particle and
  Nuclear Physics, 102, 1

\bibitem[\protect\citeauthoryear{{Makino}, {Fukushige}, {Funato} \&
  {Kokubo}}{{Makino} et~al.}{1998}]{Makino98-runaway}
{Makino} J.,  {Fukushige} T.,  {Funato} Y.,    {Kokubo} E.,  1998, New
  Astronomy, 3, 411

\bibitem[\protect\citeauthoryear{{Mei}, {Bai}, {Hiraga} \& {Kohlstedt}}{{Mei}
  et~al.}{2002}]{Mei2002}
{Mei} S.,  {Bai} W.,  {Hiraga} T.,    {Kohlstedt} D.~L.,  2002, Earth and
  Planetary Science Letters, 201, 491

\bibitem[\protect\citeauthoryear{{Noack}, {Godolt}, {von Paris}, {Plesa},
  {Stracke}, {Breuer} \& {Rauer}}{{Noack} et~al.}{2014}]{Noack2014}
{Noack} L.,  {Godolt} M.,  {von Paris} P.,  {Plesa} A.~C.,  {Stracke} B.,
  {Breuer} D.,    {Rauer} H.,  2014, \planss, 98, 14

\bibitem[\protect\citeauthoryear{{Paxton}, {Bildsten}, {Dotter}, {Herwig},
  {Lesaffre} \& {Timmes}}{{Paxton} et~al.}{2011}]{MESA1}
{Paxton} B.,  {Bildsten} L.,  {Dotter} A.,  {Herwig} F.,  {Lesaffre} P.,
  {Timmes} F.,  2011, \apjs, 192, 3

\bibitem[\protect\citeauthoryear{{Paxton}, {Cantiello}, {Arras}, {Bildsten},
  {Brown}, {Dotter}, {Mankovich}, {Montgomery}, {Stello}, {Timmes} \&
  {Townsend}}{{Paxton} et~al.}{2013}]{MESA13}
{Paxton} B.,  {Cantiello} M.,  {Arras} P.,  {Bildsten} L.,  {Brown} E.~F.,
  {Dotter} A.,  {Mankovich} C.,  {Montgomery} M.~H.,  {Stello} D.,  {Timmes}
  F.~X.,    {Townsend} R.,  2013, \apjs, 208, 4

\bibitem[\protect\citeauthoryear{{Paxton}, {Marchant}, {Schwab}, {Bauer},
  {Bildsten}, {Cantiello}, {Dessart}, {Farmer}, {Hu}, {Langer}, {Townsend},
  {Townsley} \& {Timmes}}{{Paxton} et~al.}{2015}]{MESA15}
{Paxton} B.,  {Marchant} P.,  {Schwab} J.,  {Bauer} E.~B.,  {Bildsten} L.,
  {Cantiello} M.,  {Dessart} L.,  {Farmer} R.,  {Hu} H.,  {Langer} N.,
  {Townsend} R.~H.~D.,  {Townsley} D.~M.,    {Timmes} F.~X.,  2015, \apjs, 220,
  15

\bibitem[\protect\citeauthoryear{{Paxton}, {Smolec}, {Schwab}, {Gautschy},
  {Bildsten}, {Cantiello}, {Dotter}, {Farmer}, {Goldberg}, {Jermyn}, {Kanbur},
  {Marchant}, {Thoul}, {Townsend}, {Wolf}, {Zhang} \& {Timmes}}{{Paxton}
  et~al.}{2019}]{MESA19}
{Paxton} B.,  {Smolec} R.,  {Schwab} J.,  {Gautschy} A.,  {Bildsten} L.,
  {Cantiello} M.,  {Dotter} A.,  {Farmer} R.,  {Goldberg} J.~A.,  {Jermyn}
  A.~S.,  {Kanbur} S.~M.,  {Marchant} P.,  {Thoul} A.,  {Townsend} R. H.~D.,
  {Wolf} W.~M.,  {Zhang} M.,    {Timmes} F.~X.,  2019, \apjs, 243, 10

\bibitem[\protect\citeauthoryear{{Perez-Becker} \& {Chiang}}{{Perez-Becker} \&
  {Chiang}}{2013}]{Perez-Becker13}
{Perez-Becker} D.,  {Chiang} E.,  2013, \mnras, 433, 2294

\bibitem[\protect\citeauthoryear{{Petigura}, {Rogers}, {Isaacson}, {Owen},
  {Kraus}, {Winn}, {MacDougall}, {Howard}, {Fulton}, {Kosiarek}, {Weiss},
  {Behmard} \& {Blunt}}{{Petigura} et~al.}{2022}]{Petigura2022}
{Petigura} E.~A.,  {Rogers} J.~G.,  {Isaacson} H.,  {Owen} J.~E.,  {Kraus}
  A.~L.,  {Winn} J.~N.,  {MacDougall} M.~G.,  {Howard} A.~W.,  {Fulton} B.,
  {Kosiarek} M.~R.,  {Weiss} L.~M.,  {Behmard} A.,    {Blunt} S.,  2022, \aj,
  163, 179

\bibitem[\protect\citeauthoryear{{Rappaport}, {Barclay}, {DeVore}, {Rowe},
  {Sanchis-Ojeda} \& {Still}}{{Rappaport} et~al.}{2014}]{KOI2700b-discov}
{Rappaport} S.,  {Barclay} T.,  {DeVore} J.,  {Rowe} J.,  {Sanchis-Ojeda} R.,
   {Still} M.,  2014, \apj, 784, 40

\bibitem[\protect\citeauthoryear{{Rappaport}, {Levine}, {Chiang}, {El Mellah},
  {Jenkins}, {Kalomeni}, {Kite}, {Kotson}, {Nelson}, {Rousseau-Nepton} \&
  {Tran}}{{Rappaport} et~al.}{2012}]{KIC1255-discov}
{Rappaport} S.,  {Levine} A.,  {Chiang} E.,  {El Mellah} I.,  {Jenkins} J.,
  {Kalomeni} B.,  {Kite} E.~S.,  {Kotson} M.,  {Nelson} L.,  {Rousseau-Nepton}
  L.,    {Tran} K.,  2012, \apj, 752, 1

\bibitem[\protect\citeauthoryear{{Ridden-Harper}, {Snellen}, {Keller} \&
  {Molli{\`e}re}}{{Ridden-Harper} et~al.}{2019}]{Ridden-Harper2019}
{Ridden-Harper} A.~R.,  {Snellen} I.~A.~G.,  {Keller} C.~U.,    {Molli{\`e}re}
  P.,  2019, \aap, 628, A70

\bibitem[\protect\citeauthoryear{{Roscoe}}{{Roscoe}}{1952}]{Roscoe52}
{Roscoe} R.,  1952, British Journal of Applied Physics, 3, 267

\bibitem[\protect\citeauthoryear{{Sanchis-Ojeda}, {Rappaport}, {Pall{\`e}},
  {Delrez}, {DeVore}, {Gandolfi}, {Fukui}, {Ribas}, {Stassun}, {Albrecht},
  {Dai}, {Gaidos}, {Gillon}, {Hirano}, {Holman}, {Howard}, {Isaacson}, {Jehin},
  {Kuzuhara}, {Mann}, {Marcy}, {Miles-P{\'a}ez},
  {Monta{\~n}{\'e}s-Rodr{\'\i}guez}, {Murgas}, {Narita}, {Nowak}, {Onitsuka},
  {Paegert}, {Van Eylen}, {Winn} \& {Yu}}{{Sanchis-Ojeda}
  et~al.}{2015}]{K2-22b-discov}
{Sanchis-Ojeda} R.,  {Rappaport} S.,  {Pall{\`e}} E.,  {Delrez} L.,  {DeVore}
  J.,  {Gandolfi} D.,  {Fukui} A.,  {Ribas} I.,  {Stassun} K.~G.,  {Albrecht}
  S.,  {Dai} F.,  {Gaidos} E.,  {Gillon} M.,  {Hirano} T.,  {Holman} M.,
  {Howard} A.~W.,  {Isaacson} H.,  {Jehin} E.,  {Kuzuhara} M.,  {Mann} A.~W.,
  {Marcy} G.~W.,  {Miles-P{\'a}ez} P.~A.,  {Monta{\~n}{\'e}s-Rodr{\'\i}guez}
  P.,  {Murgas} F.,  {Narita} N.,  {Nowak} G.,  {Onitsuka} M.,  {Paegert} M.,
  {Van Eylen} V.,  {Winn} J.~N.,    {Yu} L.,  2015, \apj, 812, 112

\bibitem[\protect\citeauthoryear{Sanchis-Ojeda, Rappaport, Winn, Kotson, Levine
  \& Mellah}{Sanchis-Ojeda et~al.}{2014}]{Sanchis-Ojeda_2014}
Sanchis-Ojeda R.,  Rappaport S.,  Winn J.~N.,  Kotson M.~C.,  Levine A.,
  Mellah I.~E.,  2014, The Astrophysical Journal, 787, 47

\bibitem[\protect\citeauthoryear{{Schaefer} \& {Fegley}}{{Schaefer} \&
  {Fegley}}{2009}]{schaefer_fegley2009}
{Schaefer} L.,  {Fegley} B.,  2009, \apjl, 703, L113

\bibitem[\protect\citeauthoryear{Schäling}{Schäling}{2014}]{boost2014}
Schäling B.,  2014, The boost C++ libraries.
XML Press

\bibitem[\protect\citeauthoryear{Simon \& Glatzel}{Simon \&
  Glatzel}{1929}]{Simon-Glatzel}
Simon F.,  Glatzel G.,  1929, Zeitschrift f{\"u}r anorganische und allgemeine
  Chemie, 178, 309

\bibitem[\protect\citeauthoryear{Sluiter}{Sluiter}{2012}]{SLUITER2012}
Sluiter M.,  2012, in Pereloma E.,  Edmonds D.~V.,  eds, Woodhead Publishing
  Series in Metals and Surface Engineering, Vol.~2, Phase Transformations in
  Steels.
Woodhead Publishing, pp 365--404

\bibitem[\protect\citeauthoryear{Solomatov}{Solomatov}{2007}]{SOLOMATOV-chapter}
Solomatov V.,  2007, in Schubert G.,  ed., , Treatise on Geophysics.
Elsevier, Amsterdam, pp 91--119

\bibitem[\protect\citeauthoryear{{Stixrude}, {de Koker}, {Sun}, {Mookherjee} \&
  {Karki}}{{Stixrude} et~al.}{2009}]{Middle-out-Stix2009}
{Stixrude} L.,  {de Koker} N.,  {Sun} N.,  {Mookherjee} M.,    {Karki} B.~B.,
  2009, Earth and Planetary Science Letters, 278, 226

\bibitem[\protect\citeauthoryear{{Stixrude} \& {Lithgow-Bertelloni}}{{Stixrude}
  \& {Lithgow-Bertelloni}}{2011}]{Stixtrude2}
{Stixrude} L.,  {Lithgow-Bertelloni} C.,  2011, Geophysical Journal
  International, 184, 1180

\bibitem[\protect\citeauthoryear{{Tackley}, {Ammann}, {Brodholt}, {Dobson} \&
  {Valencia}}{{Tackley} et~al.}{2013}]{Tackley13}
{Tackley} P.~J.,  {Ammann} M.,  {Brodholt} J.~P.,  {Dobson} D.~P.,
  {Valencia} D.,  2013, \icarus, 225, 50

\bibitem[\protect\citeauthoryear{{Tosi}, {Godolt}, {Stracke}, {Ruedas},
  {Grenfell}, {H{\"o}ning}, {Nikolaou}, {Plesa}, {Breuer} \& {Spohn}}{{Tosi}
  et~al.}{2017}]{Tosi2017}
{Tosi} N.,  {Godolt} M.,  {Stracke} B.,  {Ruedas} T.,  {Grenfell} J.~L.,
  {H{\"o}ning} D.,  {Nikolaou} A.,  {Plesa} A.~C.,  {Breuer} D.,    {Spohn} T.,
   2017, \aap, 605, A71

\bibitem[\protect\citeauthoryear{Tsujino, Nishihara, Nakajima, Takahashi, ichi
  Funakoshi \& Higo}{Tsujino et~al.}{2013}]{TSUJINO2013}
Tsujino N.,  Nishihara Y.,  Nakajima Y.,  Takahashi E.,  ichi Funakoshi K.,
  Higo Y.,  2013, Earth and Planetary Science Letters, 375, 244

\bibitem[\protect\citeauthoryear{Turcotte, Schubert \& Schubert}{Turcotte
  et~al.}{2002}]{turcotte2002geodynamics}
Turcotte D.,  Schubert G.,    Schubert J.,  2002, Geodynamics.
Geodynamics, Cambridge University Press

\bibitem[\protect\citeauthoryear{{Valencia}, {Sasselov} \&
  {O'Connell}}{{Valencia} et~al.}{2007}]{Valencia_Sasselov2007}
{Valencia} D.,  {Sasselov} D.~D.,    {O'Connell} R.~J.,  2007, \apj, 665, 1413

\bibitem[\protect\citeauthoryear{{Vallis}}{{Vallis}}{2006}]{Vallis2006}
{Vallis} G.~K.,  2006, {Atmospheric and Oceanic Fluid Dynamics}

\bibitem[\protect\citeauthoryear{{van Lieshout}, {Min} \& {Dominik}}{{van
  Lieshout} et~al.}{2014}]{vanLieshout14}
{van Lieshout} R.,  {Min} M.,    {Dominik} C.,  2014, \aap, 572, A76

\bibitem[\protect\citeauthoryear{{van Lieshout}, {Min}, {Dominik}, {Brogi}, {de
  Graaff}, {Hekker}, {Kama}, {Keller}, {Ridden-Harper} \& {van Werkhoven}}{{van
  Lieshout} et~al.}{2016}]{vanLieshout16}
{van Lieshout} R.,  {Min} M.,  {Dominik} C.,  {Brogi} M.,  {de Graaff} T.,
  {Hekker} S.,  {Kama} M.,  {Keller} C.~U.,  {Ridden-Harper} A.,    {van
  Werkhoven} T.~I.~M.,  2016, \aap, 596, A32

\bibitem[\protect\citeauthoryear{{van Lieshout} \& {Rappaport}}{{van Lieshout}
  \& {Rappaport}}{2018}]{disint18}
{van Lieshout} R.,  {Rappaport} S.~A.,  2018, in {Deeg} H.~J.,  {Belmonte}
  J.~A.,  eds, , Handbook of Exoplanets.
p.~15

\bibitem[\protect\citeauthoryear{Virtanen, Gommers, Oliphant, Haberland, Reddy,
  Cournapeau, Burovski, Peterson, Weckesser, Bright, {van der Walt}, Brett,
  Wilson, Millman, Mayorov, Nelson, Jones, Kern, Larson, Carey, Polat, Feng,
  Moore, {VanderPlas}, Laxalde, Perktold, Cimrman, Henriksen, Quintero, Harris,
  Archibald, Ribeiro, Pedregosa, {van Mulbregt} \& {SciPy 1.0
  Contributors}}{Virtanen et~al.}{2020}]{scipy2020}
Virtanen P.,  Gommers R.,  Oliphant T.~E.,  Haberland M.,  Reddy T.,
  Cournapeau D.,  Burovski E.,  Peterson P.,  Weckesser W.,  Bright J.,  {van
  der Walt} S.~J.,  Brett M.,  Wilson J.,  Millman K.~J.,  Mayorov N.,  Nelson
  A. R.~J.,  Jones E.,  Kern R.,  Larson E.,  Carey C.~J.,  Polat {\.I}.,  Feng
  Y.,  Moore E.~W.,  {VanderPlas} J.,  Laxalde D.,  Perktold J.,  Cimrman R.,
  Henriksen I.,  Quintero E.~A.,  Harris C.~R.,  Archibald A.~M.,  Ribeiro
  A.~H.,  Pedregosa F.,  {van Mulbregt} P.,    {SciPy 1.0 Contributors} 2020,
  Nature Methods, 17, 261

\bibitem[\protect\citeauthoryear{Walker, Longhi \& Hays}{Walker
  et~al.}{1975}]{walker1975differentiation}
Walker D.,  Longhi J.,    Hays J.~F.,  1975, in Lunar and Planetary Science
  Conference Proceedings Vol.~6, Differentiation of a very thick magma body and
  implications for the source regions of mare basalts.
pp 1103--1120

\bibitem[\protect\citeauthoryear{{Wolf} \& {Bower}}{{Wolf} \&
  {Bower}}{2018}]{RTpress}
{Wolf} A.~S.,  {Bower} D.~J.,  2018, Physics of the Earth and Planetary
  Interiors, 278, 59

\bibitem[\protect\citeauthoryear{{Zhang} \& {Rogers}}{{Zhang} \&
  {Rogers}}{2022}]{Zhang22}
{Zhang} J.,  {Rogers} L.~A.,  2022, \apj, 938, 131

\bibitem[\protect\citeauthoryear{{Zilinskas}, {van Buchem}, {Miguel}, {Louca},
  {Lupu}, {Zieba} \& {van Westrenen}}{{Zilinskas} et~al.}{2022}]{Zilinkas22}
{Zilinskas} M.,  {van Buchem} C.,  {Miguel} Y.,  {Louca} A.,  {Lupu} R.,
  {Zieba} S.,    {van Westrenen} W.,  2022, arXiv e-prints, p. arXiv:2202.04759

\bibitem[\protect\citeauthoryear{{Zuckerman}, {Melis}, {Klein}, {Koester} \&
  {Jura}}{{Zuckerman} et~al.}{2010}]{Zuckerman2010}
{Zuckerman} B.,  {Melis} C.,  {Klein} B.,  {Koester} D.,    {Jura} M.,  2010,
  \apj, 722, 725

\end{thebibliography}


\appendix
\section{Entropy change of melting}\label{app: CC}

Here we demonstrate a thermodynamically consistent way of calculating $\Delta S$, the entropy change from solid to melt, which does not require an absolute entropy scale.

We assume that the volume and entropy of the mixed solid and melt phases are additive, meaning that
\begin{equation}
    V = V_l\phi + V_s(1-\phi) \text{ and } S = S_l\phi + S_s(1-\phi) \; .\label{eq: additive}
\end{equation}
Using this definition and the Maxwell relation
\begin{equation}
    \left.\diffp{V}{T}\right|_P = -\left.\diffp{S}{P}\right|_T  \label{eq: -dSP=dVT}
\end{equation}
one finds that
\begin{equation}
    \begin{split}
        \left.\diffp{V_l}{T}\right|_P\phi + \left.\diffp{V_s}{T}\right|_P (1-\phi) + \Delta V\left.\diffp{\phi}{T}\right|_P =  \\ -\left.\diffp{S_l}{P}\right|_T \phi -\left.\diffp{S_s}{P}\right|_T (1-\phi) -\Delta S \left.\diffp{\phi}{P}\right|_T \; .
    \end{split}
\end{equation}
If the melt and solid satisfy this Maxwell relation individually or if the latent heat terms dominate, which is generally the case, then
\begin{equation}
    \Delta V \left.\diffp{\phi}{T}\right|_P = -\Delta S \left.\diffp{\phi}{P}\right|_T \; . \label{eq: dS_dV relation}
\end{equation}

We then also assume that the melt fraction, $\phi$, is simply a function of $P$ and $T$, which is true in equilibrium. This means we can use the triple product rule and write
\begin{equation}
    \left.\diffp{T}{P}\right|_\phi = - \left.\diffp{\phi}{P}\right|_T /\left.\diffp{\phi}{T}\right|_P \; . \label{eq: cyclic}
\end{equation}

Combining \autoref{eq: dS_dV relation} and \autoref{eq: cyclic} gives
\begin{equation}
    \left.\diffp{T}{P}\right|_\phi = \frac{\Delta V}{\Delta S} \label{eq: genC-C}
\end{equation}
which is a generalised version of the Clausius-Clapeyron relation. This can then be used to calculate the entropy change by rearranging it to \autoref{eq: DeltaS} in the main text.

\section{Asymmetric boundary condition} \label{app: BC}
\subsection{Simplification of conduction-convection equation at constant gravity} \label{sec: cond_conv for BC}

In this section we derive a simple ODE that can give the vertical temperature pressure structure at a given angle $\theta$, assuming the region is vertically small and angular heat fluxes are small.

The equation of heat transport by conduction and convection in one dimension (Equations \ref{eq: conduction}-\ref{eq: totalF}) is given, in full, by
\begin{equation}
    F = -k \diff{T}{P} \diff{P}{r}  - \rho l u C_P \left( \diff{T}{P} - \left.\diffp{T}{P}\right|_\text{Ad} \right) \diff{P}{r} \; .
\end{equation}

Considering first the viscous case, so $u$ is given by \autoref{eq: highviscv} this becomes 
\begin{equation}
     F = -k \diff{T}{P} \diff{P}{r}  - \frac{\rho C_P \delta g l^4}{18 \nu T} \left( \diff{T}{P} - \left.\diffp{T}{P}\right|_\text{Ad} \right)^2 \left(\diff{P}{r}\right)^2 \; . \label{eq: cond_conv visc}
\end{equation}

Using $\diff{P}{r} = -\rho g$, we can rearrange this to
\begin{equation}
    A \left(\diff{T}{P}\right)^2 + \left(k\rho g - 2 A \left.\diffp{T}{P}\right|_\text{Ad} \right) \diff{T}{P} + A\left(\left.\diffp{T}{P}\right|_\text{Ad} \right)^2  = F
\end{equation}
where
\begin{equation}
    A \equiv \frac{\rho^3 C_P \delta g^3 l^4}{18 \nu T} \; .
\end{equation}
Assuming that $g$ is constant and density changes are small enough that we can take the mixing length, which will be the distance to the surface, to be $l = \frac{P}{\rho g}$, then the root of this quadratic equation is simply \begin{equation}
    \diff{T}{P} = \frac{ 2 A \left.\diffp{T}{P}\right|_\text{Ad} - k\rho g + \sqrt{(k\rho g)^2 + 4A\left( F - k\rho g \left.\diffp{T}{P}\right|_\text{Ad}\right)}}{2A} \label{eq: dTdP const g}
\end{equation}
i.e., an ODE for $T$ and $P$in the viscous limit. 

One cannot reach such a simplification when including the inviscid case in the same way as \S\ref{sec: heat_flow}. Fortunately, for this boundary model, smoothness is less of a priority than for the Henyey scheme, where it is desired for numerical convergence. Since the change occurs due to melting/crystallisation, a sharp change is also physically reasonable. We therefore simply change $\mathrm{d}T/\mathrm{d}P$ when the material becomes sufficiently inviscid. Furthermore, the conduction term is not important in the inviscid regime, since, generally, the material becomes viscous before it becomes conductive, as shown below. Therefore, using \autoref{eq: lowviscv} the inviscid gradient is given by
\begin{equation}
     \diff{T}{P} =  \left.\diffp{T}{P}\right|_\text{Ad} + \left(\frac{16F^2T}{\delta\rho^5 l^4 C_P^2 g^4}\right)^\frac{1}{3} \; . \label{eq: dTdP const g inviscid}
\end{equation}

The switch from viscid to inviscid occurs when the Reynolds number
\begin{equation}
    Re = \frac{u_\textit{visc} l }{\nu}
\end{equation}
becomes higher than a critical value of 9/8. This critical value is simply when the viscous and inviscid velocities (Equations \ref{eq: highviscv} and \ref{eq: lowviscv}) are equal. 

Thus, at any angle with an outward heat flux $F(\theta)$ we can work out the temperature--pressure structure using Equations \ref{eq: dTdP const g} and \ref{eq: dTdP const g inviscid}.

The statement that conduction is not important in the inviscid region can be demonstrated as follows.
The domination of conduction or convection is governed by the Peclet number 
\begin{equation}
    Pe = \frac{F_\text{conv}}{F_\text{cond}} = \frac{u l \rho C_P}{k} \label{eq: Peclet}
\end{equation}
where a high Peclet number corresponds to convection dominating heat transport. The second equality comes from considering when the temperature gradient is much higher than adiabatic. 

If one considers the ratio of the Peclet and Reynolds numbers in the viscous limit
\begin{equation}
    \frac{Pe}{Re} = \frac{\nu \rho C_P}{k}
\end{equation}
one sees that for any reasonable values the Peclet number is larger than the Reynolds number, meaning the fluid first becomes viscous and then conductive. 

\subsection{Fitting of luminosity-temperature function} \label{sec: fit BC}
To aid the convergence of the Henyey scheme it is desirable for the $T_0(L)$ relation to be smooth. However, the calculation described in \S\ref{sec: BCs} can produce numerical variation, particularly in the near-constant $T_0$ section (see \autoref{fig: T0_func}). 

We opted to fit a function to our result, motivated by limits to the equations' solutions. 

The solutions to \autoref{eq: dTdP const g} have a large temperature increase in the conductive region close to the surface, followed by a less steep, close to adiabatic increase once convection can dominate. The temperature $T_0$ can thus be approximated as 
\begin{equation}
    T_0 \approx T_s + \frac{F}{k \rho g} \min(P_\text{crit},P_0)
\end{equation}
where $\frac{F}{k \rho g}$ is the conductive gradient, and $P_\text{crit}$ is the point when conduction dominates. We can define this point using the Peclet number (\autoref{eq: Peclet}), which we can write as
\begin{equation}
    Pe  = \left( \frac{\rho^2 C_p \alpha g F l^4}{18 \eta k^2} \right)^{\frac{1}{2}} \; .
\end{equation}
Taking $l = \frac{P}{\rho g}$ again gives
\begin{equation}
    T_0 \sim T_s + \left(\frac{18 k Pe_\text{crit}^2}{C_P \alpha \rho^2}\right)^\frac{1}{4} F^\frac{3}{4}g^{-\frac{1}{4}} \eta^\frac{1}{4} 
\end{equation}
for the case where $P_\text{crit} < P_0$.
For low temperatures, $P_0 < P_\text{crit}$ so we take
\begin{equation}
    T_0 = T_s + A_1 F^{a_1} g^{b_1} \; , \; a_1\approx 1 \, , \, b_1\approx -1 \label{eq: BC_fitting 1}
\end{equation}
with $A_1$ a constant.
Once the system becomes partially melted the exponential part of the viscosity $\eta \propto \exp(-\alpha_\eta \phi)$ (see \autoref{eq: full_viscosity}) becomes the most important so
\begin{equation}
    T_0 \sim C\exp(-\lambda T_0) F^a_2 g^b_2
\end{equation}
with $C , \lambda , a_2 , b_2$ constants. We, therefore, fit the function
\begin{equation}
    T_0 = A_2 + a_2 \ln{F} + b_2 \ln{g} \label{eq: BC_fitting 2}
\end{equation}
to this region.

Around $\phi \approx \phi_c$ the viscosity, and so the flux, is a very strong function of temperature, meaning that temperature is approximately constant for a large range of fluxes, this region we approximate as a constant $T_\text{crit}$.

For high temperatures when the fluid is molten $\eta \sim \eta_l\left(\frac{\phi - \phi_c}{1 - \phi_c }\right)^{2.5}$. Since $\phi$ is linear in $T$ (for a given $P$, \autoref{eq: melt_frac}) this is not far from a power law in $T$ so we take
\begin{equation}
    T_0 = T_s + A_3 F^{a_3} g^{b_3} \; , \; a_3\approx \frac{3}{4} \, , \, b_3\approx -\frac{1}{4} \; . \label{eq: BC_fitting 3}
\end{equation}

We combine all these into a master formula of
\begin{equation}
    T_0 = \left(\left[\left\{f_1^{-\beta} + f_2^{-\beta}\right\}^\frac{\gamma}{\beta}  +  T_\text{crit}^{-\gamma} \right]^{-\frac{\alpha}{\delta}} + f_3^\alpha \right)^\frac{1}{\alpha} \label{eq: fit_BC}
\end{equation}
where $\alpha,\beta,\gamma$ are smoothing constants and $f_1$, $f_2$ and $f_3$ are given by Equations \ref{eq: BC_fitting 1}, \ref{eq: BC_fitting 2} and \ref{eq: BC_fitting 3}.

We fit the coefficients including the smoothing values using {\texttt{ scipy curve\_fit}}. The only value we do not fit is $T_\textit{crit}$ which we set to  1778.6. Coefficients for different substellar temperatures are shown in \autoref{tab: fit_params}.

As can be seen, $a_1$ and $b_1$ are very close to our predicted values of 1 and -1. Meanwhile, $a_3$ and $b_3$, show a not insignificant deviation from our predicted values of 0.75 and -0.25. This is not surprising, as we entirely neglected any viscosity dependence. The values of $\alpha$, $\beta$, $\gamma$ are high, reflecting sharp transitions between regions.

Our fit produces fractional errors on the calculated $T_0$ of $\lesssim 1.5\%$ which is probably well within errors in the physical parameters and heat transport assumptions.

\begin{table}
\resizebox{\columnwidth}{!}{
\begin{tabular}{|l|c|c|c|c|c|}\hline
\multirow{2}{*}{\textbf{Parameter}} & \multicolumn{5}{c|}{\textbf{Substellar tempertature [K]}}                  \\ \cline{2-6}
                                    & \textbf{2070} & \textbf{2190} & \textbf{2320} & \textbf{2460} & \textbf{2600} \\ \hline
$A_1$                               & \num{6.79e+04}      & 6.87e+04      & \num{6.86e+04}      & \num{6.92e+04}      & \num{6.96e+04}      \\\hline
$a_1$                               & 1.04          & 1.04          & 1.04          & 1.04          & 1.04          \\\hline
$b_1$                               & -1.04         & -1.04         & -1.04         & -1.04         & -1.04         \\\hline
$A_2$                               & 1860      & 1870      & 1880      & 1880      & 1880      \\\hline
$a_2$                               & 81.1          & 82.8          & 84.2          & 85.3          & 86            \\\hline
$b_2$                               & 17.2          & 13.9          & 13.3          & 14.7          & 12.4          \\\hline
$A_3$                               & 0.41          & 0.734         & 0.66          & 0.38          & 0.201         \\\hline
$a_3$                               & 0.522         & 0.485         & 0.498         & 0.545         & 0.597         \\\hline
$b_3$                               & -0.369        & -0.326        & -0.298        & -0.281        & -0.273        \\\hline
$\alpha$                            & 6.98          & 7.23          & 7.79          & 7.99          & 8.21          \\\hline
$\beta$                             & 9.94          & 10.3          & 10.4          & 10.2          & 10          \\\hline
\end{tabular}
}
\caption{Best fit parameters for our boundary condition function \autoref{eq: fit_BC}.}\label{tab: fit_params}
\end{table}


\bsp	
\label{lastpage}
\end{document}